\documentclass[a4paper, 11pt, twoside]{thesis}  
\graphicspath{{Figures/}}  

\usepackage{verbatim}  
\usepackage[utf8]{inputenc}

\usepackage[T1]{fontenc}
\usepackage{lmodern}

\usepackage{amsmath}
\usepackage{amsthm}
\usepackage{mathtools}
\usepackage{pgfplots}
\pgfplotsset{compat=newest}
\usepackage{tikz}
\usetikzlibrary{arrows,matrix,positioning,fit,decorations,backgrounds,calc}
\usepackage{qtree}
\usepackage{siunitx}
\usepackage{enumitem}
\usepackage{cases}
\usepackage{chngcntr}
\usepackage[multiple,symbol*,perpage]{footmisc}
\usepackage[section]{placeins}
\usepackage{relsize}
\usepackage[normalem]{ulem}
\usepackage{mathdots}
\usepackage{float}
\makeindex
\usepackage[totoc]{idxlayout}

\let\cleardoublepage\clearpage

\hypersetup{urlcolor=blue, colorlinks=true, hypertexnames=true}  

\usepackage{etoolbox}
\AtBeginEnvironment{figure}{}{}{}
\makeatletter

\preto{\@verbatim}{\topsep=0pt \partopsep=0pt }

\def\imod#1{\allowbreak\mkern5mu({\operator@font mod}\,\,#1)}

\def\tombstone{\tag*{\qedsymbol}}
\def\texttombstone{\hfill\qedsymbol}

\DefineFNsymbols*{chrissymbols}[math]{\dagger\ddagger\mathsection\mathparagraph\S\P\|}
\setfnsymbol{chrissymbols}

\counterwithout{figure}{chapter}

\newcommand{\eqnum}{\leavevmode\hfill\refstepcounter{equation}\textup{\tagform@{\theequation}}}

\newcommand\numberthis{\addtocounter{equation}{1}\tag{\theequation}}

\tikzset{%
  matrixbox/.style={rectangle,rounded corners,draw,thin,inner xsep=-2pt}
}
\tikzset{%
  matrixboxtwo/.style={rectangle,rounded corners,draw,thin,inner ysep=7pt}
}

\makeatother

\def\cpp{C{}\texttt{++}}

\def\code#1{%
{%
\texttt{#1}%
}%
}

\def\DKSS{De, Kurur, Saha and Saptharishi}
\def\OMUL{\code{OMUL}}
\def\KMUL{\code{KMUL}}
\def\T3MUL{\code{T3MUL}}
\def\QMUL{\code{QMUL}}
\def\SMUL{\code{SMUL}}
\def\DMUL{\code{DKSS\_MUL}}
\def\N{\mathbb{N}}
\def\Z{\mathbb{Z}}
\def\F{\mathbb{F}}
\def\Zpc{\Z/p^z\Z}
\def\Zp{\Z/p\Z}
\def\R{\mathcal{R}}
\def\S{\mathcal{S}}
\def\P{\mathcal{P}}

\def\half{\frac{1}{2}}

\newcommand{\ndiv}{\nmid}
\newcommand{\divides}{\mid}

\newtheoremstyle{mystyle}  
  {}           
  {0pt}        
  {\itshape}   
  {0pt}        
  {\bfseries}  
  {.}          
  { }          
  {}           
\theoremstyle{mystyle}
\newtheorem*{mydef}{Definition}
\newtheorem*{mythm}{Theorem}
\allowdisplaybreaks[1]

\begingroup\lccode`~=`_
\lowercase{\endgroup
  \protected\def~{\ifmmode\sb\else\textunderscore\fi}
}
\catcode`_=\active

\begin{document}

\frontmatter     

\fancyhead[RE,LO]{}
\fancyhead[LE,RO]{\thepage}

\pagestyle{empty}  

\title      {Fast Multiplication of Large Integers:\\ Implementation and Analysis of the DKSS Algorithm}
\authors    {Christoph Lüders}
\authtown   {Bonn}
\authemail  {
\includegraphics{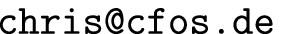}
}
\addresses  {\groupname\\\deptname\\\univname}  
\date       {\today}
\subject    {}
\supervisor {Prof.~Dr.~Michael Clausen}

\maketitle

\setstretch{1.0}  

\pagestyle{empty}
\cleardoublepage

\fancyhead[RE,LO]{Quotes}
\fancyhead[LE,RO]{\thepage}

\null\vfill

Ford: \textit{``What do you get if you multiply six \ldots~by nine --- \emph{by nine?} Is that it?''} \\
Arthur: \textit{``That's it. Six by nine: forty-two! I always said that there is something
fun\-da\-men\-tally wrong about the universe.%
''} 
\begin{flushright}
The Hitchhiker's Guide to the Galaxy radio series, episode 6
\end{flushright}
\bigskip
\bigskip
\bigskip

\textit{``Warum hat er es denn so eilig?''}
\begin{flushright}
N.~N.\ about Arnold Schönhage and his fast multiplication
\end{flushright}

\vfill\vfill\vfill\vfill\vfill\vfill\vfill\vfill\null
\cleardoublepage

\pagestyle{empty}
\cleardoublepage

\chapter{Acknowledgments}
\fancyhead[RE,LO]{Acknowledgments}
\fancyhead[LE,RO]{\thepage}

First and foremost, I wish to express my sincerest gratitude to my advisor Prof.\ Dr.\ Michael Clausen
for his help, understanding, advice and encouragement.
It was a pleasure to work under his guidance.

Furthermore, I wish to thank Karsten Köpnick and Nicolai Dubben
for their proofreading and fruitful discussions.
Their comments and questions were greatly appreciated.

I thank Prof.\ Dr.\ Arnold Schönhage for inspiring conversations
many years ago when I first started to get interested in long numbers and also
for his friendly comments on this thesis.

I am grateful to the authors of \LaTeX{} for their excellent typesetting system and
the authors of PGF/TikZ and PGFPlots for their packages to produce beautiful graphics.
Furthermore, I thank the many contributors on \url{tex.stackexchange.com}
for the fast and excellent help in times of need.

I thank Martin Winkler and our mutual business for understanding and support of my studies,
especially when time got tight.

Many others have supported me, I just want to mention
Dirk Eisenack, Heidi Förster, Goran Rasched and Susanne Röhrig.
I am happy and grateful for their interest in, suggestions for and support of my work.

Also, I thank Anne-Sophie Matheron for asking me one too many times why I
didn't finish my diploma.

Lastly, I thank my family and especially my mother for her boundless faith in me.

\cleardoublepage

\pagestyle{fancy}  

\lhead{\emph{Contents}}  
\tableofcontents  

\lhead{\emph{List of Figures}}  
\listoffigures  

\clearpage  
\lhead{\emph{Symbols & Notation}}  
\listofnomenclature{lll}  
{
$\mathbb{R}$               & field of real numbers  \\
$\mathbb{C}$               & field of complex numbers  \\
$\Z$                       & ring of all integers: $0, \pm 1, \pm 2, \ldots$  \\
$\N$                       & $\{ x \in \Z \; | \; x > 0 \}$  \\
$\N_0$                     & $\{ x \in \Z \; | \; x \ge 0 \}$  \\
$[a:b]$                    & $\{ x \in \Z \; | \; a \le x \le b \}$, for $a$, $b \in \Z$  \\
$a/bc \cdot d$             & $(a/(bc)) \cdot d$  \\
$a \divides b$             & $a \in \Z$ divides $b \in \Z$  \\
$a \ndiv b$                & $a \in \Z$ does not divide $b \in \Z$  \\
$(a,b)$                    & greatest common divisor of $a,$ $b \in \Z$  \\
$\lfloor a \rfloor$        & $\max \{ x \in \Z \; | \; x \le a \}$, floor of $a \in \mathbb{R}$  \\
$\lceil a \rceil$          & $\min \{ x \in \Z \; | \; x \ge a \}$, ceiling of $a \in \mathbb{R}$  \\
$\log x$                   & logarithm of $x$ to base $2$  \\
$\log^a x$                 & $(\log(x))^a$  \\
$\exp^n_b(x)$              & $n$-times iterated exponentiation of $x$ to base $b$, cf.~page~\pageref{iterexp}  \\
$\log^* x$                 & iterated logarithm of $x$ to base 2, cf.~page~\pageref{logstar}  \\
$a$ has degree-bound $n$   & $\deg(a) < n$, $a$ a polynomial, $n \in \N$  \\
$g(n) \in O(f(n))$         & $\exists c \ge 0, n_0 \in \N_0: \forall n \ge n_0: |g(n)| \le c \cdot f(n)$  \\

}

\mainmatter   
\pagestyle{fancy}  

\edef\avgsmulc{0.3159}
\edef\evenexp{246347190125373091264841517557606054462633639158644889995199340572211391820569991344480586839243247279808688429619098327164549721389008802552643321856}
\edef\evenexpexp{149}
\edef\eventwoexpexp{498}
\edef\avgdmulc{1.2727}

\chapter{Introduction}
\label{chapter1}
\fancyhead[RE,LO]{Chapter 1. \emph{Introduction}}
\fancyhead[LE,RO]{\thepage}

Multiplication of integers is one of the most basic arithmetic operations.
Yet, if numbers get larger,
the time needed to multiply two numbers increases as well.
The naive method to multiply requires $c \cdot N^2$ bit-operations to multiply numbers
with $N$ digits, where $c$ is some constant.%
\footnote{Usually, the constant is omitted and instead of $c \cdot N^2$ we write $O(N^2)$.
The constant $c$ is hidden in the $O(\ldots)$.}
For large numbers this process soon becomes too slow and faster means are desirable.

Fortunately, in the 1960s methods were discovered that lowered the number of operations
successively until in 1971 Schönhage and Strassen \cite{SS1971} found a technique that only requires
$O(N \cdot \log N  \cdot \log \log N)$ bit-operations.%
\footnote{The logarithm function to base~10, $\log_{10} N$, is approximately the number of decimal digits
of $N$. So if $N$ is multiplied by 10, the logarithm just increases by 1.
This is to show how slowly it is growing.}
This algorithm was the asymp\-toti\-cally fastest known method to multiply until
in 2007 Fürer \cite{Fuerer2007} found an even faster way.
\emph{Asymptotically} means that the algorithm was the fastest,
provided numbers are long enough. Elaborate algorithms
often involve some costs for set-up that only pay off if the inputs are long enough.%
\footnote{Think of finding names in a stack of business cards: if you sort the cards first,
you can find a name quickly, but it is only worth the effort if you search
for a certain number of names.}

Fürer's algorithm inspired \DKSS{} to their multiplication method \cite{De2008}, published in 2008,
and a follow-up paper \cite{De2013}, the latter being discussed in this thesis
and which I call \emph{DKSS multiplication}.
Both Fürer's and DKSS' new algorithms require $N \cdot \log N \cdot 2^{O(\log^* N)}$ bit-operations, where $\log^*N$
(pronounced ``log star'') is the number of times the logarithm function has to be applied to get a value~$\le 1$.

However, Fürer conjectured that his new method only becomes faster than
Schönhage and Strassen's algorithm for ``astronomically large numbers''
\cite[sec.~8]{Fuerer2009}.
Feeling unhappy about this vague assessment, I implemented the DKSS algorithm and
compared it to Schönhage and Strassen's method to see if or when any improvement in speed
could be achieved in practice. Both algorithms use only integer operations, in contrast to
Fürer's algorithm that uses floating point operations.

\bigskip

The ability to multiply numbers with millions or billions of digits is not only
academically interesting, but bears much practical relevance.
For example, number theoretical tasks like primality tests require fast multiplication of potentially
very large numbers. Such calculations can be performed nowadays with computer algebra systems like
Magma, Maple, Mathematica, MATLAP, or Sage.
Calculation of $\pi$ or $e$ to billions of digits or computing billions of
roots of Riemann's zeta function are other fields that requires
fast large number multiplication \cite[sec.~8.0]{Gathen2013}.

Also, fast multiplication is an important building block of a general library
for arithmetical operations on long numbers, like
the GNU Multiple Precision Arithmetic Library \cite{Granlund2014}.
Addition and subtraction are not hard to implement and many of the more complex tasks ---
like inversion, division, square root, greatest common divisor --- revert back
to multiplication, cf.\ \cite{Gaudry2007}. Once these operations are implemented for integers, they can
be used to provide arbitrary-precision arithmetic for floating point numbers
that attenuate rounding problems, cf.\ \cite{Ghazi2010}.

\index{Kronecker-Schönhage\enskip substitution}%
Another big application for multiplication of long numbers is polynomial multiplication with integer coefficients,
since it can be reduced to one huge integer multiplication through Kronecker-Schönhage substitution
\cite[sec.~2]{Schoenhage1982}.
If (multivariate) polynomials are of high degree, the resulting integers can become very long and
fast means for multiplication are essential. Factoring of polynomials is also an important field
of activity, see \cite{Gaudry2007}.

\bigskip

All elaborate multiplication methods use some sort of \emph{fast Fourier transform} (FFT) at their core.
The main idea behind all FFT multiplication methods is to break a long number into smaller pieces
and interpret those pieces as coefficients of a polynomial.
Since a polynomial of degree less than $2n$ is uniquely determined by its sample values
for $2n$ pairwise different sample points, two polynomials of degree less than $n$ can be multiplied like this:%
\footnote{Since the resulting polynomial is the product of its two factors, it has degree $2n-2$.
Therefore, at least $2n-1$ sample points are needed to recover the result.}
\begin{enumerate}
\item Evaluate both polynomials at the same $2n$ sample points,
\item multiply the sample values pairwise, and
\item interpolate the polynomial from the sample value products.
\end{enumerate}
The FFT is ``fast'', since it computes $n$ sample values with only $O(n \cdot \log n)$
operations, which is an enormous advance from the naive approach and its $O(n^2)$ operations.
This method was already known by Gauss in 1805 \cite{Heideman1985},
but rediscovered by Cooley and Tukey in 1965 \cite{Cooley1965}
and then revolutionized computation.

The method by Schönhage and Strassen breaks numbers of $N$~bits into pieces of length $O(\sqrt{N})$ bits.
Furthermore, it is cleverly designed to take advantage of the binary nature of today's
computers: multiplications by~2 and its powers are particularly simple and fast to perform.
This is why it has not only held the crown of the asymptotically fastest
multiplication algorithm for over 35~years, but is also in widespread practical use today.

The new DKSS multiplication has a better asymptotic time bound, but its structure
is more complicated. This elaborated structure allows input numbers to be broken
into pieces only $O((\log N)^2)$~bits small. However, the arithmetic operations are more
costly. The purpose of this thesis is to see if or when DKSS multiplication
becomes faster than Schönhage-Strassen multiplication in practical applications.

\bigskip

Chapter~\ref{chapter2} of this thesis presents an overview of multiplication algorithms from the naive
method to techniques that provide a good trade-off if numbers are
of medium length (like Karatsuba's method in Section~\ref{kmul}).
The fast Fourier transform is introduced in Section~\ref{fft} and is followed by
a detailed description of Schönhage and Strassen's procedure in Section~\ref{smul}.
All methods were implemented and their run-times are determined theoretically,
measured in practice and illustrated graphically.
Schönhage and Strassen's algorithm is more thoroughly analyzed in respect
of its run-time, memory consumption and possible areas for improvement.

In Chapter~\ref{chapter3} the DKSS algorithm is explained in detail and its run-time
is analyzed theoretically. Section~\ref{dkss:diffs}
describes the differences between my implementation and the paper \cite{De2013}.

Chapter~\ref{chapter4} presents details of the implementation and illustrates its run-time (Section~\ref{dkss:impl:exectime}),
memory requirements (Section~\ref{dkss:impl:memory}) and source code complexity (Section~\ref{dkss:impl:codesize})
in comparison to Schönhage and Strassen's method both in numbers and graphics.
Section~\ref{dkss:impl:augur} estimates the crossover point at which both algorithms
become equally fast.

Lastly, Chapter~\ref{chapter5} sums up the results and shows
possible areas for improvement together with an assessment of their potential.

\bigskip

In this version of my thesis the typesetting has been modified to produce a more
concise layout and some minor errors have been corrected.

\chapter{Overview of Established Algorithms}
\label{chapter2}
\fancyhead[RE,LO]{Chapter 2. \emph{Overview of Established Algorithms}}
\fancyhead[LE,RO]{\thepage}

This chapter covers the well established algorithms to multiply large numbers,
starting with the naive method. Methods for medium-sized numbers are
discussed, the fast Fourier transform is introduced
and Schönhage-Strassen multiplication is presented in detail.
But first, some basic remarks about storage of large numbers and
memory allocation for temporary storage are necessary.

\section{Representation of Numbers}

I assume my implementation is running on a binary computer and
the machine has a native \emph{word size} of $w$ bits, so it can
hold nonnegative integer values $0 \ldots 2^w-1$ in its general purpose registers.
We call this unit a computer \emph{word}.
Today, the most common word
sizes are 32 and 64 bits, therefore a machine register can hold integers between 0 and $2^{32}-1 = \num{4294967295}$ or
$2^{64}-1 = \num{18446744073709551615}$, respectively.

If we want to do calculations with numbers that exceed the aforementioned range, we must
use some multiple precision representation for them. If we call $W \coloneqq 2^w$ the \emph{wordbase}\index{Wordbase}, we can write any
nonnegative integer $a < W^n$ as $a = \sum_{i=0}^{n-1} a_i W^i$, with $a_i \in [0 : W-1]$.
We can view this as representing $a$ with $n$ words or ``digits'' in base $W$.

In my implementation a nonnegative number $a < W^n$ is represented by an array of $n$ words
as $a = (a_0, a_1, \ldots, a_{n-2}, a_{n-1})$. The words are ordered with increasing
indices in main memory. This ordering is called \emph{little-endian}.
It was a design choice to use this ordering: cache prefetching used to
work better in forward loops, which are often used due to carry propagation.
Modern CPUs seem to have remedied this problem.

The same ordering is used by Schönhage et al.\ \cite[p.~7]{Schoenhage1994} as well as
in GMP, \emph{The GNU Multiple Precision Arithmetic Library} \cite[sec.~16.1]{Granlund2014} and MPIR, a prominent GMP fork \cite{MPIR}.
Interestingly, Zuras \cite{Zuras1994} describes that storing numbers as big-endian worked
better on his compiler and architecture.

The \emph{i-code} in \cite[p.~6]{Schoenhage1994} stores the length $n$ of the number after the most significant
word in main memory. In contrast, my implementation keeps the length as a separate
integer and provides both pointer to array and length as arguments on function calls.

Please note that we can usually pad any number with zero words on the upper end
without influencing the result of operations (except for possible zero-padding
of the result).  It is a small waste of memory and processing time, but
can simplify implementation of algorithms, for example, if an algorithm
expects the length of a number to be even.

Negative numbers are represented as the two's complement of their absolute value.
I followed the example of the i-code from \cite{Schoenhage1994} in this design decision.
It seems like a sensible choice, since execution time of simple operations like addition and subtraction
benefit from this representation, whereas more elaborate operations like multiplication
can afford the slight increase in execution time if negative numbers are being handled.

If negative numbers are handled and padding takes place, they have to be padded with
all binary ones, that is, words with binary value $-1$.  The most significant bit acts as sign bit if
a number is interpreted as a signed value.

\section{Memory Management}
\label{memmgt}

For all but the most basic functions we will need some temporary memory. To make routines
fast, it is important that storage can be allocated and freed quickly.
This forbids the use of the regular C-style \code{malloc()} or {\cpp} \code{new} (which is
just a wrapper for the former). C-style \code{malloc()} is designed to allow memory of
different sizes to be allocated and freed at random and still maintain low fragmentation;
many implementations are even thread-safe.

Since lifetime of temporary storage in our algorithms ends when a called function returns, we can
use a stack-like model for temporary memory, which greatly simplifies
the design of the allocator, makes it fast and doesn't need any locking.
Plus, it has the added benefit of good cache locality.
\index{Region-based memory management}
This is known as \emph{region-based memory management}.
In my code, this allocator is called \code{tape_alloc}.

To keep allocated memory continuous, every time memory is needed the allocator allocates more than is requested
and records the total amount of memory allocated. When afterwards all memory is freed and later on a new allocation
request is made, the allocator will right away allocate the total amount of memory used last time.
The idea is that since algorithms often involve multiple calculations that handle long numbers
in the same size-range, upcoming memory requirements will be as they were in the past.

Schönhage et al.\ implemented their algorithms on a
hypothetical Turing machine called \emph{TP} with six variable-length tapes
and a special assembly language-like instruction set called \emph{TPAL}
(see \cite{Schoenhage1994} and \cite{SchoeTP}). Of course,
this machine has to be emulated on a real computer, so TPAL instructions are translated
to C or assembly language for the target machine. Thus the tape-like structure of memory is retained.

The GMP library allocates temporary memory on the stack with \code{alloca()}. This
should be fast and thread-safe, since no locking is required.

\section{Ordinary Multiplication}
\index{Ordinary multiplication (\OMUL{})}

All of us have learned to multiply with pencil and paper in school. This is often referred to
as \emph{ordinary multiplication} or \emph{grade school multiplication}.
The implementation of it is called \OMUL{} (this name and others are inspired
by \cite{Schoenhage1994}).

Suppose we want to multiply two nonnegative integers $a$ and $b$ with lengths of $n$ and $m$~words, respectively,
to compute the product $c \coloneqq ab$ with length $n+m$.
To do that we have to multiply each $a_i$, $i \in [0 : n-1]$ with each $b_j$, $j \in [0 : m-1]$ and add the product to $c_{i+j}$,
which has to be set to zero before we start. Plus, there has to be some carry propagation.

In Python 3.x, our \OMUL{} algorithm looks like this.%
\footnote{The coding style is very un-pythonic and should only serve for explanation.}
I have left out the carry propagation
here, since this example only serves to show the principle. The {\cpp} example will be more specific.

\lstset{aboveskip=0.5cm}
\begin{lstlisting}[language=Python]
def omul(a, b):
   c = [0] * (len(a) + len(b))         # initialize result with zeros
   for i in range(0, len(a)):          # cycle over a
      for j in range(0, len(b)):       # cycle over b
         c[i+j] += a[i] * b[j]         # elementary mul and add
   return c
\end{lstlisting}

This Python implementation hides an important implementation detail: If a multiple precision number is
made up of words and these are the same size as a processor register, then the product of two
such words will be twice the size of a processor register!
Our code must be able to handle this double-sized result.
This is not a problem in the Python code above, since Python's
\code{int} type is multiple precision by itself. A similar function in {\cpp} shows more of that detail:

\begin{lstlisting}[language=C++]
void omul(word* c, word* a, unsigned alen, word* b, unsigned blen) {
   memset(c, 0, (alen+blen) * sizeof(word));          // set result to zero
   for (unsigned i=0; i<alen; ++i) {                  // loop over a[i]'s
      word carry = 0;                                 // for overflow
      unsigned j = 0;
      while (j < blen) {                              // loop over b[j]'s
         carry = muladdc(c[i+j], a[i], b[j], carry);
         ++j;
      }
      c[i+j] = carry;
   }
}
\end{lstlisting}

The type \code{word} is a placeholder for an unsigned integer type with the size of a processor word or smaller.
The interesting part happens in the function \code{muladdc()}: \code{a[i]} and \code{b[j]}
get multiplied, the input carry \code{carry} and the already computed result \code{c[i+j]} are added to the product, the lower part
of the result is written back to memory (into \code{c[i+j]}) and the higher part of the
result is saved in \code{carry} to be handled in the next iteration.

We have not yet addressed the problem of the double-sized multiplication result. We have two
choices here: either use a \code{word} type that is only half the processor word size, so
the product can be stored in a full processor word, or use some special function to get both
the high and low part of a full sized multiplication in two separate variables. Luckily, modern compilers offer
an intrinsic function for that and compile good code from it.
The other option is still available, but takes about 60~\% more time here for inputs of the same bit-size.%
\footnote{All timings are expressed in processor cycles and were done on an Intel Core i7-3770 CPU in
64-bit mode running Windows 7. Appendix~\ref{tech} describes the test setup in detail.}

For a 64-bit \code{word} type in Microsoft Visual {\cpp}, the \code{muladdc()} function looks like this:

\begin{lstlisting}[language=C++]
typedef unsigned __int64 uint64;                // keep it short
uint64 muladdc(uint64& mem, uint64 a, uint64 b, uint64 carry_in) {
   uint64 hiprod;
   uint64 lowprod = _umul128(a, b, &hiprod);    // intrinsic function
   hiprod += addc(mem, lowprod, carry_in);
   return hiprod;                               // carry out
}

uint64 addc(uint64& mem, uint64 v, uint64 carry_in) {
   uint64 r1 = mem + v;
   uint64 carry_out = r1 < v;                   // overflow?
   uint64 r2 = r1 + carry_in;
   carry_out += r2 < carry_in;                  // overflow?
   mem = r2;
   return carry_out;
}
\end{lstlisting}

Again, we have to wrestle with word size limitation when handling
overflow from addition in \code{addc()}. Unfortunately, Microsoft's
{\cpp} compiler doesn't offer a way to read the processor's carry flag. So, we have to do an additional comparison
of the result with one of the inputs to determine overflow
\cite[p.~29]{Warren2002}. The resulting code is surprisingly fast, despite the superfluous comparison.

The total run-time of \OMUL{} is easily determined: we have to do $nm$ word-to-doubleword multiplications,
since each $a_i$ has to be multiplied by each $b_j$. The number of additions depends on the implementation:
the Python version has $nm$ additions, but they are at least triple-size, since the carries accumulate.
The {\cpp} version has four word-sized additions per multiplication.

In either case, the total run-time is $O(nm)$, and assuming $m=n$ it is $O(n^2)$.

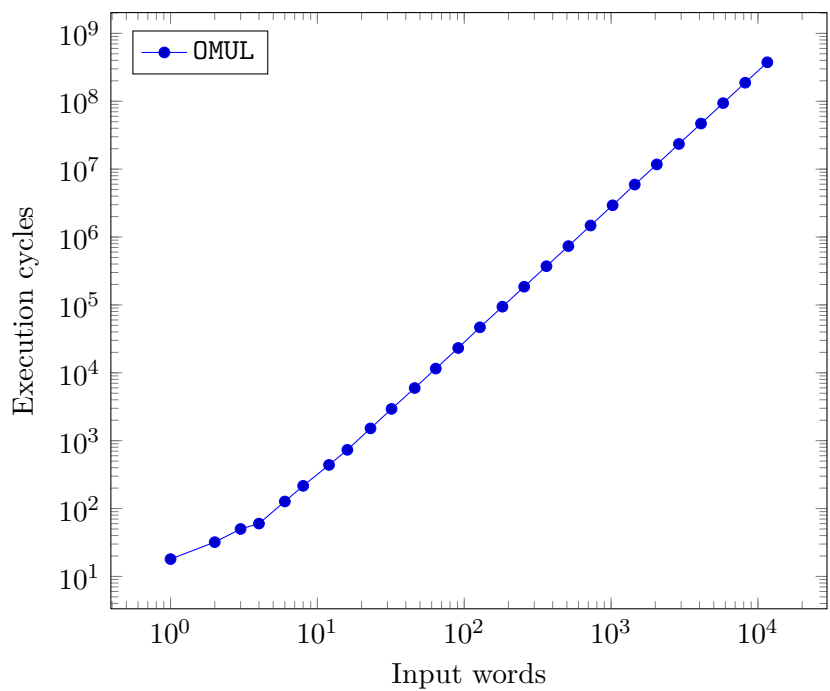
\begin{figure}[b]
\bigskip
\centering

\begin{tikzpicture}
\begin{loglogaxis}[
   width=11cm,
   xlabel={Input words},
   ylabel={Execution cycles},
   legend pos=north west]

\addplot table[x index=0, y index=1] {omul-speed.txt};
\addlegendentry{\OMUL{}}

\end{loglogaxis}
\end{tikzpicture}

\caption{Execution time of \OMUL{}}
\label{omulgraph}
\end{figure}

This is the ``classical'' time bound and even in 1956 it was still conjectured to be optimal, since no one had found
a faster way to multiply for more than four millennia \cite{Karatsuba1995}.

Figure~\ref{omulgraph} shows a double-logarithmic graphical display of execution times in processor cycles for different input sizes. Observe the slight
bend at the beginning, which shows the constant costs of calls and loop setup. Apart from that the graph is
very straight, which shows that caching has no big influence, even though the highest input sizes well exceed
the level 1 and level 2 cache sizes on the test machine.

\FloatBarrier
\section{Karatsuba Multiplication}
\index{Karatsuba multiplication (\KMUL{})}
\label{kmul}

Let $a$, $b < W^{2n}$ be two nonnegative integers, that is, both numbers consist of maximum $2n$ words.
We are looking for a faster way to multiply both numbers to get their product $c = ab < W^{4n}$.

We can ``cut'' both numbers in half, that is, express them as
\[a = a_0 + a_1 W^n  \;\;\;\; \text{and} \;\;\;\; b = b_0 + b_1 W^n, \numberthis \label{kcut} \]
with $a_0$, $a_1$, $b_0$, $b_1 < W^n$. The classical approach to calculate the full product from its four half-sized inputs is
\begin{align*}
ab &= (a_0 + a_1 W^n)(b_0 + b_1 W^n) \\
   &= a_0 b_0 + (a_0 b_1 + a_1 b_0) W^n + a_1 b_1 W^{2n}.  \numberthis \label{oprod}
\end{align*}
This way, we can break down a single $2n$-word multiplication into four $n$-word multiplications.
Unfortunately, we don't gain any speed by this.

\bigskip
In 1960 Karatsuba found a faster way to multiply long numbers \cite{Karatsuba1962}. The following slightly
improved version is due to Knuth \cite[p.~295]{Knuth1997}. The implementation of it is called \KMUL{}.

First, we compute the following three $n$-word multiplications
\begin{align*}
P_0 &= a_0 b_0   \\
P_1 &= (a_0-a_1) (b_0-b_1)  \\
P_2 &= a_1 b_1
\end{align*}
and use these three ``small'' products to recover the full product with only some
extra additions and subtractions plus shifts
(multiplications by powers of $W$):
\begin{align*}
ab = {} & P_0 (1+W^n) - P_1 W^n + P_2 (W^n+W^{2n})   \numberthis \label{kprod} \\
   = {} & a_0 b_0 (1+W^n) - (a_0-a_1) (b_0-b_1) W^n + a_1 b_1 (W^n+W^{2n})    \\
   = {} & a_0 b_0 + a_0 b_0 W^n - a_0 b_0 W^n + a_0 b_1 W^n + a_1 b_0 W^n - a_1 b_1 W^n + {}  \\
        & a_1 b_1 W^n + a_1 b_1 W^{2n}  \\
   = {} & a_0 b_0 + (a_0 b_1 + a_1 b_0) W^n + a_1 b_1 W^{2n}.   \tombstone
\end{align*}
It looks like more work, but it is a real improvement. Since ordinary multiplication runs in $O(n^2)$, saving
multiplications at the cost of additions, subtractions and shifts, which can be done in linear time, is a good deal in itself.
But if we use Karatsuba's algorithm recursively, we can even achieve a time bound of $O(n^{\log 3}) \approx O(n^{1.585})$.

We are going to prove this bound by induction.
Denote $T(n)$ the time it takes to multiply two $n$-word numbers.
We know that we can reduce a $2n$-word multiplication to three $n$-word multiplications and
some operations with linear run-time. Furthermore, we have to assign some cost to $T(1)$. So
\begin{align*}
T(1)  &= c,  \\
T(2n) &= 3T(n) + 2cn.
\end{align*}
We are going to show that
\[T(n) = 3cn^{\log 3} - 2cn.\]
This proof originates from \cite[p.~63]{AHU1974}.
It is easy to check the induction basis: $T(1) = 3c\cdot1^{\log 3}-2c \cdot 1 = c$. Next, we have to check the induction step:
\begin{align*}
T(2n) &= 3T(n) + 2cn \\
      &= 3(3cn^{\log 3} - 2cn) + 2cn \\
      &= 3c(3n^{\log 3}) - 6cn + 2cn \\
      &= 3c(3n^{\log 3}) - 2c(2n) \\
      &= 3c(2^{\log 3}n^{\log 3}) - 2c(2n) \\
      &= 3c(2n)^{\log 3} - 2c(2n).   \tombstone
\end{align*}
If we implement this procedure, we first compute the three products $P_0$, $P_1$, $P_2$ and
then use \eqref{kprod} to shift and
add up the small products to get the full product. That means,
we need some temporary storage for the small products and for the two factors
that make up $P_1$.

To compute the two factors $(a_0-a_1)$ and $(b_0-b_1)$ we would like to
avoid working with negative numbers, to keep things simple.
To that end I use a knack (borrowed from GMP) and compare the minuend and subtrahend of the subtraction,
always subtract the smaller from the larger and keep the sign bit in an extra variable.
The implementation accommodates for the sign bit later when it re-assembles the three sub-products.

The mentioned ideas look like this when coded in {\cpp}:

\begin{lstlisting}[language=C++]
void kmul(word* r, word* a, unsigned alen, word* b, unsigned blen) {
   if (alen < blen) {                              // b must not be longer than a
      swap(a, b),                                        // swap pointers
      swap(alen, blen);
   }
   if (blen < kmul_thresh) {                             // inputs too short?
      omul(r, a, alen, b, blen);                         // use omul
      return;
   }
   unsigned llen = blen / 2;                             // low part length
   unsigned ahlen = alen - llen;                         // a high part length
   unsigned bhlen = blen - llen;                         // b high part length

   // compute r0 = a0 * b0: this will lie in 'r' on index 0..llen-1
   kmul(r, a, llen, b, llen);
   // compute r2 = a1 * b1: this will lie in 'r' on index 2*llen..alen+blen-1
   kmul(r+2*llen, a+llen, ahlen, b+llen, bhlen);

   // allocate temporary space for differences and third mul
   tape_alloc tmp(4*ahlen + 1);
   word* sa = tmp.p;
   word* sb = tmp.p + ahlen;
   word* ps = tmp.p + 2*ahlen;

   // subtract values for later multiplication
   bool asign = compu_nn(a+llen, ahlen, a, llen) < 0;    // asign set if a1 < a0
   if (asign) subu(sa, ahlen, a, llen, a+llen, ahlen);   // a0 - a1 > 0
   else subu(sa, ahlen, a+llen, ahlen, a, llen);         // a1 - a0 >= 0

   bool bsign = compu_nn(b+llen, bhlen, b, llen) < 0;    // bsign set if b1 < b0
   if (bsign) subu(sb, ahlen, b, llen, b+llen, bhlen);   // b0 - b1 > 0
   else subu(sb, ahlen, b+llen, bhlen, b, llen);         // b1 - b0 >= 0

   // multiply both absolute differences
   unsigned plen = 2*ahlen + 1;                          // there can be a carry
   kmul(ps, sa, ahlen, sb, ahlen);
   ps[plen-1] = 0;

   // compute middle result
   if (asign == bsign) subu_on_neg(ps, plen, r, 2*llen); // ps = r0 - ps
   else addu_on(ps, plen, r, 2*llen);                    // ps += r0
   addu_on(ps, plen, r + 2*llen, ahlen + bhlen);         // ps += r2
   // add the final temp into the result
   addu_on(r+llen, ahlen + blen, ps, plen);
}
\end{lstlisting}

The code checks if input sizes suggest \OMUL{} will be faster and if so,
calls it instead. This is because \KMUL{} is \emph{asymptotically} faster than \OMUL{}, but not so for small
input lengths. Obviously, \KMUL{} is more complicated than \OMUL{}, as it uses several calls to add and subtract,
conditional branches and temporary memory. All this takes its time compared to a very streamlined
double-loop structure of \OMUL{} that modern processors are really good at executing.

\begin{figure}[b]
\bigskip
\centering

\begin{tikzpicture}
\begin{loglogaxis}[
   width=11cm,
   xlabel={Input words},
   ylabel={Execution cycles},
   legend pos=north west]

\addplot table[x index=0, y index=1] {omul-speed.txt};
\addlegendentry{\OMUL{}}
\addplot table[x index=0, y index=1] {kmul-speed.txt};
\addlegendentry{\KMUL{}}

\end{loglogaxis}
\end{tikzpicture}

\caption{Execution time of \KMUL{}}
\label{kmulgraph}
\end{figure}

To achieve maximum performance we have to find the input length where \KMUL{} starts to be faster
than \OMUL{}. This is called the \emph{crossover point} or
\emph{threshold value}. The crossover point depends on the processor architecture, memory speed and
efficiency of the implementation. To find the crossover point we have to benchmark both algorithms
against one another at various input lengths.

\begin{figure}
\bigskip
\centering

\begin{tikzpicture}
\begin{axis}[
   width=11cm,
   xlabel={Input words},
   ylabel={Execution cycles},
   scaled ticks=false,
   legend pos=north west]

\addplot table[x index=0, y index=1] {omul-small-speed.txt};
\addlegendentry{\OMUL{}}
\addplot table[x index=0, y index=1] {kmul-small-speed.txt};
\addlegendentry{\KMUL{}}

\end{axis}
\end{tikzpicture}

\caption{Execution time of \KMUL{} (close-up)}
\label{kmulgraphtwo}
\end{figure}
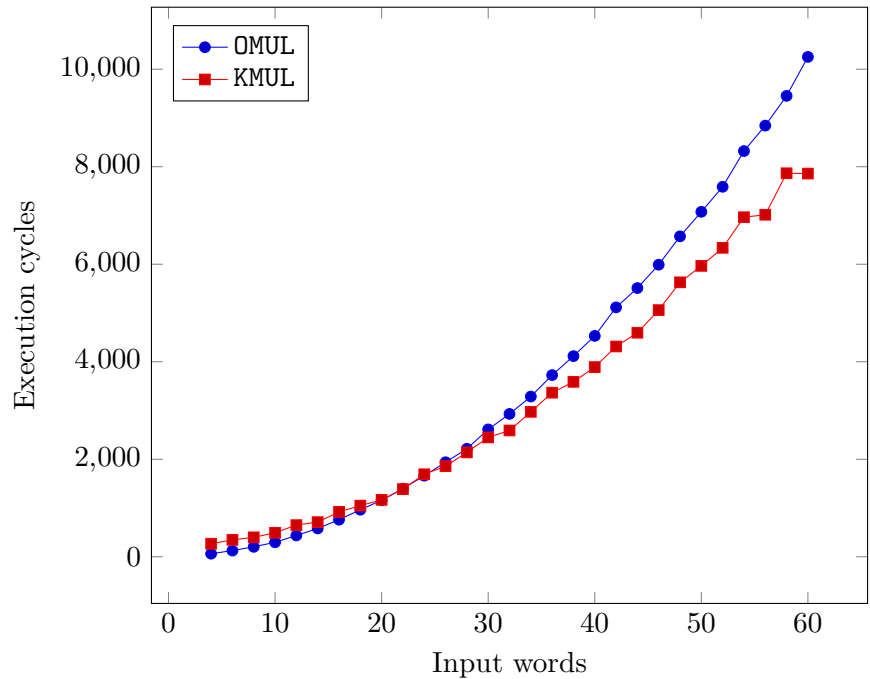

Figure~\ref{kmulgraph} shows the timings of \OMUL{} and \KMUL{}. We can see that KMUL is faster
the longer the inputs are (with an input length of \num{10000} words \KMUL{} is about nine times faster than \OMUL{}),
but in the low ranges it is slower than \OMUL{}.

To have a better look at the crossover point, Figure~\ref{kmulgraphtwo} has linear scaling
and shows only input sizes up to 60 words. From the graph we can see the crossover
point is at about 24 words input length, that is, about 460 decimal digits.

\FloatBarrier
\section{Toom-Cook Multiplication}
\index{Toom-Cook multiplication (\T3MUL{})}

Let us take a broader look at Karatsuba's algorithm:
it follows from the fundamental theorem of algebra that
any polynomial $a(x)$ of degree $< k$ is determined by its values at $k$ distinct points.
In the case of Karatsuba's algorithm, if we substitute $W^n$ in \eqref{kcut} with the indeterminate $x$
we get input polynomials of degree one: $a(x) = a_0 + a_1 x$ and $b(x) = b_0 + b_1x$.
If we multiply both, the result $c(x) \coloneqq a(x)b(x)$ is a polynomial of degree two.

What we did in Karatsuba multiplication can be understood as evaluating both polynomials
$a(x)$ and $b(x)$ at points $\{ 0, -1, \infty \}$.%
\footnote{By abuse of notation $a(\infty)$ means $\lim_{x \to \infty} a(x)/x$ and gives the highest coefficient.}%
\footnote{Other distinct points of evaluation would have done as well. For example, Karatsuba's
original paper used $\{ 0, 1, \infty \}$.}
Then we multiplied the results pointwise and inter\-polated
to regain the polynomial $c(x)$. To regain the integer result we evaluated $c(x)$ at $x = W^n$.

We can generalize this technique: evaluate polynomials of degree $< k$
at $2k-1$ distinct points, multiply pointwise and interpolate.
The time bound of this scheme is $O(n^{\log_k (2k-1)})$, so
for $k=3$, $4$, $5$ it is approximately $O(n^{1.465})$, $O(n^{1.404})$ and $O(n^{1.365})$, respectively.
This method is due to Toom \cite{Toom1963} and Cook \cite{Cook1966}.

The points for evaluation can be freely chosen (as long as they are distinct),
but it is not obvious which choice leads to the simplest
formulas for evaluation and interpolation.
Zuras \cite{Zuras1994} and Bodrato \cite{Bodrato2006} offer good solutions.

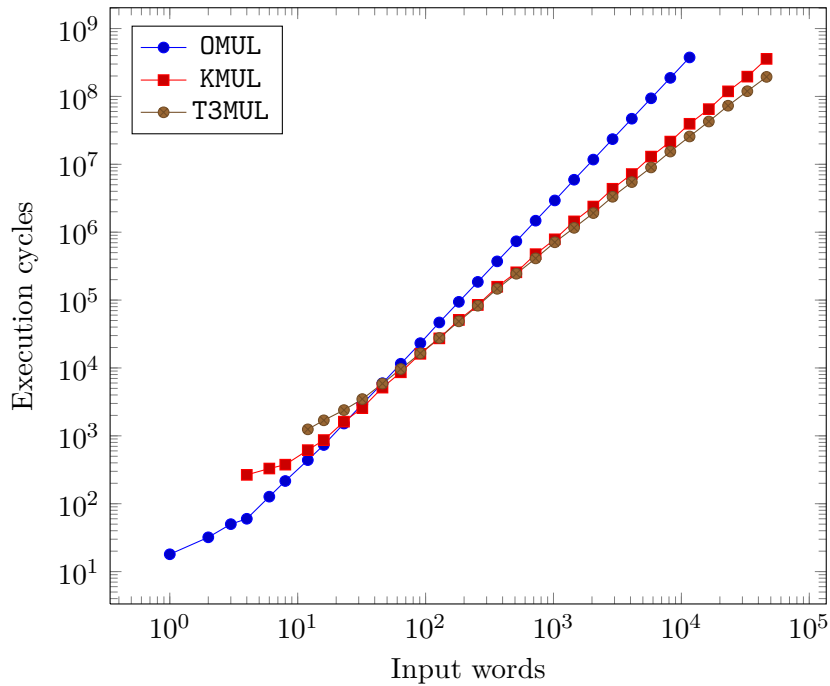
\begin{figure}
\bigskip
\centering

\begin{tikzpicture}
\begin{loglogaxis}[
   width=11cm,
   xlabel={Input words},
   ylabel={Execution cycles},
   legend pos=north west]

\addplot table[x index=0, y index=1] {omul-speed.txt};
\addlegendentry{\OMUL{}}
\addplot table[x index=0, y index=1] {kmul-speed.txt};
\addlegendentry{\KMUL{}}
\addplot table[x index=0, y index=1] {t3mul-speed.txt};
\addlegendentry{\T3MUL{}}

\end{loglogaxis}
\end{tikzpicture}

\caption{Execution time of \T3MUL{}}
\label{t3mulgraph}
\end{figure}

I implemented the Toom-Cook 3-way method from \cite{Bodrato2006} and called it \T3MUL{}.
Figure~\ref{t3mulgraph} shows a graph of execution time vs.\ input length.
The crossover point of \T3MUL{} and \KMUL{} is at about 100 words or 2000 decimal digits.

Unfortunately, the exponent in the time bound drops slowly as $k$ increases and the number of linear operations (additions,
subtractions and divisions by constants) rises quickly with $k$. This leads to ever higher crossover points.
Furthermore, each new $k$-way method has to be set in code individually. This calls for a more general
solution.

\FloatBarrier

\section{The Fast Fourier Transform}
\label{fft}
\index{Fast Fourier transform}

We are going to have a look at the fast Fourier transform (or \emph{FFT}) which was \mbox{(re-)}dis\-covered
in 1965 by Cooley and Tukey \cite{Cooley1965}. By choosing to evaluate the polynomial at certain special points
it allows us to do the evaluation very quickly.

This is just a short description of the fast Fourier transform as far as it concerns us now. A good
introduction can be found in Cormen et al.\ \cite[ch.~30]{Cormen2009},
Clausen and Baum \cite{Clausen1993} cover the topic from a group-theoretic standpoint
and Duhamel and Vetterli \cite{Duhamel1990} present a good overview of
the plethora of different FFT algorithms.

\bigskip
Let $R$ be a commutative ring with unity and let $n$ be a power of~2.%
\footnote{Please note that $n$ no longer designates an input length in words.}
The number $n$ is called the \emph{FFT length}.
Let $\omega_n$ be a primitive $n$-th root of unity in $R$, that is, $\omega_n^n=1$ and $\omega_n^k \neq 1$ for $k \in [1 : n-1]$.
We simply write $\omega$ instead of $\omega_n$, if the value of $n$ is clear from the context.
Furthermore, let $a(x)=\sum_{j=0}^{n-1}a_jx^j$ be a polynomial over $R$
with degree-bound $n$.\footnote{Any integer $n > \deg(a)$ is called a \emph{degree-bound} of $a$.}

For example,
$\mathbb{R}$ contains only a primitive 2nd root of unity, namely $-1$, but no higher orders.
But $\mathbb{C}$ does: $\omega_n = e^{2\pi i/n}$
is a primitive $n$-th root of unity in $\mathbb{C}$.

Another example is the quotient ring $\Z/n\Z$: it can be identified with the integers from $0$ to $n-1$, where all
operations are executed modulo $n$. $\Z/n\Z$ can contain up to $n-1$ roots of unity.
In the case of $n=17$, $\omega=3$ is a primitive 16\textsuperscript{th} root of unity.

We want to evaluate $a(x)$ at $n$ distinct, but otherwise arbitrary points. If we choose to
evaluate $a(x)$ at $\omega^k$, $k \in [0 : n-1]$, we can design the evaluation particularly efficient.
Because $\omega$ is a \emph{primitive} $n$-th root of unity,
we know $\omega^0$, $\omega^1$, \ldots, $\omega^{n-1}$ to be pairwise different.

We can re-sort $a(x)$ in even and odd powers and rewrite it as
\begin{align*}
a(x)  &= a_0 + a_1x + a_2x^2 + a_3x^3 + \ldots  \\
      &= a_0 + a_2x^2 + \ldots + a_1x + a_3x^3 + \ldots  \\
      &= \underbrace{a_0 + a_2x^2 + \ldots}_{\eqqcolon e(x^2)} + \: x (\underbrace{a_1 + a_3x^2 + \ldots}_{\eqqcolon o(x^2)})  \\
      &= e(x^2) + x o(x^2),
\end{align*}
where $e(x)$ and $o(x)$ are polynomials with half the degree-bound as $a(x)$. Since $n$ is a power of~2, we can
proceed recursively with this divide-and-conquer approach until the degree-bound of both polynomials $e(x)$ and $o(x)$ is one, that is, both
consist only of a constant.

We can evaluate $a(x)$ at $\omega^k$, $k \in [0 : n/2-1]$ and get
\[ a(\omega^k) = e(\omega^{2k}) + \omega^k o(\omega^{2k}). \]
But note what we get if we evaluate $a(x)$ at $\omega^{k+n/2}$, $k \in [0 : n/2-1]$:
\begin{align*}
a(\omega^{k+n/2})  &= e((\omega^{k+n/2})^2) + \omega^{k+n/2} o((\omega^{k+n/2})^2) \\
                   &= e(\omega^{2k+n}) + \omega^{k+n/2} o(\omega^{2k+n}) \\
                   &= e(\omega^{2k}) - \omega^k o(\omega^{2k}),
\end{align*}
since $\omega^{n/2}=-1$ and $\omega^n=1$.

If we have already computed $e(\omega^{2k})$ and $o(\omega^{2k})$ we save time by calculating both $a(\omega^k)$ and $a(\omega^{k+n/2})$ side by side:
\begin{align*}
a(\omega^k)        &= e(\omega^{2k}) + \omega^k o(\omega^{2k}) \\
a(\omega^{k+n/2})  &= e(\omega^{2k}) - \omega^k o(\omega^{2k}).
\end{align*}
This is the concept that makes the fast Fourier transform efficient.
After solving both halves of the problem
we can calculate two results in $O(1)$ additional time.%
\footnote{The simultaneous calculation of sum and difference is called a
\emph{butterfly operation}\index{Butterfly operation} and
the factors $\omega^k$ in front of $o(\omega^{2k})$ are often called \emph{twiddle factors}\index{Twiddle factors}.}

There are different types of FFTs and the one just described is called a \emph{Cooley-Tukey FFT}\index{Fast Fourier transform!Cooley-Tukey}
of length $n$. More precisely, it is a radix-2 decimation in time FFT.
See \cite{Duhamel1990} for other types of FFTs.

\bigskip
We can write this algorithm as a recursive function in Python. The computation of the
actual root of unity has been left out of this example to keep it independent of the
ring~$R$.

\begin{lstlisting}[language=Python]
def fft(a):                                  # a is a list
   n = len(a)                                # degree-bound of a
   if n <= 1: return a
   even = fft(a[0::2])                       # slice up even numbered values
   odd = fft(a[1::2])                        # ... and odd
   r = [0] * n                               # fill list with n dummies
   for k in range(0, n//2):                  # n//2 means integer divide
      w = root_of_unity(n, k)                # n-th root to k-th power
      r[k] = even[k] + w * odd[k]
      r[k+n//2] = even[k] - w * odd[k]
   return r
\end{lstlisting}

\begin{figure}[b]
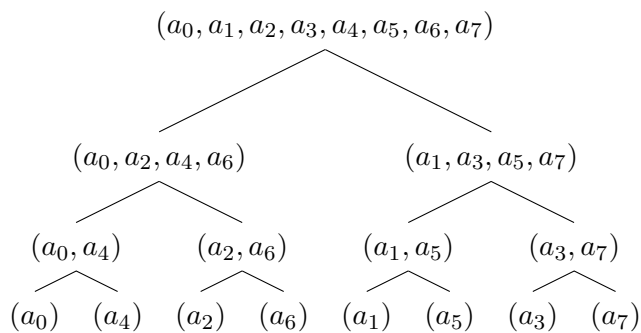

\bigskip
\Tree[.{$(a_0, a_1, a_2, a_3, a_4, a_5, a_6, a_7)$}
   [.{$(a_0, a_2, a_4, a_6)$}
      [.{$(a_0, a_4)$}
         {$(a_0)$} {$(a_4)$} ]
      [.{$(a_2, a_6)$}
         {$(a_2)$} {$(a_6)$} ]]
   [.{$(a_1, a_3, a_5, a_7)$}
      [.{$(a_1, a_5)$}
         {$(a_1)$} {$(a_5)$} ]
      [.{$(a_3, a_7)$}
         {$(a_3)$} {$(a_7)$} ]]]
\caption{Splitting an array into even and odd positions}
\label{fftsplitone}
\bigskip
\end{figure}

Since at each recursion level the input list is split into values with even and odd indices, we get the structure
shown in Figure~\ref{fftsplitone}, if we assume a start with eight input values.

\index{Bit-reversal}
Notice the ordering of the indices at the lowest level: the values are at an index
which is the \emph{bit-reversed} input index.  ``Bit-reversed'' here means only reversing the
bits that are actually used in indexing: in the last example we had eight values, hence we needed 3 bits for indexing. Accordingly, the
bit-reversed index of, for example, $a_3 = a_{011_b}$ is $110_b = 6$.

The bit-reversal is a consequence of the splitting of the array into even and odd indices.
Since even indices have the lowest bit set to zero, all ``left'' members of the output array have the highest bit of the index
set to zero, whereas all ``right'' members have odd indices and have the highest bit set to one. This repeats itself
through all levels.

We use this observation to decrease the memory footprint:
the \code{fft()} function listed above uses temporary memory at each level, first to split up the input in
even and odd indexed values and then to create the list of return values. We would like to
save those allocations. Luckily, that is possible.
The following design and the term ``shuffle'' is taken from Sedgewick \cite[ch.~41]{Sedgewick1992}.
\index{Fast Fourier transform!coefficient shuffling}

\begin{figure}[b]
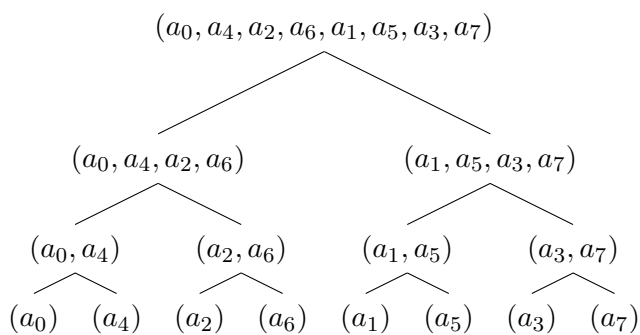

\bigskip
\Tree[.{$(a_0, a_4, a_2, a_6, a_1, a_5, a_3, a_7)$}
   [.{$(a_0, a_4, a_2, a_6)$}
      [.{$(a_0, a_4)$}
         {$(a_0)$} {$(a_4)$} ]
      [.{$(a_2, a_6)$}
         {$(a_2)$} {$(a_6)$} ]]
   [.{$(a_1, a_5, a_3, a_7)$}
      [.{$(a_1, a_5)$}
         {$(a_1)$} {$(a_5)$} ]
      [.{$(a_3, a_7)$}
         {$(a_3)$} {$(a_7)$} ]]]
\caption{Halving the already shuffled array}
\label{fftsplittwo}
\bigskip
\end{figure}

If we reorder the input list according to its bit-reversed indices, all even
indexed values are in the first half and all odd indexed values
in the second half. This saves us the creation of function arguments for the recursive calls.
All we need to hand over to the lower levels is the position and length in the array they should work on, since the values are already in the right order.
Then the tree of function arguments looks like Figure~\ref{fftsplittwo}.

We don't even need extra storage for the return values! We can use
the memory of the input parameters and overwrite it with the return values;
the input parameters are no longer needed after the function has calculated the return values from them.

If we put all this into code, our Python function looks like this:

\begin{lstlisting}[language=Python]
def bit_rev(x, b):                           # reverse b lower bits of x
   return sum(1<<(b-1-i) for i in range(0, b) if (x>>i) & 1)

def shuffle(a):                              # shuffle input list a
   r = []                                    # empty list
   b = (len(a)-1).bit_length()               # bits used for indexing
   pos = [bit_rev(n, b)
          for n in range(0, len(a))]         # list of new positions
   for i in pos:                             # cycle through list of positions
      r.append(a[i])                         # ... and build return list
   return r

def fft_eval(a, pos, n):                     # work on a[pos..pos+n-1]
   half = n//2                               # integer divide
   if half > 1:
      fft_eval(a, pos, half)                 # even part
      fft_eval(a, pos+half, half)            # odd part
   for k in range(0, half):
      w = root_of_unity(n, k)                # n-th root to k-th power
      t = w * a[pos+half+k]                  # multiply only once
      a[pos+half+k] = a[pos+k] - t           # must use this order
      a[pos+k] = a[pos+k] + t                # ... to avoid overwriting
   return

def fft_inplace(a):
   aa = shuffle(a)                           # create reordered a
   fft_eval(aa, 0, len(aa))                  # fft works in-place
   return aa
\end{lstlisting}

\label{fft:cost}
\index{Fast Fourier transform!run-time}
Let us analyze the number of arithmetic operations of the algorithm above.
We have assumed that $n$ is a power of~2. With each level the length of the input is halved until $n=1$;
this leads to $\log n$ levels of recursion. Furthermore, while the number $n$ gets halved with each level,
both halves are worked on, so all values are cycled over (see Figure~\ref{fftsplittwo}).
Since two return values are calculated with
three arithmetic operations (two additions and one multiplication),
the arithmetic cost per level is $3n/2$, which leads to a total cost for the
whole operation of $T(n)=3n/2 \cdot \log n$.

\pagebreak 
We can prove the run-time more formally (inspired by \cite[pp.~77--78]{Sedgewick1992}).
Obviously, $T(1)=0$. Then the total arithmetic cost is
\begin{align*}
T(n)              = {} & 2T(n/2) + 3n/2  \\
                  = {} & 2(2T(n/4) + 3n/4) + 3n/2 \\
                  = {} & 4T(n/4) + 3n/2 + 3n/2 \\
                    {} & \vdots \\
                  = {} & 2^{\log n} T(n/2^{\log n}) + 3n/2 \cdot \log n \\
                  = {} & n T(1) + 3n/2 \cdot \log n \\
                  = {} & 3n/2 \cdot \log n .   \label{fftcost}\numberthis 
\end{align*}

\section{FFT-based Polynomial Multiplication}
\label{fftpolymul}
\index{Fast Fourier transform!polynomial}

Now that we have introduced the fast Fourier transform and proved its run-time, let us see how
we can use it to multiply two polynomials rapidly.

Let $R$ be a commutative ring with unity and let $n$ be a power of~2.
Let $\omega$ be a primitive $n$-th root of unity in $R$.
Furthermore, let $a(x)=\sum_{j=0}^{n/2-1}a_jx^j$ and $b(x)=\sum_{j=0}^{n/2-1}b_jx^j$ be polynomials over $R$.

Please note that the polynomials $a(x)$ and $b(x)$ have a
degree-bound of $n/2$.
Since we are about to compute $c(x) \coloneqq a(x)b(x)$ we need to choose the number of sample points
$n$ as $n > \deg(c) = \deg(a)+\deg(b)$. To keep notation simple, we let $a_j = b_j = 0$ for $j \in [n/2 : n-1]$.

We evaluate both input polynomials at sample points $\omega^k$, $k \in [0 : n-1]$ to get
sample values $\widehat{a}_k \coloneqq a(\omega^k)$ and $\widehat{b}_k \coloneqq b(\omega^k)$.
Then, we multiply the sample values pairwise to get
the $\widehat{c}_k \coloneqq \widehat{a}_k \widehat{b}_k$.
But how to retrieve the result polynomial $c(x)$ from the $\widehat{c}_k$?
We will see how to accomplish that with ease if
$R$, $n$ and $\omega$ meet two additional requirements:
\begin{itemize}
\item $\omega^k-1$, $k \in [1 : n-1]$, must not be a zero divisor in $R$, and   \eqnum\label{ass1}
\item $n$ must be a unit in $R$, meaning $n$ is invertible.                     \eqnum\label{ass2}
\end{itemize}

We return to these requirements later.
Assuming that they hold, we can prove that the same algorithm can be used on the $\widehat{c}_k$ to regain the $c_j$
that was used to compute the $\widehat{a}_k$ and $\widehat{b}_k$ in the first place! That is to say: the Fourier transform
is almost self-inverse, except for ordering of the coefficients and scaling.

Let us see what happens if we use the $\widehat{a}_k = a(\omega^k) = \sum_{j=0}^{n-1} a_j \omega^{kj}$
as coefficients of the polynomial $\widehat{a}(x) \coloneqq \sum_{k=0}^{n-1}\widehat{a}_kx^k$ and
evaluate $\widehat{a}(x)$ at $\omega^{\ell}$, $\ell \in [0 : n-1]$, to compute
$\widehat{\widehat{a}}_\ell \coloneqq \widehat{a}(\omega^\ell)$. We get what is called an \emph{inverse transform}\index{Fast Fourier transform!inverse}:
\begin{align*}
\widehat{\widehat{a}}_\ell &= \widehat{a}(\omega^{\ell})  \\
                        &= \sum_{k=0}^{n-1} \widehat{a}_k \omega^{\ell k} \\
                        &= \sum_{k=0}^{n-1} \big( \sum_{j=0}^{n-1} a_j\omega^{kj} \big) \omega^{\ell k} \\
                        &= \sum_{j=0}^{n-1} \sum_{k=0}^{n-1} a_j\omega^{(j+\ell)k} \\
                        &= \sum_{j=0}^{n-1} a_j \sum_{k=0}^{n-1} (\omega^{j+\ell})^k \\
                        &= n \cdot a_{(-\ell)\bmod n} .   \tombstone
\end{align*}
The last line holds due to the sum of the geometric series:
\begin{numcases}{\sum_{k=0}^{n-1} (\omega^{j+\ell})^k =}
\frac{\omega^{(j+\ell)n}-1}{\omega^{j+\ell}-1} = 0   &  $\text{if } j+\ell \not\equiv 0 \imod n$,  \label{geodiv}\\
\sum_{k=0}^{n-1} 1 = n                               &  $\text{if } j+\ell \equiv 0 \imod n$.
\end{numcases}

Now we see why \eqref{ass1} is required:
if $\omega^k -1$, for $k \in [1 : n-1]$, is a zero divisor we are not allowed to do the division in \eqref{geodiv}.
Furthermore, to remove the factor $n$ in front of every $a_{(-\ell)\bmod n}$ we need \eqref{ass2}.

If we want to get the $\widehat{\widehat{a}}_\ell$ in the same order as the original $a_j$,
we can simply evaluate $\widehat{a}(x)$ at $\omega^{-\ell}$ instead of $\omega^{\ell}$.
This is called a \emph{backwards transform}\index{Fast Fourier transform!backwards}.

To summarize: we can evaluate the $\widehat{c}_k$ at points $\omega^{-\ell}$ to retrieve
$n \cdot c_\ell$, divide by $n$ and have thus recovered the coefficients of our desired product polynomial.

\bigskip
The overall arithmetic cost of this polynomial multiplication method is three FFTs
in $O(n \log n)$ plus $n$ multiplications of pairs of sample values in $O(n)$
plus $n$ normalizations in $O(n)$.
The FFTs dominate the total cost, so it is $O(n \log n)$.

\section{Modular FFT-based Multiplication}
\label{fftmodmul}
\index{QMUL@\QMUL{}}
\index{Fast Fourier transform!modular}

We can put last section's method into action and design a fast multiplication algorithm
for long numbers using the quotient ring $R = \Z/p\Z$, with prime $p$. This is sometimes called
a \emph{number theoretic transform}\index{Number theoretic transform} or \emph{NTT}.
According to \cite[p.~306]{Knuth1997} this method goes back to Strassen in 1968.

We want to multiply nonnegative integers $a$ and $b$ to get the product $c \coloneqq ab$.
We are free to choose an arbitrary $p$ for our calculations, as long as last section's
requirements are met. Furthermore, our choice of $p$ should be well suited for
implementation. If we choose $p$ to be prime it means that the ring $\Z/p\Z$ is
even a field, so we are sure to meet requirements \eqref{ass1} and \eqref{ass2}.
This field is denoted $\F_p$.

For the FFTs we need roots of unity of sufficiently high degree.
Let $p>3$.
Since $p$ is prime, we know that $\F_p^* = \{ 1, 2, 3, \ldots, p-1 \}$ is a
cyclic multiplicative group of order $p-1$. We call $g \in \F_p^*$ a \emph{generator} of $\F_p^*$ if its powers
$g^j$, $j \in [0 : p-2]$, create the whole group.
Also, $g$ is a primitive $(p-1)$-th root of unity. Note that $g^{p-1} = g^0 = 1$.

$\F_p^*$ contains $p-1$ elements. Since $p-1$ is even,
we can find integers $u$, $v>1$, such that $p-1=uv$. Without loss of generality, let $v$ be a power of~2.
We know that $g^{p-1}=1$, hence $g^{uv}=(g^u)^v=1$.
Since $v$ divides $p-1$, we know $g^u$ is a $v$-th primitive root of
unity in $\F_p^*$.

\bigskip
Let us see how to reduce a long multiplication to polynomial multiplication:
we have to distribute the bits of input numbers $a$ and $b$ to coefficients of polynomials $a(x)$ and $b(x)$.
In Karatsuba's algorithm we did cut the input numbers in two pieces of $n$ words each, or $wn$ bits,
where $w$ is the word size and $W=2^w$ is the wordbase.
Accordingly, evaluating polynomial $a(x)$ at $W^n$ yielded number $a$. Now we are going to
cut the input numbers into pieces of $r$ bits. But how to choose $r$?

The larger $r$ is, the less coefficients we get, that is, the lower the degree of the polynomial.
In consequence, this can lead to smaller FFT lengths, which are faster to compute.
This is why we want to choose $r$ as large as possible.

If we multiply polynomials $a(x)=\sum_{j=0}^{n/2-1}a_jx^j$ and $b(x)=\sum_{k=0}^{n/2-1}b_k x^k$
to get product $c(x) \coloneqq a(x)b(x) = \sum_{\ell=0}^{n-2}c_\ell x^\ell$ with $c_\ell = \sum_{j+k=\ell} a_j b_k$,
observe that $c_\ell$ can contain up to $n/2$ summands.
By construction $a_j$, $b_k < 2^r$, hence $c_\ell < \frac{n}{2}(2^r)^2 = n 2^{2r-1}$.
But $c_\ell$ must also be less than $p$.
Hence, our choice of $p$ must make sure that
\[ p \geq n 2^{2r-1}. \numberthis\label{pchoice} \]

For practical reasons, we want to choose a prime $p$ that can be handled easily by the target machine's processor,
hence I chose $p$ to be almost as big as the wordbase, so it can still be stored in one machine word.
``Almost'' means $\lfloor \log p \rfloor = w-1$ to maximize the use
of available bits per word.

The above mentioned constrains led me to choose the following parameters:%
\footnote{I reproduce the numbers here, since it required some effort to calculate them. If
one wants to do FFT with modular arithmetic, one must first find a suitable prime modulus $p$ and
a matching primitive $n$-th root of unity. So here they are.}

{\centering
\begin{tabular}{cclc}
Word size (bits)  &  Modulus $p$                   & Composition of $p$ & Generator $g$   \\
\hline
8                 &  \num{193}                     & $3 \cdot 2^6 + 1$      & 5               \\
16                &  \num{40961}                   & $5 \cdot 2^{13} + 1$   & 3               \\
32                &  \num{3489660929}              & $13 \cdot 2^{28} + 1$  & 3               \\
64                &  \num{10232178353385766913}    & $71 \cdot 2^{57} + 1$  & 3               \\
\end{tabular}

}
\bigskip

\pagebreak 
From these numbers we can calculate the respective primitive $n$-th root of unity $\omega$:

{\centering
\begin{tabular}{ccc}
Word size (bits)  & Order $n$ of primitive root $\omega$    & Primitive $n$-th root $\omega$       \\
\hline
8                 & $2^6$                                   & $5^3 = \num{125}$                    \\
16                & $2^{13}$                                & $3^5 = \num{243}$                    \\
32                & $2^{28}$                                & $3^{13} = \num{1594323}$             \\
64                & $2^{57}$                                & $3^{71} = \num{3419711604162223203}$ \\
\end{tabular}

}
\bigskip

Now that we have chosen $p$, we can use \eqref{pchoice} to calculate the maximum $r$ for a given FFT length~$n$:
\begin{align*}
n2^{2r-1}            & \leq p                                     \\
\log(n2^{2r-1})    & \leq \log p                              \\
\log n + 2r-1      & \leq \log p                              \\
2r                   & \leq \log p - \log n + 1               \\
r                    & \leq \frac{1}{2}(\log p - \log n + 1)
\end{align*}
Choosing $r$ determines the degree of the polynomials and hence $n$, which in turn can have an influence
on $r$. So, we might have to cycle several times over this formula to find the largest $r$ and smallest $n$.

Please note that this also imposes an upper bound on the length of input numbers this
algorithm can handle:
\begin{align*}
\log n + 2r-1      & \leq \log p                              \\
\log n             & \leq \log p - 2r + 1                     \\
\log n             & \leq w - 2r                          && \text{(since }\lfloor \log p \rfloor = w-1 \text{)}      \\
\log n             & \leq w - 2                           && \text{(} r \text{ has to be at least 1)}\\
n                    & \leq 2^{w - 2}             \\
n                    & \leq W/4.
\end{align*}
This determines the maximum FFT length. In this case, $r$ was 1 bit and hence the maximum output
length is $W/4$ bits or $W/32$ bytes. Choosing a larger $r$ only makes matters worse.
The maximum FFT length might be even less than that, since the order of $\omega$ limits
the FFT length as well.

Now that we have chosen the necessary parameters, we can attend to the implementation.
A Python version of the main routine looks pretty straightforward. I termed this function
\QMUL{}, alluding to QuickMul by Yap and Li \cite{qmul2000}.

\begin{lstlisting}[language=Python]
def qmul(a, b):
   p, n, w, r = select_param(a, b)     # p: prime modulus, n: FFT length
                                       # w: n-th root, r: bits per coefficient
   al = split_input(a, p, r, n)        # split inputs into n parts, ...
   bl = split_input(b, p, r, n)        # ... each maximum r bits long

   al = shuffle(al)                    # shuffle inputs
   bl = shuffle(bl)

   fft_eval(al, 0, n, w)               # evaluate inputs at roots of unity
   fft_eval(bl, 0, n, w)

   cl = []                             # empty list
   for i in range(0, n):               # multiply pointwise
      cl.append(al[i] * bl[i])         # append to list

   cl = shuffle(cl)                    # shuffle result
   fft_eval(cl, 0, n, w)               # evaluate result

   inv = modinv(n, p)                  # normalize result
   for i in range(0, n):
      cl[i] *= inv

   c = reasm(cl, r)                    # reassemble result
   return c
\end{lstlisting}

The functions \code{fft_eval()} and \code{shuffle()} have already been shown. Functions
\code{reasm()} and \code{split_input()} are new: they cut up the input numbers and add up the
coefficients of the resulting polynomial to a number, respectively. To find the proper number
of bits per coefficient \code{r} and compute the FFT length \code{n} and matching root \code{w} function \code{select_param()} is used.

The actual implementation I used for benchmarking was done in \cpp. To give an impression of the code,
the following function is a simplified version of the evaluation. The actual code is more complicated,
since I use {\cpp} templates to unroll the last five levels of the FFT. This saves some
call and loop overhead at the cost of code size.

\label{qmullisting}
\begin{lstlisting}[language=C++]
void qmul_evaluate(word* p, unsigned i, unsigned lg) {
   unsigned n = 1 << lg;                     // number of values
   unsigned half = n/2;                      // half of them
   if (half > 1) {
      qmul_evaluate(p, i, lg-1);             // even part
      qmul_evaluate(p, i+half, lg-1);        // odd part
   }

   // w^0=1: no multiplication needed
   word t = p[i+half];
   p[i+half] = modsub(p[i], t);
   p[i] = modadd(p[i], t);

   // handle w^k, k>0
   word* pw = pre_w[lg];                     // use precomputed roots
   for (unsigned k=1; k<half; ++k) {
      word t = modmul(pw[k], p[i+half+k]);
      p[i+half+k] = modsub(p[i+k], t);
      p[i+k] = modadd(p[i+k], t);
   }
}
\end{lstlisting}

Functions \code{modadd()} and \code{modsub()} are pretty straightforward and I don't include them here,
but \code{modmul()} is more complicated. It takes two 64-bit numbers as inputs
and multiplies them modulo $p$. We recall that there is an intrinsic compiler function to do the multiplication,
but the result has 128 bits and has to be reduced modulo $p$. To accomplish that, we could
use a 128-by-64-bit division, but it is quite slow and takes up to 75 cycles.

\index{Division by a constant}
Warren \cite[pp.~178--188]{Warren2002} shows a solution how to replace a division by a constant
with a multiplication and a shift. Since the input has 128 bits, the multiplication has to be 128-bit wide
as well. But this only results in floor division. To get the rest from division we must do one more multiplication and one subtraction.
In total, we have to do six 64-bit multiplications, plus some additions and shifts to do the
one modular multiplication. Benchmarking shows that the whole modular multiplication
can be done like that in about 10 cycles.

To save the time needed to calculate the roots of unity, I use arrays of precomputed
roots \code{pre_w[lg]}. These powers are independent of the inputs and are reused every time
\QMUL{} is used.

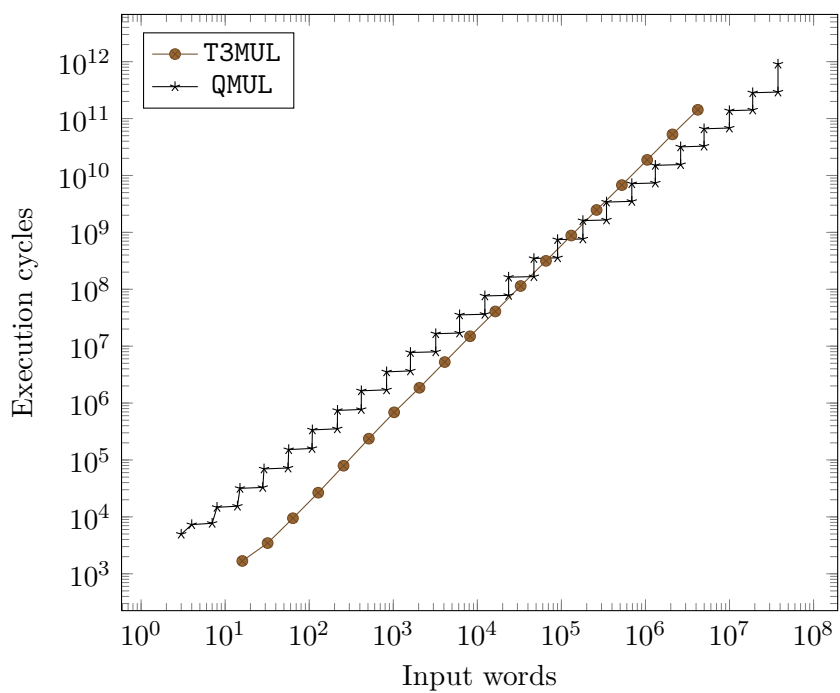
\begin{figure}
\bigskip
\centering

\begin{tikzpicture}
\begin{loglogaxis}[
   width=11cm,
   xlabel={Input words},
   ylabel={Execution cycles},
   legend pos=north west]

\pgfplotsset{cycle list shift=2}
\addplot table[x index=0, y index=1] {t3mul-large-speed.txt};
\addlegendentry{\T3MUL{}}
\addplot table[x index=0, y index=1] {qmul-large-speed.txt};
\addlegendentry{\QMUL{}}

\end{loglogaxis}
\end{tikzpicture}

\caption{Execution time of \QMUL{}}
\label{qmulgraph}
\end{figure}

A graph of execution cycles of \QMUL{} in comparison to \T3MUL{} is presented in Figure~\ref{qmulgraph}.
Please note that this graph covers a wider range of sizes than the previous graphs.
We see that our new algorithm is asymptotically faster than the hitherto used \T3MUL{} implementation.

The stair-like shape of the graph is a result of the FFT: if input numbers get too
large, the FFT depth must be increased by one level, thereby doubling the number of evaluation
points. From these graphs we can say that \QMUL{} starts to outperform \T3MUL{} for inputs
with a length of about \num{110000} words or more, that is, about \num{2100000} decimal digits.

\bigskip
So we found an algorithm with a good asymptotic cost, but it only starts to pay off
if inputs are quite long. Why is that so? What are the weak points of \QMUL{}?
\begin{itemize}
\item The modular multiplication and reductions are expensive. Six word-sized multiplications are not cheap.
\item The FFT length is large and hence many extra bits room for the sum of the coefficient products must be left free.
   Since the unit of operation is only a processor word, this ``eats up'' quite some percentage of its size.
   Plus, it implies a large FFT depth as well.
\item The maximum length for long numbers is limited to $W/32$ bytes,
even if larger numbers could be handled by the machine.
\end{itemize}

The following celebrated algorithm will address all of the weak points listed above.

\section{Modular Schönhage-Strassen Multiplication}
\index{Schönhage-Strassen\enskip multiplication (\SMUL{})}
\label{smul}

The idea of Schönhage-Strassen multiplication (of which my implementation is called \SMUL{}) is
to perform FFTs in rings of the form
$\Z/(2^K+1)\Z$, which are sometimes called \emph{Fermat rings}\index{Fermat ring}.
In Fermat rings, $2$ is a primitive $2K$-th root of unity. This fact can be exploited
to speed up multiplications by roots of unity during the FFTs: multiplications by powers of~2
can be implemented as shifts followed by a modular reduction and thus take only $O(K)$ time.
This is the cornerstone of the efficiency of Schönhage and Strassen's multiplication.

Since all roots of unity are powers of~2, we don't need to precompute them as in \QMUL{},
but can just keep track of shift counts.
Furthermore, modular reductions are simple
and can be done with just another long number addition and/or subtraction.
In this context, a shift followed by a modular reduction is called a \emph{cyclic shift}\index{Cyclic shift}.

\SMUL{} always computes the product modulo $2^N+1$, where $N$ can be chosen.
If we multiply input numbers $a$ and $b$ to get $c \coloneqq ab$, we have to
provide an $N$ that is big enough to hold the full result.
\SMUL{} reduces a multiplication with $N$ bits
to smaller multiplications of $K \approx \sqrt{N}$ bits, in contrast to a reduction to word size as in \QMUL{}.%
\footnote{A reduction to word size is usually not possible in \SMUL{}, because the FFT length is not sufficiently
large to cut input numbers in parts so small, since there are not enough roots of unity.}
If the size of pointwise multiplications exceeds a certain threshold, \SMUL{} is used recursively,
otherwise a simpler algorithm takes over.

This algorithm was first published by Schönhage and Strassen in 1971 \cite{SS1971} and provided
results modulo $2^N+1$, where $N$ itself is a power of~2. A later version published
by Schönhage \cite{Schoenhage1982} relaxes the requirement to ``suitable numbers'' of the form $N=\nu 2^n$, $\nu \in [n-1 : 2n-1]$.
For the implementation we can relax the requirement even more: Section \ref{smulproc} lists the details.

We introduce some notation:
to compute the product $c$ of nonnegative numbers $a$ and $b$, we do FFTs in
the ring $R \coloneqq \Z/(2^K+1)\Z$.
We use a Cooley-Tukey FFT and thus the FFT length $n$ has to be a power of~2.
Since we can choose $R$ (and hence $K$) to suit our needs,
we choose $K=r 2^m$, with positive integers $r$ and $m$.
Our choice of $K$ and $m$ will in turn determine $N$, where $N=s 2^m$, with positive integer $s$.
This $s$ is the number of input bits per coefficient.

It it easy to see that 2 is a primitive $2K$-th root of unity: since $2^K+1 \equiv 0$, we have $2^K \equiv -1$
and hence $2^{2K} \equiv 1$. Furthermore, it is obvious that for $u \in [1 : K-1]$ we get $2^u \not\equiv\pm 1$.
For $K+v \eqqcolon u \in [K+1:2K-1]$ we see that $2^u = 2^{K+v} = 2^K 2^v = -2^v \not\equiv 1$.

Because the FFT length is a power of~2, we need a primitive root of unity of the same order.
Since~2 is a primitive root of unity of order $2K=2r2^m$, it holds that
$1 \equiv 2^{2K} = 2^{2r2^m} = (2^{2r})^{2^{m}}$. This makes $\omega \coloneqq 2^{2r}$ a primitive $2^{m}$-th root of unity
and the FFT length $n \coloneqq 2^{m}$.
We deliberately chose an even exponent for $\omega$, since we will be needing $\sqrt{\omega}$ later.

\subsection{Invertibility of the Transform}
\label{smul:inverse}

For the existence of the inverse FFT requirements \eqref{ass1} and \eqref{ass2} have to be met.
Since $2^K+1$ may not be prime, we cannot rely on our argument from Section \ref{fftmodmul},
so we must show that the requirements are met here, too:
\begin{itemize}
\item With $\omega=2^{2r}$, $\omega^j-1$, $j \in [1 : n-1]$, must not be a zero divisor in $\Z/(2^K+1)\Z$, and \eqnum\label{ssass1}
\item $n=2^{m}$ must be a unit in $\Z/(2^K+1)\Z$. \eqnum\label{ssass2}
\end{itemize}

To prove \eqref{ssass1} we need some identities about the \emph{greatest common divisor} (gcd):
Let $a$, $b$ and $u$ be positive integers and $(a,b)$ signify the greatest common divisor of $a$ and $b$. Then
the following identities hold:
\begin{align}
(a, b) &= (b,a),                                   \label{gcd1} \\
(a, b) &= (a-b,b) \text{, if } a \ge b,            \label{gcd1b} \\
(ua, ub) &= u(a,b),                                \label{gcd1a} \\
(ua, b) &= (u,b)(a,b) \text{, if }(u, a)=1,        \label{gcd3} \\
(ua, b) &= (a,b) \text{, if }u \ndiv b,            \label{gcd2} \displaybreak[1] \\
(2^a-1, 2^b-1) &= 2^{(a,b)}-1,                     \label{gcd4} \\
(2^a-1,2^a+1)  &= 1,                               \label{gcd2pm1} \\
(2^a-1, 2^b+1) &= \frac{2^{(a,2b)}-1}{2^{(a,b)}-1} \label{gcd5} .
\end{align}

Identities \eqref{gcd1} -- \eqref{gcd2} are well known, so we don't prove them here.

We prove \eqref{gcd4} by induction on $a+b$.
We assume without loss of generality that $a \geq b$.
The induction basis is easily checked: $(2^1-1, 2^1-1)=1=2^{(1,1)}-1$.

Now we show the induction step: we assume $(2^\alpha-1, 2^\beta-1) = 2^{(\alpha,\beta)}-1$, for $\alpha+\beta < a+b$.
We use \eqref{gcd1b} and get
\begin{align*}
(2^{a}-1, 2^b-1)   &= (2^{a}-1-(2^b-1), 2^b-1)  \\
                     &= (2^{a}-2^b, 2^b-1)  \\
                     &= (2^b(2^{a-b}-1), 2^b-1)  \\
                     &= (2^{a-b}-1, 2^b-1)  && \text{(by }\eqref{gcd2}\text{)}  \\
                     &= 2^{(a-b,b)}-1  && \text{(by IH)}  \\
                     &= 2^{(a,b)}-1.  && \text{(by }\eqref{gcd1b}\text{)}  \tombstone
\end{align*}

To prove \eqref{gcd2pm1} we use \eqref{gcd1b} and see that
\begin{align*}
(2^b-1,2^b+1)  &= (2^b-1,2^b+1-(2^b-1)) \\
               &= (2^b-1,2) \\
               &= 1.  \tombstone
\end{align*}
To prove \eqref{gcd5} we use the well known difference of squares $a^2-b^2=(a+b)(a-b)$ and
apply it to our case, where it yields $2^{2b}-1 = (2^b+1)(2^b-1)$. It holds that
\begin{align*}
2^{(a,2b)}-1      &= (2^a-1, 2^{2b}-1)  && \text{(by \eqref{gcd4})}  \\
                  &= (2^a-1, (2^b+1)(2^b-1)) \\
                  &= (2^a-1, 2^b+1)(2^a-1, 2^b-1)  && \text{(by \eqref{gcd3} and \eqref{gcd2pm1})} \\
                  &= (2^a-1, 2^b+1)(2^{(a,b)}-1) \\
\intertext{Divide by $2^{(a,b)}-1$ and get}
\frac{2^{(a,2b)}-1}{2^{(a,b)}-1} &= (2^a-1, 2^b+1).  \tombstone
\end{align*}

Recalling that $\omega=2^{2r}$, $n=2^{m}$ and $K=r2^m$ we can now prove \eqref{ssass1}
by showing that $(\omega^j-1, 2^K+1)=1$, for $j \in [1 : n-1]$.
Thus all $\omega^j-1$ are units and therefore no zero divisors.
\begin{align*}
(\omega^j-1, 2^K+1)     &= ((2^{2r})^j-1, 2^{r2^m} \! +1) \\
                        &= (2^{2rj}-1, 2^{r2^m} \! +1) \\
                        &= \frac{2^{(2rj,2r2^m)}-1}{2^{(2rj,r2^m)}-1}  && \text{(by \eqref{gcd5})}  \\
                        &= \frac{2^{2r(j,2^{m})}-1}{2^{2r(j,2^{m-1})}-1}.  && \text{(by \eqref{gcd1a})}  \\
\intertext{Since $j<2^{m}$ it is clear that $(j,2^{m})=(j,2^{m-1})$. Hence}
(\omega^j-1, 2^K+1)     &= \frac{2^{2r(j,2^{m-1})}-1}{2^{2r(j,2^{m-1})}-1} = 1.  \tombstone
\end{align*}

Still open is \eqref{ssass2}. For $n=2^{m}$ to be a unit in $\Z/(2^K+1)\Z$,
there must exist an $i$ with $2^{m}i \equiv 1 \equiv 2^{2K}$. Obviously, $i = 2^{2K-m}$ works.   \texttombstone

\subsection{Convolutions}

\hspace*{0pt}\refstepcounter{dummy}\index{Convolution}%
Schönhage-Strassen multiplication always computes results modulo $2^N+1$.
If it is used recursively to compute the pointwise products this comes in handy, since it
allows multiplications where the results are in $[0 : 2^K]$ without performing a modular reduction.
This lowers the FFT length and thus the execution time by a factor of two.
We will now see how to use convolutions to accomplish this.

If $a(x)=\sum_{i=0}^{n-1} a_i x^i$ and $b(x)=\sum_{j=0}^{n-1} b_j x^j$ are two polynomials with coefficients $a_i$, $b_j \in R$,
then the coefficients of their product $c(x) \coloneqq a(x)b(x)=\sum_{k=0}^{2n-1}c_k x^k$
are given by the \emph{(acyclic) convolution formula}
\[ c_k=\sum_{i+j=k}a_i b_j. \numberthis \label{defconv} \]

\begin{figure}[h]
\bigskip
\centering
\begin{tikzpicture}
   \begin{axis}[
         name=plot1,
         height=8cm,
         width=8cm,
         xtick={0,...,3},
         ytick={0,...,3},
         xticklabel={$a_{\pgfmathprintnumber[int trunc]{\tick}}$},
         yticklabel={$b_{\pgfmathprintnumber[int trunc]{\tick}}$}
   ]

   \addplot[blue,mark=square*] coordinates {(0,0)}                         node[pin={[pin distance=0cm]0:{\color{blue}$c_0$}}] {};
   \addplot[blue,mark=square*] coordinates {(0,1) (1,0)}                   node[pos=0.5,pin={[pin distance=0cm]180:{\color{blue}$c_1$}}] {};
   \addplot[blue,mark=square*] coordinates {(0,2) (1,1) (2,0)}             node[pos=0.5,pin={[pin distance=0cm]180:{\color{blue}$c_2$}}] {};
   \addplot[blue,mark=square*] coordinates {(0,3) (1,2) (2,1) (3,0)}       node[pos=0.5,pin={[pin distance=0cm]180:{\color{blue}$c_3$}}] {};
   \addplot[blue,mark=square*] coordinates {(1,3) (2,2) (3,1)}             node[pos=0.5,pin={[pin distance=0cm]180:{\color{blue}$c_4$}}] {};
   \addplot[blue,mark=square*] coordinates {(2,3) (3,2)}                   node[pos=0.5,pin={[pin distance=0cm]180:{\color{blue}$c_5$}}] {};
   \addplot[blue,mark=square*] coordinates {(3,3)}                         node[pin={[pin distance=0cm]180:{\color{blue}$c_6$}}] {};
   \end{axis}
\end{tikzpicture}
\caption{Convolution of two polynomials}
\label{acyclicconvolutiondiagram}
\end{figure}
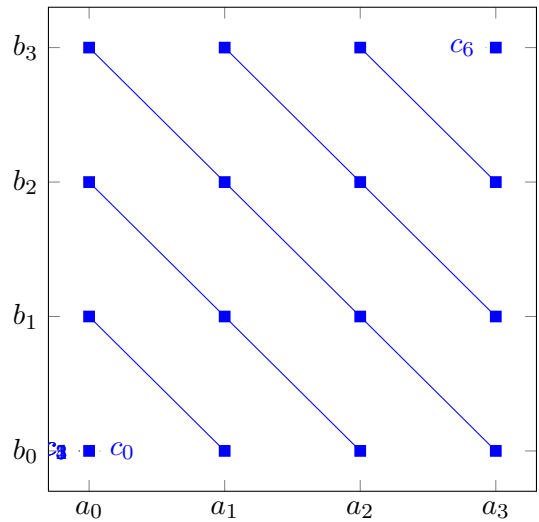

Figure~\ref{acyclicconvolutiondiagram} shows the product of two polynomials $a$ and $b$ both with degree three.
The lines from top-left to bottom-right are convolutions with the dots being products of the individual coefficients.
For each convolution the sum of the indices of the coefficient products is constant.
As Gilbert Strang put it: ``You smell a convolution when [the indices] add to [k]'' \cite{StrangV22}.

In the process of the FFT as laid out in Section \ref{fftpolymul}, two input polynomials
are evaluated at $\omega^i$, $i \in [0:n-1]$, where $\omega$ is a primitive $n$-th root of unity.
Afterwards, the sample values are multiplied pointwise and transformed backwards to get the product polynomial.

Define the mapping
\begin{gather*}
   \phi: R[x] \to R^n,  \\
   a(x) \mapsto \big( a(\omega^0, \ldots, a(\omega^{n-1}) \big).
\end{gather*}
The kernel of $\phi$ is the ideal generated by $\prod_{i=0}^{n-1}(x-\omega^i)$.
Since $\omega^n=1$, surely $(\omega^i)^n=1$ holds as well, for $i \in [0:n-1]$.
So the polynomial $x^n-1$ yields zero for each $x=\omega^i$, hence it has
$n$ distinct roots and the $n$ linear factors $x-\omega^i$.  From that we conclude that
$\prod_{i=0}^{n-1}(x-\omega^i) = x^n-1$ and hence that the kernel of $\phi$
is the ideal generated by $x^n-1$.

This means that polynomial multiplication that uses the mapping $\phi$ always gives results
modulo $x^n-1$. This is called the \emph{cyclic} convolution of two polynomials.
Given the aforementioned polynomials $a(x)$ and $b(x)$ it produces the product polynomial
$c(x)$ with coefficients
\[ c_k = \sum_{\substack{i+j \equiv k \\ \imod n}} a_{i} b_{j}.  \numberthis\label{cycconv} \]

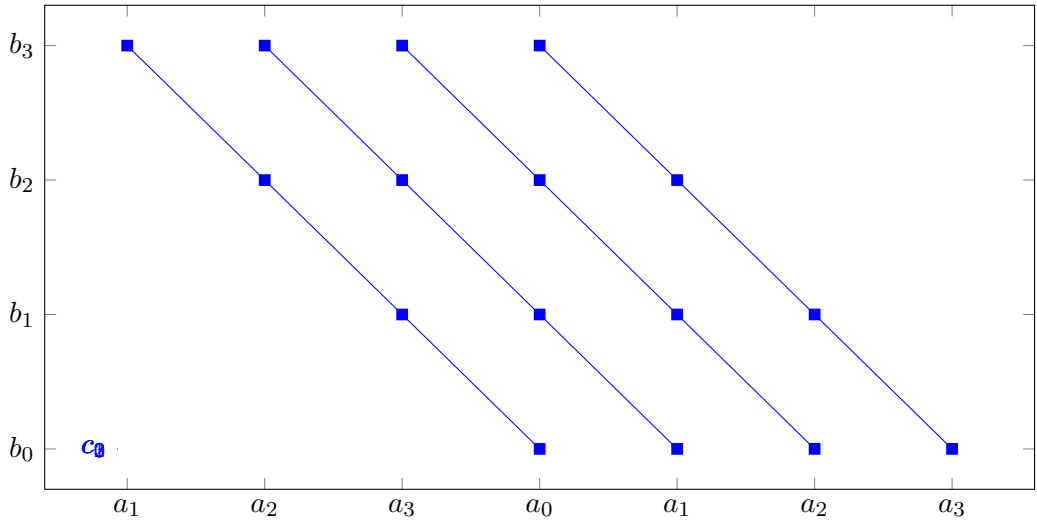
\begin{figure}[h]
\bigskip
\centering
\begin{tikzpicture}
   \begin{axis}[
         name=plot1,
         height=8cm,
         width=14.6cm,
         xtick={0,...,7},
         ytick={0,...,3},
         xticklabel={$a_{\pgfmathparse{int(mod(\tick,4))}\pgfmathresult}$},
         yticklabel={$b_{\pgfmathprintnumber[int trunc]{\tick}}$}
   ]

   \addplot[blue,mark=square*] coordinates {(1,3) (2,2) (3,1) (4,0)} node[pos=0.5,pin={[pin distance=0cm]180:{\color{blue}$c_0$}}] {};
   \addplot[blue,mark=square*] coordinates {(2,3) (3,2) (4,1) (5,0)} node[pos=0.5,pin={[pin distance=0cm]180:{\color{blue}$c_1$}}] {};
   \addplot[blue,mark=square*] coordinates {(3,3) (4,2) (5,1) (6,0)} node[pos=0.5,pin={[pin distance=0cm]180:{\color{blue}$c_2$}}] {};
   \addplot[blue,mark=square*] coordinates {(4,3) (5,2) (6,1) (7,0)} node[pos=0.5,pin={[pin distance=0cm]180:{\color{blue}$c_3$}}] {};
   \end{axis}
\end{tikzpicture}
\caption{Cyclic convolution of two polynomials}
\label{cyclicconvolutiondiagram}
\end{figure}

Figure~\ref{cyclicconvolutiondiagram} shows the cyclic convolution of two polynomials of degree three.
Here, the upper half of coefficients ``wraps around'' and is added to the lower half.
This is why it is sometimes called a \emph{wrapped} convolution.

We now know that a cyclic convolution gives us results modulo $x^n-1$.
Can we get results modulo $x^n+1$? Schönhage shows us we can.

Since $i$, $j $, $k < n$ we can write \eqref{cycconv} as
\[ c_k = \sum_{i+j = k} a_{i} b_{j} \; + \sum_{i+j = n+k} a_{i} b_{j}.  \numberthis\label{cycconv2} \]
The second sum contains the higher half product coefficients that wrap around and are \emph{added} to the
lower half coefficients, since $x^n \equiv 1$.
But if we want results modulo $x^n+1$, it holds that $x^n \equiv -1$, hence what we are looking for
is a way to compute
\[ c_k = \sum_{i+j = k} a_{i} b_{j} \; - \sum_{i+j = n+k} a_{i} b_{j}.  \numberthis\label{ncycconv} \]

Schönhage's idea is to weight each of the coefficients $a_i$ and $b_j$ prior to the cyclic convolution in such a way
that for $i+j=n+k$ and $k < n$ it holds that $\theta^n a_i b_j = -a_i b_j$, for some $\theta \in R$ that we will specify immediately.
This puts the desired minus sign in front of the second term in \eqref{ncycconv}.

Choose the weight $\theta$ as follows:
let $\theta$ be a primitive $n$-th root of $-1$, that is, $\theta^n=-1$ and hence $\theta^2=\omega$.
To compute \eqref{ncycconv}, we use \eqref{cycconv}, but weight the inputs like
\[  \widetilde{a}_i \coloneqq \theta^i a_i \quad\text{ and }\quad \widetilde{b}_j \coloneqq \theta^j b_j  \numberthis\label{weightcoeff}  \]
and apply the proper ``counterweight'' $\theta^{-k}$ to the whole sum, so we get
\begin{align*}
c_{k} &= \theta^{-k} \!\!\!\! \sum_{\substack{i+j \equiv k \\ \imod n}} \widetilde{a}_i \widetilde{b}_j  \numberthis\label{weightresult}  \\
      &= \theta^{-k} \big( \!\!\! \sum_{i+j = k} \widetilde{a}_i \widetilde{b}_j \; + \!\! \sum_{i+j = n+k} \widetilde{a}_i \widetilde{b}_j \big)  \\
      &= \theta^{-k} \!\! \sum_{i+j = k} \theta^i a_i \theta^j b_j \; + \; \theta^{-k} \!\!\!\! \sum_{i+j = n+k} \theta^i a_i \theta^j b_j  \\
      &= \theta^{-k} \!\! \sum_{i+j = k} \theta^{k} a_i b_j \; + \; \theta^{-k} \!\!\!\! \sum_{i+j = n+k} \theta^{n+k} a_i b_j  \\
      &= \sum_{i+j = k} a_i b_j \; + \; \theta^{n} \!\!\!\! \sum_{i+j = n+k} a_i b_j  \\
      &= \sum_{i+j = k} a_i b_j \; - \sum_{i+j = n+k} a_i b_j.  \tombstone
\end{align*}

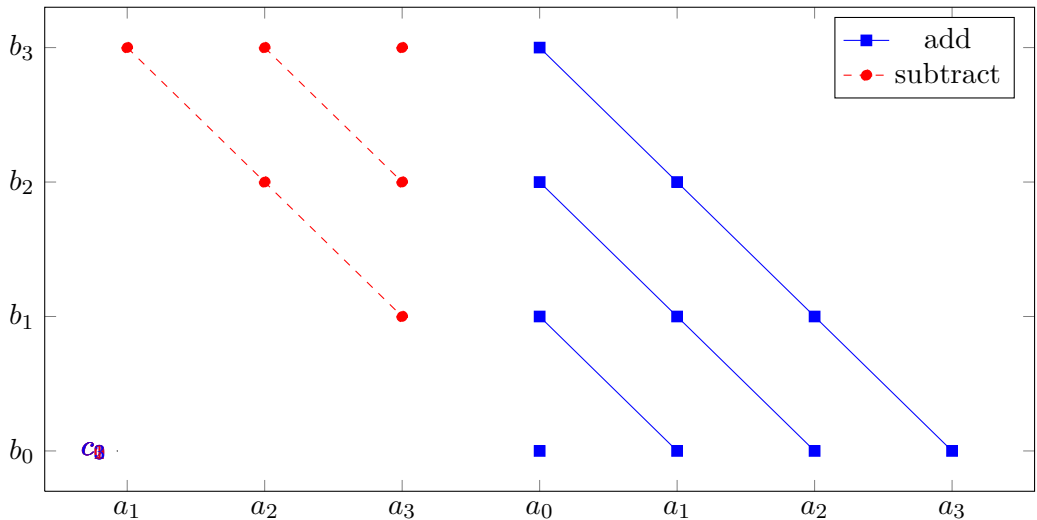
\begin{figure}[t]
\bigskip
\centering
\begin{tikzpicture}
   \begin{axis}[
         name=plot1,
         height=8cm,
         width=14.6cm,
         xtick={0,...,7},
         ytick={0,...,3},
         xticklabel={$a_{\pgfmathparse{int(mod(\tick,4))}\pgfmathresult}$},
         yticklabel={$b_{\pgfmathprintnumber[int trunc]{\tick}}$}
   ]

   \addplot[blue,mark=square*] coordinates {(4,0)} node[pos=0.5,pin={[pin distance=0cm]180:{\color{blue}$c_0$}}] {};
   \addplot[red,mark=*,dashed] coordinates {(1,3) (2,2) (3,1)} node[pos=0.5,pin={[pin distance=0cm]180:{\color{red}$c_0$}}] {};
   \addplot[blue,mark=square*] coordinates {(4,1) (5,0)} node[pos=0.5,pin={[pin distance=0cm]180:{\color{blue}$c_1$}}] {};
   \addplot[red,mark=*,dashed] coordinates {(2,3) (3,2)} node[pos=0.5,pin={[pin distance=0cm]180:{\color{red}$c_1$}}] {};
   \addplot[blue,mark=square*] coordinates {(4,2) (5,1) (6,0)} node[pos=0.5,pin={[pin distance=0cm]180:{\color{blue}$c_2$}}] {};
   \addplot[red,mark=*,dashed] coordinates {(3,3)} node[pos=0.5,pin={[pin distance=0cm]180:{\color{red}$c_2$}}] {};
   \addplot[blue,mark=square*] coordinates {(4,3) (5,2) (6,1) (7,0)} node[pos=0.5,pin={[pin distance=0cm]180:{\color{blue}$c_3$}}] {};
   \legend{add, subtract}
   \end{axis}
\end{tikzpicture}
\caption{Negacyclic convolution of two polynomials}
\label{negacyclicconvolutiondiagram}
\end{figure}

This is called a \emph{negacyclic}\index{Convolution!negacyclic} or \emph{negative wrapped} convolution. Figure~\ref{negacyclicconvolutiondiagram}
shows a diagram of it.
Please note that $\theta$ is in $\Z/(2^K+1)\Z$ as well a power of~2, so weighting can be done by a cyclic shift.

According to \eqref{ncycconv}, the $c_k$ can become negative.
Yet, we are looking for nonnegative $c'_k \equiv c_k \imod{2^K+1}$ with $c'_k \in [0 : 2^K]$.
If $c_k < 0$, we can find $c'_k \coloneqq c_k + 2^K + 1$.

\FloatBarrier
\subsection{The Procedure}
\label{smulproc}

We are now all set to describe the whole procedure:
given nonnegative integers $a$ and $b$ find their product $c \coloneqq ab$ modulo $2^N+1$.

Since the product is computed modulo $2^N+1$, we must choose $N$ big enough for the full product $c$.
If we choose $N \ge \lceil \log a \rceil + \lceil \log b \rceil$ this is surely the case.

Denote $R$ the ring $\Z/(2^K+1)\Z$, for some $K=r2^m$. Let $n \coloneqq 2^{m}$ be the FFT length and
let $s \coloneqq \lceil N / n \rceil$ be the bit length of input coefficients cut from $a$ and $b$.
Then our choice of parameters has to meet the following constraints:
\begin{itemize}
\item $R$ must contain a primitive $n$-th root of unity $\omega$ that is an even power of~2.
   $(2^{2x})^n \equiv 1 \equiv 2^{2K}$ leads to the sufficient condition
   \[ n \divides K. \numberthis\label{reqnthroot} \]
\item $R$ must be big enough to hold the convolution sums.
   Because of \eqref{ncycconv}, the $c_k \in [-n2^{2s}+1 : n2^{2s}-1]$, so the
   total range has size $2n2^{2s}-1$.
   Hence select $K$ so that $2^K+1 > 2n2^{2s}-1 = 2^{m+2s+1}-1$.
   It is sufficient to select
   \[  K \ge m + 2s + 1.  \numberthis\label{reqm2s}  \]
\item For increased speed, we might want to choose a larger $K$ that contains a higher power of~2.
   We will perform benchmarking later to find out if it pays off.
\end{itemize}

These constraints lead to values for the FFT length $n \coloneqq 2^{m}$,
the number of input bits per coefficient $s \coloneqq \lceil N/n \rceil$,
and $K = r2^m \ge m+2s+1$. This in turn forces a new, maybe slightly higher value for
$N \coloneqq s2^{m}$, and determines $\omega \coloneqq 2^{2r}$ and $\theta \coloneqq 2^{r}$.
Given those parameters, we can proceed like we did
with \QMUL{} in Section \ref{fftmodmul}, but with some alterations:
\begin{enumerate}
\item Split both input numbers $a$ and $b$ into $n$ coefficients of $s$ bits each. Use at least $K+1$~bits to store them,
   to allow encoding of the value $2^K$.\label{smulsplit}
\item Weight both coefficient vectors according to \eqref{weightcoeff} with
   powers of $\theta$ by performing cyclic shifts on them. \label{smulweightab}
\item Shuffle the coefficients $a_i$ and $b_j$. \label{smulshuffle}
\item Evaluate $a_i$ and $b_j$. Multiplications by powers of $\omega$ are cyclic shifts.  \label{smulevalab}
\item Do $n$ pointwise multiplications $c_k \coloneqq a_k b_k$ in $\Z/(2^K+1)\Z$.
   If \SMUL{} is used recursively, provide $K$ as parameter. Otherwise, use
   some other multiplication function like \T3MUL{} and reduce modulo $2^K+1$ afterwards.  \label{smulmodmul}
\item Shuffle the product coefficients $c_k$.  \label{smulshufflec}
\item Evaluate the product coefficients $c_k$.  \label{smulevalc}
\item Apply the counterweights to the $c_k$ according to \eqref{weightresult}.
   Since $\theta^{2n} \equiv 1$ it follows that $\theta^{-k} \equiv \theta^{2n-k}$.
   \label{smulweightc}
\item Normalize the $c_k$ with $1/n \equiv 2^{-m}$ (again a cyclic shift).  \label{smulnorm}
\item Add up the $c_k$ and propagate the carries. Make sure to properly handle negative coefficients.  \label{smulcarryprop}
\item Do a reduction modulo $2^N+1$.  \label{smulreduce}
\end{enumerate}

If \SMUL{} is used recursively, its input parameter $N$
cannot be chosen freely. The calling \SMUL{} provides its parameter $K$ as the input parameter $N$ of the called \SMUL{}.

\bigskip
I implemented some optimizations to the procedure outlined above to save execution time:\label{smul:opts}
\begin{itemize}
\item Steps \ref{smulsplit}, \ref{smulweightab} and \ref{smulshuffle} can be combined. Furthermore, since some high part
   of $a$ and $b$ is virtually zero-padded, initialization of that part can be done quickly.
\item Steps \ref{smulweightc} and \ref{smulnorm} can be combined.
\item On the outermost \SMUL{}, where $N$ can be chosen, we don't need to do a negacyclic transform. This
   lets us skip the weighting of $a_i$, $b_j$ and $c_k$ in Steps \ref{smulweightab} and \ref{smulweightc}.
   We don't check for negative coefficients in Step \ref{smulcarryprop} and don't need the reduction in Step \ref{smulreduce}.
   Furthermore, we don't need $\theta = \sqrt{\omega}$ and thus can extend the FFT length by another factor of~2. The sufficient condition for selecting $K$ relaxes to $n \divides 2K$.
\item The cyclic shifts often shift by multiples of the word size $w$, where a word-sized copy is faster
   than access to individual bits.
\end{itemize}

\subsection{Run-time Analysis}
\label{smul:runtime}
Let us analyze the cost of \SMUL{} to compute a product modulo $2^N+1$ and call this cost $T(N)$.

According to \eqref{reqnthroot}, it is clear that $K \ge 2^m$, but we will show the time bound
for the condition
\[ K = 2^m.  \numberthis\label{kand2m}  \]
This means that we might have to choose a $K$ that is larger than required by \eqref{reqnthroot} and
\eqref{reqm2s}, but our choice only increases $K$ by at most a factor of~2.

Furthermore, according to \eqref{reqm2s}, $K \ge m+2s+1$, where $s=\lceil N/2^m \rceil$.
Surely we will find a suitable $K = 2^m \le 2(m+2s+1)$. So for sufficiently large values of $N$
\begin{align*}
m+2N/2^m+1  &\le K      \le 2m + 4N/2^m + 2  \\
2N/K        &\le K      \le 5N/K  \\      
2N          &\le K^2    \le 5N.  \numberthis\label{k2mandn}  \\
\sqrt{2N}   &\le K       \le \sqrt{5N}.  \numberthis\label{ksqrtn}
\end{align*}

Steps \ref{smulsplit}, \ref{smulshuffle}, \ref{smulshufflec} and \ref{smulcarryprop} have obviously cost $O(2^m K)$.
The same applies to Steps \ref{smulweightab}, \ref{smulweightc} and \ref{smulnorm},
since the cost of cyclic shifts modulo $2^K+1$ is $O(K)$ as well.
By the same argument Step \ref{smulreduce} has cost $O(N)$.

According to \eqref{fftcost}, the FFT evaluation costs $O(n \log n)$, with $n = 2^m$,
but we have to take into account that in contrast to \eqref{fftcost}, multiplications by roots of unity
don't cost $O(1)$ here, but $O(K)$, so the cost of evaluation in Steps \ref{smulevalab} and \ref{smulevalc}
is $O(m 2^m)O(K)$.
That leaves Step \ref{smulmodmul}, where we have $2^m$ multiplications modulo $2^K+1$, so the cost
of that is $2^m T(K)$.

If we add everything up, we get for the total cost
\begin{align*}
T(N)  = {}   & O(2^m K) + O(N) + O(m 2^m)O(K) + 2^m T(K).  \\
\intertext{Using \eqref{kand2m} and \eqref{k2mandn} we get}
T(N)  = {}   & O(N) + O(m N) + K T(K)  \\
      = {}   & O(m N) + K T(K).  \\
\intertext{By \eqref{ksqrtn} we know that $2^m = K \le \sqrt{5N}$, hence $m \le \frac{1}{2} \log{(5N)}$. Ergo}
T(N)  = {}   & O(N \log N) + O(\sqrt{N}) T(\sqrt{5N}).  \\
\intertext{Unrolling the recursion once leads to}
T(N)  = {}   & O(N \log N) + O(\sqrt{N}) \big (O(\sqrt{5N} \log \sqrt{5N}) + O(\sqrt[4]{5N}) T(\sqrt[4]{5^3 N}) \big )  \\
      = {}   & O(N \log N) + O(\sqrt{N}) \big (O(\sqrt{N} \log N) + O(\sqrt[4]{N}) T(\sqrt[4]{5^3 N}) \big )  \\
      = {}   & O(N \log N) + O(N \log N) + \underbrace{O(\sqrt[4]{N^3}) T(\sqrt[4]{5^3 N})}_{\eqqcolon \Delta}.  \\
\intertext{After $\log \log N$ steps the remaining summand $\Delta \le O(N)$:}
T(N)  = {}   & O(N \log N) + \underbrace{O(N \log N) + \ldots}_{\log \log N \text{ times}} + \: O(N)  \\
      = {}   & O(N \log N) + O(N \log N) \log \log N + O(N)  \\
      = {}   & O(N \cdot \log N \cdot \log \log N).  \numberthis\label{sml:est}  
\end{align*}
To see why it takes $\log \log N$ steps, observe that
the order of the root doubles with each recursion step. Hence, after
$\log \log N$ steps the order has reached $2^{\log \log N} = \log N \eqqcolon \lambda$. So the
remaining summand $\Delta \le O(\sqrt[\lambda]{N^{\lambda-1}})T(\sqrt[\lambda]{5^{\lambda-1}N})
\le O(N)T(5 \sqrt[\lambda]{N})$. Lastly, $\sqrt[\lambda]{N} = N^{1/\log N} = 2$ and hence
$\Delta \le O(N)T(10) \le O(N)$.

Until the discovery of Fürer's algorithm \cite{Fuerer2007} in 2007 this was the lowest known time bound
for a multiplication algorithm.

\bigskip
Now let us look at memory requirements. Memory needed for all $2^m$ coefficients
of one input number is $2^mK$ bits.  According to \eqref{reqm2s} with $s=\lceil N/2^m \rceil$
it holds that $K \ge m + 2 \lceil N/2^m \rceil + 1$.
Hence memory requirements in bits for one polynomial are
\begin{align*}
2^m K    & \ge    2^m \cdot (m + 2 \lceil N/2^m \rceil + 1)  \\
         & \ge    2^m \cdot (m + 2 N/2^m + 1)  \\
         & \ge    2^m \cdot 2 N/2^m  \\
         & \ge    2 N.
\end{align*}
Temporary memory is required for both input polynomials, but for the resulting polynomial
storage of one of the input polynomials can be reused. \SMUL{} needs some memory for the multiplication
of sample points, but this is only of the size of one coefficient, that is, $K$ bits
and doesn't change the order of the approximation.
Hence, if $N$ denotes the bit length of the product and $M_{\SMUL{}}(N)$ denotes
total memory required by \SMUL{}, it holds that
\[  M_{\SMUL{}}(N) \approx 4 N \text{ bits}.  \numberthis\label{smul:mem}  \]

Figure~\ref{compmem} shows measured memory requirements, but note that in that table $N$ refers
to the bit length of one \emph{input}, where in this section $N$ denotes the bit length of the
\emph{product}.

\FloatBarrier
\subsection{Benchmarking}
\label{smul:speed}

Apart from optimizing the implementation on higher (see page~\pageref{smul:opts}) and lower
levels (like assembly language subroutines) benchmarking shows that
we can save quite some execution time by finding the fastest FFT length $n$
from all possible values.

\begin{figure}
\bigskip
\centering

\begin{tikzpicture}
\begin{semilogxaxis}[
   width=11cm,
   xlabel={Result words},
   ylabel={$\log$ of FFT length},
   legend pos=north west,
   legend cell align=left
]

\addplot[color=green] table[x index=0, y index=1] {smul-fft.txt};
\addlegendentry{\SMUL{} FFT length}

\addplot[domain=94:85991425, blue] { ln(sqrt(x*64*2)) / ln(2) };  
\addlegendentry{$\sqrt{N}$}

\end{semilogxaxis}
\end{tikzpicture}

\caption{\SMUL{} FFT length vs.\ input length}
\label{smulfftwiggle}
\end{figure}
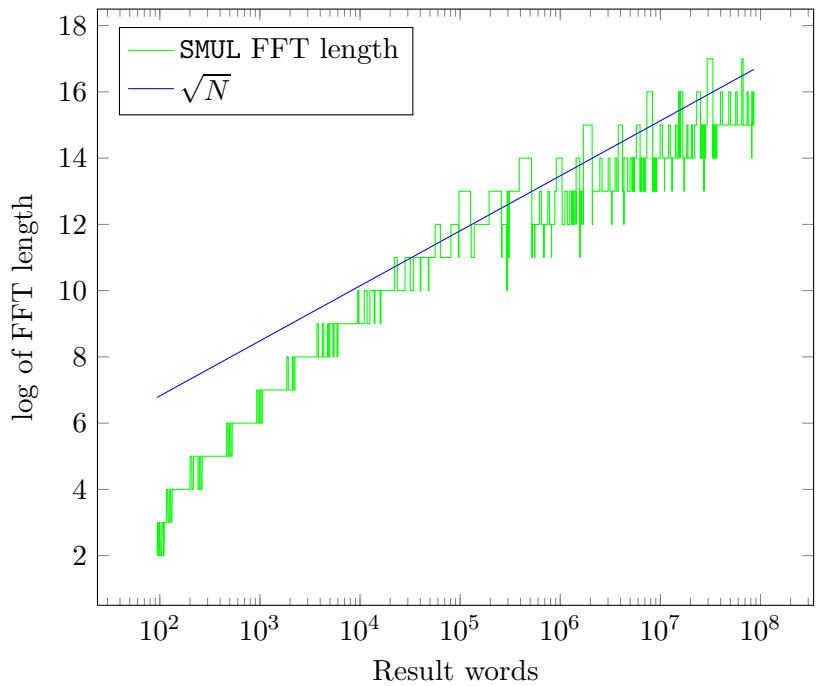

For this, we measure execution cycles for multiplications with different
possible FFT lengths.  In principle, larger FFT lengths lead to faster multiplications,
but the largest possible FFT length is usually not the fastest.
Larger FFT lengths lead to smaller coefficient sizes,
but more operations on the coefficients. On the other hand, the value of the
primitive $n$-th root $\omega$ might allow byte- or even word aligned
(or even better SSE-word aligned) shifts, which can be implemented faster than
general bit-shifts. The smaller the FFT length, the better the alignment for
the cyclic shifts.

Maybe even more importantly, certain values of $K$ that contain high powers of~2
allow for larger FFT lengths in the recursively called \SMUL{}.
So sometimes larger $K$ work much faster, even if the FFT length stays unchanged.

As a result, the fastest FFT length switches several times until it settles for
a higher value. Figure~\ref{smulfftwiggle} gives an impression of this.
The $\sqrt{N}$ graph is printed for orientation, since $2^m \ge \sqrt{2N/r}$.

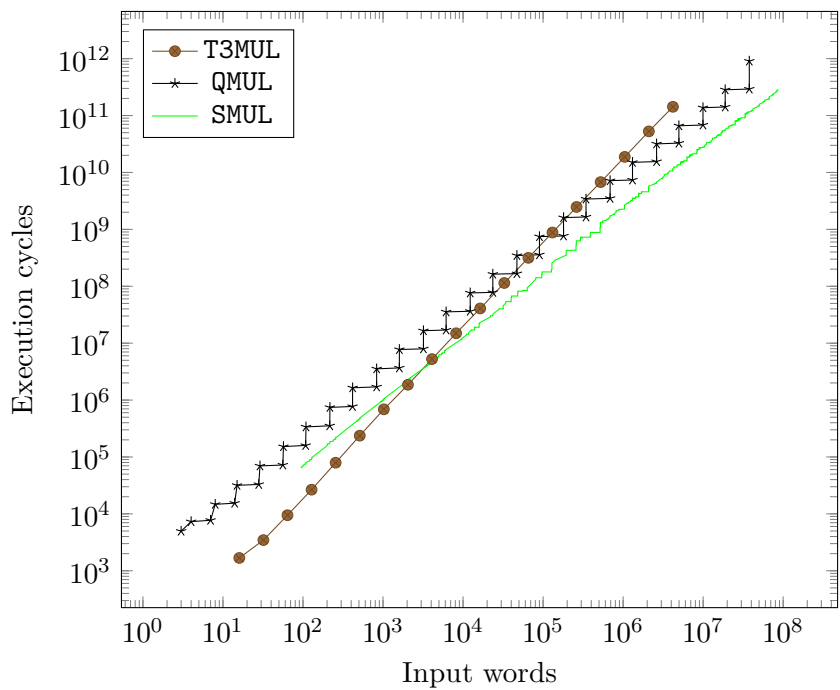
\begin{figure}
\bigskip
\centering

\begin{tikzpicture}
\begin{loglogaxis}[
   width=11cm,
   xlabel={Input words},
   ylabel={Execution cycles},
   legend pos=north west
]

\pgfplotsset{cycle list shift=2}
\addplot table[x index=0, y index=1] {t3mul-large-speed.txt};
\addlegendentry{\T3MUL{}}
\addplot table[x index=0, y index=1] {qmul-large-speed.txt};
\addlegendentry{\QMUL{}}

\addplot[color=green] table[x index=0, y index=1] {smul-speed-short.txt};
\addlegendentry{\SMUL{}}

\end{loglogaxis}
\end{tikzpicture}

\caption{Execution time of \SMUL{}}
\label{smulgraph}
\end{figure}

The graph of execution cycles vs.\ input lengths is shown in Figure~\ref{smulgraph}.
We can see that it is well below the \QMUL{} graph, but intersects with the \T3MUL{} graph
at about 2500 words, that is, about \num{47500} decimal digits. Furthermore,
we observe a certain ``bumpiness'' of the graph,
which is a result of the changing FFT lengths and ring sizes. Yet, it is much smoother
than the \QMUL{} graph.

\begin{figure}
\bigskip
\centering

\begin{tikzpicture}
\begin{semilogxaxis}[
   width=11cm,
   xlabel={Input bits},
   ylabel={Constant $\sigma$},
   legend pos=north west
]

\addplot[color=green] table[x index=0, y index=1] {smulc-graph.txt};
\addlegendentry{\SMUL{} constant $\sigma$}
\node[pin={[pin distance=1cm]270:Level~1 cache size},draw=black] at (axis cs:32768,0.245) {};
\node[pin={[pin distance=1cm]90:Level~2 cache size},draw=black] at (axis cs:262144,0.245) {};
\node[pin=130:Level~3 cache size,draw=black] at (axis cs:8388608,0.28) {};
\addplot[domain=33554432:5503451200, blue] { \avgsmulc };  
\addlegendentry{average $\sigma$}

\end{semilogxaxis}
\end{tikzpicture}

\caption{\SMUL{} run-time constant $\sigma$}
\label{smulcgraph}
\end{figure}
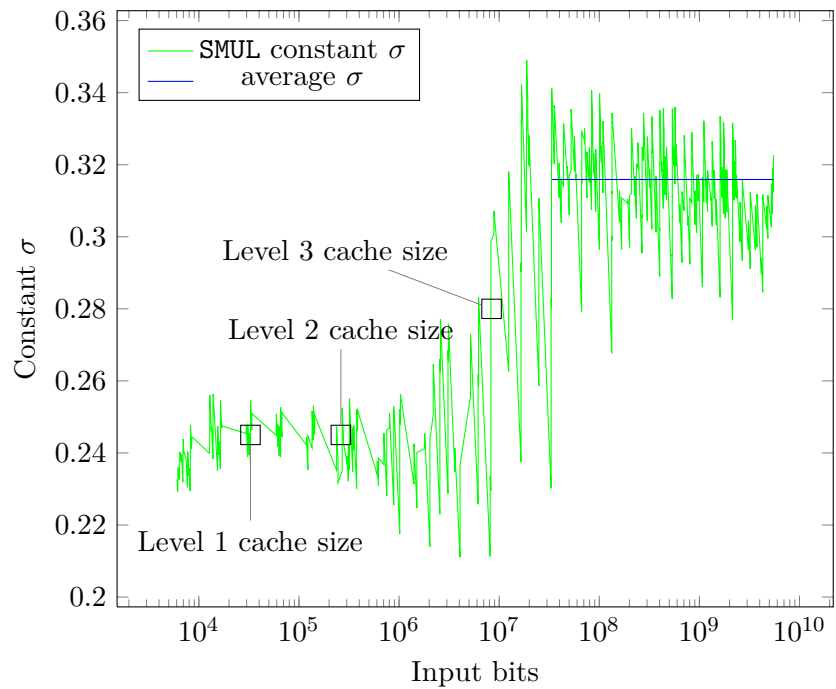

Lastly, we try to model the run-time according to its theoretical value \eqref{sml:est} for large values of $N$.
If we write the run-time with an explicit constant, then
\[  T_\sigma(N) \le \sigma \cdot N \cdot \log N \cdot \log \log N.  \numberthis\label{smul:modeltime}  \]
Dividing measured execution cycles
by $N \cdot \log N \cdot \log \log N$ to calculate $\sigma$ leads to the graph depicted in Figure~\ref{smulcgraph}.
Please note that here $N$ is the length of the product in \emph{bits}.
Interestingly, this graph seems to have two plateau-like sections.

The first plateau ranges roughly from \num{12800} to \num{8000000} input bits and the second
plateau starts at about \num{32000000} input bits. Since \SMUL{} requires about $4 N$ bits of
temporary memory, the above numbers indicate a plateau from 12~KB to 8~MB and
another starting from 32~MB temporary memory. This corresponds quite nicely with the cache sizes
of the test machine (see Appendix~\ref{tech}). Such an influence on the run-time constant $\sigma$ is
no longer visible only after
the required temporary memory is some orders of magnitude larger than the cache size. In our
case that would be starting from about 32~MB temporary memory.

Since after level~3 there are no other caching mechanisms, we can average $\sigma$ for
input sizes above 32~Mbits (since a word is 64~bits, this equals 512~Kwords) and that
leads to an average $\sigma \approx \avgsmulc{}$.

\FloatBarrier
\subsection{Outlook}

There are some possible improvements which might lower execution time that I have so far not implemented or tested.
Namely, this is:

\begin{itemize}
\item I benchmark different FFT lengths $n$ to find the fastest one, but I could benchmark different
   ring sizes $K$ as well. It is sometimes profitable to use larger values for $K$, if $K$ contains higher powers of~2.
\item Schönhage's $\sqrt{2}$ knack%
   \footnote{\cite{Gaudry2007} call it $\sqrt{2}$ \emph{trick}, but Schönhage interjects
      that ``trick'' sounds too much like a swindle, so I call it a \emph{knack} instead.}%
   \cite[p.~36, exercise 18]{Schoenhage1994}. We can increase the transform length by a factor of~2 by
   noticing that $\xi = 2^{3K/4}-2^{K/4}$ is a primitive $4K$-th root of unity, since
   $\xi^2 \equiv 2 \pmod{2^K+1}$ and $2^{2K} \equiv 1 \pmod{2^K+1}$.
   In \cite{Gaudry2007}, the authors mention a 10~\% speed increase. That paper also contains some other
   promising fields of optimization.
\item The current implementation cuts input numbers into coefficients only at word boundaries.
   Maybe cutting them at bit boundaries
   might lower $K$ slightly for some input lengths.

\item The procedure outlined in \cite[pp.~502--503]{Crandall2005} saves one bit in $K$ by selecting
   \mbox{$K \ge m+2s$} and then uses a sharper bound for the $c_k$ than I did by noticing that
   each $c_k$ can only have $k+1$ positively added summands, see Figure~\ref{negacyclicconvolutiondiagram}.

\end{itemize}

\chapter{The DKSS Algorithm}
\label{chapter3}
\fancyhead[RE,LO]{Chapter 3. \emph{The DKSS Algorithm}}
\fancyhead[LE,RO]{\thepage}

This chapter describes DKSS multiplication,
especially how it employs the fast Fourier transform,
and analyzes its execution time theoretically. Finally, differences between my implementation
and the DKSS paper are described.

\section{Overview}

Schönhage and Strassen's algorithm for fast multiplication of large numbers
(implemented as \SMUL{}, see Section~\ref{smul}) uses the
ring $R = \Z/(2^K+1)\Z$ and exploits the fact that $2$ is a primitive $2K$-th root of unity
in $R$. This permits the crucial speed-up in the fast Fourier transform: multiplications
by powers of the root of unity can be realized as cyclic shifts and are thus considerably cheaper.
An $N$-bit number is broken down into numbers that are $O(\sqrt N)$ bits long and
when sample values are multiplied, the same algorithm is used recursively.

The DKSS algorithm (its implementation is called \DMUL{} here)
keeps this structure, but extends it further.  Where \SMUL{} used the ring
$\Z/(2^K+1)\Z$ with $2$ as root of unity, DKSS multiplication uses the polynomial quotient ring
$\R \coloneqq \P[\alpha]/(\alpha^m+1)$.  Since $\alpha^m \equiv -1$, $\alpha$ is a primitive $2m$-th root of
unity and again multiplications by powers of the root of unity can be done as cyclic shifts.
Underlying $\R$ is the ring $\P \coloneqq \Zpc$, where $p$ is a prime number and $z$ is a constant.
This ``double structure'' can be exploited in the FFT and allows to break down an $N$-bit input
number into numbers of $O(\log^2 N)$ bits.

In their paper \cite{De2013}, \DKSS{} describe the algorithm without any assumption about
the underlying hardware. Since we are interested in an actual implementation,
we can allow ourselves some simplifications, namely the precomputation of the prime $p$,
and as a consequence drop their concept of $k$-variate polynomials by setting $k=1$.
Section \ref{dkss:diffs} explains the differences
between my implementation and the original paper in more detail.

\section{Formal Description}
\index{DKSS multiplication!formal description}
\label{dkss:desc}

We want to multiply two nonnegative integers $a$, $b < 2^N$, $N \in \N$ to obtain their product $c \coloneqq ab < 2^{2N}$.
As usual, we convert the numbers into polynomials over a ring (denoted~$\R$),
use the fast Fourier transform to transform their coefficients, then multiply the sample values and transform backwards
to gain the product polynomial. From there, we can easily recover the resulting integer product.

Denote $\R \coloneqq \P[\alpha]/(\alpha^m+1)$.
As usual, we identify $\R$ with the set of all polynomials in $\P[\alpha]$ which are of degree less than
$m$ and where polynomial multiplication is done modulo $(\alpha^m+1)$.
Polynomial coefficients are in $\P$ and are called \emph{inner} coefficients.
Furthermore, define $\P \coloneqq \Zpc$, where $p$ is a prime number and $z$ is a
constant chosen independently of the input. We will see how to choose $p$ shortly.

Input numbers $a$ and $b$ are encoded as polynomials $a(x)$ and $b(x) \in \R[x]$ with degree-bound $M$.
That is, $a(x)$ and $b(x)$ are polynomials over $\R$ whose coefficients are themselves polynomials over $\P$.
Call the coefficients of $a(x)$ and $b(x)$ \emph{outer} coefficients.

This outline shows how to multiply $a$ and $b$. The following Sections
\ref{dkss:sec:choosem}~-- \ref{dkss:sec:carryadd} contain the details.
\begin{enumerate}
\item \label{dkss:step:choosem}  
Choose integers $m \ge 2$ and $M \ge m$ as powers of~2, such that $m \approx \log N$ and $M \approx N/\log^2 N$.
We will later perform FFTs with length $2M$, while $m$ is the degree-bound of elements of $\R$.
For simplicity of notation, define $\mu \coloneqq 2M/2m$.

\item \label{dkss:step:findp}  
Let $u \coloneqq \lceil 2N/Mm \rceil$ denote the number of input bits per inner coefficient.
Find a prime $p$ with $2M \divides p-1$ and $p^z \ge Mm2^{2u}$.
The prime power $p^z$ is the modulus of the elements of $\P$.

\item \label{dkss:step:findrho}  
From parameters $M$, $m$ and $p$ compute a principal (see Section~\ref{dkss:sec:findp} for definition) $2M$-th root of unity%
\footnote{We simply write $\rho$ instead of $\rho(\alpha)$, keeping in mind that $\rho$ itself is a polynomial in $\alpha$.}
$\rho \in \R$ with the additional property that $\rho^{2M/2m}=\alpha$.
This property plays an important part in Step \ref{dkss:step:forwfft}.

\item \label{dkss:step:encode}  
Encode $a$ and $b$ as polynomials $a(x)$, $b(x) \in \R[x]$ with degree-bound $M$.
To accomplish that, break them into $M$ blocks with $um/2$ bits in each block.
Each such block describes an outer coefficient.
Furthermore, split those blocks into $m/2$~blocks of $u$ bits each,
where each block forms an inner coefficient in the lower-degree half of a polynomial.
Set the upper $m/2$ inner coefficients to zero. Finally, set the upper $M$ outer coefficients to zero
to stretch $a(x)$ and $b(x)$ to degree-bound~$2M$.

\item \label{dkss:step:forwfft}  
Use root $\rho$ to perform a length-$2M$ fast Fourier transform of $a(x)$ and $b(x)$
to gain $\widehat{a}_i \coloneqq a(\rho^i) \in \R$, likewise $\widehat{b}_i$.
Use the special structure of $\R$ to speed up the FFT.

\item \label{dkss:step:compmul}  
Multiply components $\widehat{c}_i \coloneqq \widehat{a}_i \widehat{b}_i$.
Note that $\widehat{a}_i$, $\widehat{b}_i \in \R$ are themselves polynomials.
Their multiplication is reduced to integer multiplication and
the DKSS algorithm is used recursively.

\item \label{dkss:step:backwfft}  
Perform a backwards transform of length $2M$ to gain the product polynomial $c(x) \coloneqq a(x)b(x)$.

\item \label{dkss:step:carryadd}  
Evaluate the inner polynomials of the product polynomial $c(x)$ at $\alpha = 2^u$
and the outer polynomials at $x = 2^{um/2}$ to recover the integer result $c = ab$.
\end{enumerate}

\subsection{Choosing $M$ and $m$}
\label{dkss:sec:choosem}

Choose $m \ge 2$ and $M \ge m$ as powers of 2, such that $m \approx \log N$ and $M \approx N/\log^2 N$.
For the run-time analysis, the bounds $M = O(N/\log^2 N)$ and $m = O(\log N)$ are more convenient.

\subsection{Finding the Prime $p$}
\label{dkss:sec:findp}

We use the following definition
that captures the requirements for the existence of the inverse FFT transform
(cf.\ Sections \ref{fftpolymul} and \ref{smul:inverse}):

\begin{mydef}
\hspace*{0pt}\refstepcounter{dummy}\index{Principal root of unity}%
Let $R$ be a commutative ring with unity.
A primitive $n$-th root of unity $\zeta \in R$ is called \emph{principal} if and only if
$\sum_{i=0}^{n-1} (\zeta^j)^i = 0$, for $j \in [1 : n-1]$, and
$n$ is coprime to the characteristic of $R$.
\end{mydef}

Since numbers are encoded as polynomials with degree-bound $M$ (Step \ref{dkss:step:encode})
and then multiplied, the result has a degree-bound of $2M$,
so we need a principal $2M$-th root of unity for the FFTs.  If $p \coloneqq h \cdot 2M + 1$ is prime
for some $h \in \N$ (primes of this form are called \index{Proth prime}\emph{Proth primes}),
we can compute a principal $2M$-th root of unity $\omega$ in $\Zpc$.
Section \ref{dkss:sec:findrho} shows how it is done.

Why is $p^z \ge Mm2^{2u}$ required? Since both
$a(x)$ and $b(x)$ have degree-bound $M$, each outer coefficient of their
product $c(x)$ is the sum of up to $M$ outer coefficient products. Each of these products
is the sum of up to $m/2$ inner coefficient products, with each
factor $< 2^u$ by construction.  So the inner coefficients can take values as high as
$\half Mm(2^u-1)^2$.
If we choose $p^z \ge Mm2^{2u}$, we are on the safe side.

But does a prime of the form $p = h \cdot 2M+1$ exist for all $M$?  We can answer that
in the affirmative with the help of the following

\index{Linnik's theorem}
\begin{mythm}[Linnik \cite{Linnik1944a}, \cite{Linnik1944b}]
For any pair of coprime positive integers $d$ and $n$, the least prime $p$ with $p \equiv d \imod n$
is less than $\ell n^L$, where $\ell$ and $L$ are positive constants.%
\footnote{Over the years, progress has been made in determining the size of \emph{Linnik's constant} $L$.
A recent work by Xylouris \cite{Xylouris2011} shows that $L \le 5$.}
\end{mythm}

We want to show the existence of a prime $p$ with
$p \equiv 1 \imod{2M}$, but also require $p^z \ge Mm2^{2u}$.
Since Linnik's Theorem makes only a statement about the \emph{first} prime,
we must check that this prime to a constant power matches the requirement.
An easy calculation shows that $(2M+1)^6 \ge Mm2^{2u}$.
\label{zsix}
As $p$ is of the form $p=h \cdot 2M+1$, we see that for every $h \in \N$ and every $z \ge 6$
this means that $p^z \ge Mm2^{2u}$.
With the size condition resolved, we use Linnik's theorem to show that $p < \ell (2M)^L$.
\texttombstone

To get an estimate of $p$ in terms of $N$, we recall that $M = O(N / \log^2 N)$ and see that
\[  p < \ell (2M)^L = O \Big( \Big( \frac{N}{\log^2 N} \Big)^L \Big) = O \Big( \frac{N^L}{\log^{2L} N} \Big).  \numberthis\label{dkss:peqo} \]

In the implementation I tested candidate primes $p$ for primality by using
the Lucas-test \cite[sec.~4.1]{Crandall2005} that allows for fast deterministic primality
testing if the full factorization of $p-1$ is known.
$p-1$ is a power of~2 times a small factor, because $p = h \cdot 2M+1$,
so this test is well suited here.

With that in mind, we can precompute values for $p$ for all possible lengths $N$,
since our supported hardware is 64-bit and hence $N < 2^{67}$ and (assuming $M \approx N/\log^2 N$) $M < 2^{55}$.

\subsection{Computing the Root of Unity $\rho$}
\label{dkss:sec:findrho}

In Step \ref{dkss:step:findp} we computed a prime $p = h \cdot 2M+1$, $h \in\N$.
Now we want to find a $2M$-th root of unity $\omega$ in $\Zpc$.
A generator $\zeta$ of $\F_p^* = \{ 1, 2, \ldots, p-1 \}$ has order
$p-1 = h \cdot 2M$. Hence $\zeta$ is a primitive $(p-1)$-th root of unity
and $\zeta^h$ a primitive $2M$-th root of unity in $\Zp$. In fact, both $\zeta$ and $\zeta^h$ are even principal.
The following theorem allows us
to find roots in $\Z/p^s\Z$ for integer values $s\ge 2$:

\index{Hensel lifting}
\begin{mythm}[Hensel Lifting {\cite[sec.~2.6]{Niven1991}}]
Let $f\in\Z[x]$ and let $\zeta_s\in\Z$ be a solution to $f(x) \equiv 0 \imod{p^s}$, such that
$f'(\zeta_s)$ is a unit in $\Z/p\Z$. Then
\mbox{$\zeta_{s+1} \coloneqq \zeta_s - f(\zeta_s) / f'(\zeta_s)$} solves $f(x) \equiv 0 \imod{p^{s+1}}$
and furthermore $\zeta_{s+1} \equiv \zeta_s \imod{p^s}$.
\end{mythm}

Finding a primitive $(p-1)$-th root of unity in $\Z/p^z\Z$ means solving $f(x) = x^{p-1}-1$.
We can use Hensel lifting, because $f'(\zeta_s)=(p-1)\zeta_s^{p-2}$ is a unit in $\Zp$,
since $p-1 \not= 0$ and $\zeta_s^{p-2} \equiv \zeta^{p-2} \not\equiv 0 \imod p$.
If we start with $x=\zeta$ as solution to $f(x) \equiv 0 \imod{p}$,
then repeated lifting yields a $(p-1)$-th root of unity $\zeta_z$ in $\Zpc$.
Hence $\omega \coloneqq \zeta_z^h$ is a $2M$-th root of unity in $\Zpc$.
To see that $\omega$ is also primitive, let $j \in [1 : 2M-1]$.
Then $\omega^j = \zeta_z^{hj} \equiv \zeta^{hj} \not\equiv 1 \imod{p}$,
as $\zeta$ is a primitive $(p-1)$-th root of unity in $\Z/p\Z$.

To prove that $\omega$ is even principal note that the characteristic of $\R$ is $p^z$,
so $\omega$ has to be coprime to $p^z$, that is, coprime to $p$.
But $\omega = \zeta_z^h \equiv \zeta^h \not\equiv 0 \imod{p}$,
so $\omega$ is not a multiple of $p$. Hence $\omega$ and $p^z$ are coprime.
Furthermore, it holds for $j \in [1 : 2M-1]$ that
\[  \sum_{i=0}^{2M-1} (\omega^j)^i = \frac{1-\omega^{j2M}}{1-\omega^j}  = \frac{1-(\omega^{2M})^j}{1-\omega^j} \equiv 0 \imod{p^z}, \]
because $\omega$ is a primitive $2M$-th root of unity in $\Zpc$.
\texttombstone

\bigskip
We are looking for a principal $2M$-th root of unity $\rho \in \R$ with the additional
property $\rho^{2M/2m} = \alpha$.
Since $\R = \P[\alpha]/(\alpha^m+1)$, $\alpha$ is a principal $2m$-th root of unity.
Denote $\gamma \coloneqq \omega^{2M/2m}$, a principal $2m$-th root of unity in $\P$.
Observe that $\gamma^{i}$ is a root of $\alpha^m+1 = 0$, for an odd $i \in [1 : 2m-1]$,
since $(\gamma^{i})^{m} = (\gamma^{m})^i = (-1)^i = -1$.
Because the $\gamma^{i}$ are pairwise different
it follows that
\[  \prod_{\substack{i=1 \\ i \text{ odd}}}^{2m-1} (\alpha - \gamma^{i}) = \alpha^m+1.  \]

\index{Chinese remainder theorem}
\begin{mythm}[Chinese Remainder Theorem {\cite[sec.~2.11]{Fischer2011}}]
If $R$ is a commutative ring with unity and $I_1$, \ldots, $I_k$ are ideals of $R$,
which are pairwise coprime (that is, $I_i + I_j = R$, for $i \not= j$), then
the mapping
\begin{gather*}
\phi: R \to R/I_1 \times \ldots \times R/I_k,  \\
x \mapsto (x+I_1, \ldots, x+I_k)
\end{gather*}
is surjective and $\ker \phi = I_1 \cdot \ldots \cdot I_k$.
Especially, $\phi(x) = \phi(x')  \Leftrightarrow  x - x' \in I_1 \cdot \ldots \cdot I_k$ and
\[  R/(I_1 \cdot \ldots \cdot I_k)  \cong  R/I_1 \times \ldots \times R/I_k.  \]
\end{mythm}%
If the ideals $I_i$ are generated by $(\alpha - \gamma^{i})$, they are pairwise coprime,
since $\gamma^{i} - \gamma^{j}$ is a unit in $\R$, for $i \not= j$, see \eqref{dkss:lagrangeunits} below.
So $\alpha \in \P[\alpha]/(\alpha^m+1)$ is isomorphic to the \mbox{$k$-tuple} of remainders
$(\gamma, \gamma^3, \ldots, \gamma^{2m-1}) \in \prod_{i}(\P[\alpha]/I_i)$.
We are looking for a $\rho$ satisfying $\rho^{2M/2m}=\alpha$, but we already know that $\omega^{2M/2m}=\gamma$,
hence $(\omega, \omega^3, \ldots, \omega^{2m-1})$ is the tuple of remainders isomorphic to $\rho$.
To regain $\rho \in \P[\alpha]/(\alpha^m+1)$ we use the next

\index{Lagrange interpolation}
\begin{mythm}[Lagrange Interpolation]
Let R be a commutative ring with unity. Given a set of
$k$ data points $\{ (x_1, y_1)$, \ldots, $(x_k, y_k) \}$ with $(x_i, y_i) \in R \times R$,
where the $x_i$ are pairwise different and $x_i-x_j$ is a unit for all $i\ne j$,
there exists a polynomial $L(x)$ of degree less than $k$ passing
through all $k$ points $(x_i, y_i)$. This polynomial is given by
\[  L(x) \coloneqq \sum_{i=1}^k y_i \ell_i(x), \quad \text{ where } \quad \ell_i(x) \coloneqq \prod_{\substack{j=1 \\ j \not= i}}^k \frac{x - x_j}{x_i - x_j}.  \]
\end{mythm}

In our case we know that $\rho(\alpha) \cong (\omega, \omega^3, \ldots, \omega^{2m-1})$,
so it follows that the set of data points is $\{ (\gamma, \omega)$, $(\gamma^3, \omega^3)$, \ldots, $(\gamma^{2m-1}, \omega^{2m-1}) \}$
and hence
\[  \rho(\alpha) \coloneqq \sum_{\substack{i=1 \\ i \text{ odd}}}^{2m-1} \omega^i \ell_i(\alpha), \quad \text{ where } \quad \ell_i(\alpha) \coloneqq \prod_{\substack{j=1 \\ j \not= i \\ j \text{ odd}}}^{2m-1} \frac{\alpha - \gamma^{j}}{\gamma^{i} - \gamma^{j}}.  \numberthis\label{dkss:lagrange}  \]
The inverses to $\gamma^{i} - \gamma^{j}$ exist.  To see why,
observe that an element of $\Zpc$ is a unit if and only if it is not divisible by $p$.
But
\begin{align*}
\begin{split}
\gamma^{i} - \gamma^{j}       &= \zeta_z^{i(p-1)/2m} - \zeta_z^{j(p-1)/2m}   \\
                              &\equiv \zeta^{i(p-1)/2m} - \zeta^{j(p-1)/2m} \imod{p}  \\
                              &\not\equiv 0 \imod{p},
\end{split} \label{dkss:lagrangeunits} \numberthis
\end{align*}
because $\zeta$ is a primitive $(p-1)$-th root of unity and $i$, $j \in [1:2m-1]$ and since $i \not= j$
the two exponents of $\zeta$ are different.    \texttombstone

\subsection{Distribution of Input Bits}
\label{dkss:sec:encode}

We want to encode a nonnegative integer $a<2^N$ as polynomial over $\R[x]$ with degree-bound $M$.
We already calculated $u=\lceil 2N/Mm \rceil$, the number of bits per inner coefficient.
First, $a$ is split into $M$ blocks of $um/2$ bits each, starting at the lowest bit position.
Each of these blocks encodes one outer coefficient.
Since $Mum/2 \ge N$, we might need to zero-pad $a$ at the top.

Then, each of the $M$ outer coefficient blocks is broken into $m/2$ blocks, each $u$ bits wide. They form the
inner coefficients. Since the inner coefficients describe a polynomial with degree-bound $m$,
the upper half of the coefficients is set to zero.

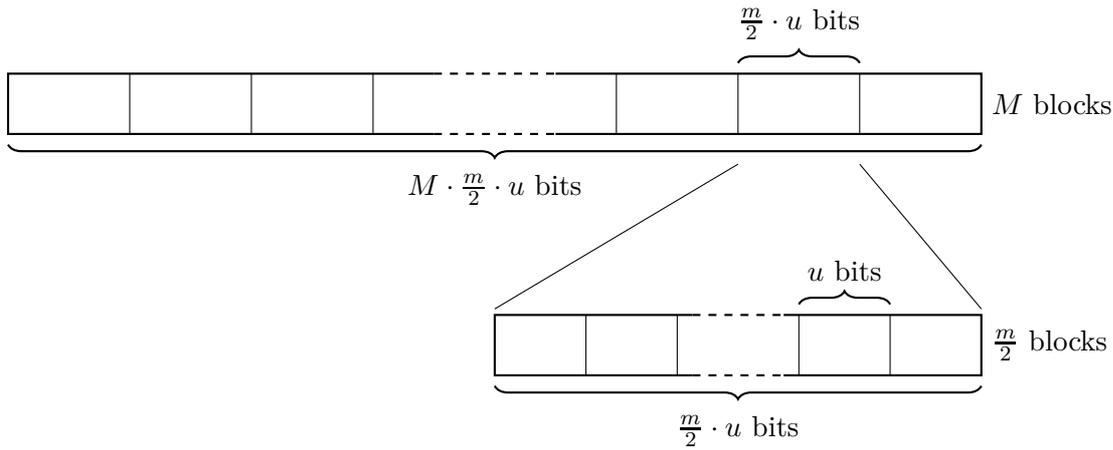
\begin{figure}
\begin{tikzpicture}[scale=0.8]

\draw[thick] (7,1) -- (0,1) -- (0,0) -- (0,0) -- (7,0) ;
\draw[thick] (9,0) -- (16,0) -- (16,1) -- (9,1) ;
\draw[thick, dashed] (7,0) -- (9,0) ;
\draw[thick, dashed] (7,1) -- (9,1) ;
\draw[thin] ( 2,0) -- ( 2,1) ;
\draw[thin] ( 4,0) -- ( 4,1) ;
\draw[thin] ( 6,0) -- ( 6,1) ;
\draw[thin] (10,0) -- (10,1) ;
\draw[thin] (12,0) -- (12,1) ;
\draw[thin] (14,0) -- (14,1) ;
\node[right] at (16,0.5) {$M$ blocks} ;

\draw [thick, decoration={brace, amplitude=5pt, raise=4pt}, decorate] (12,1) -- (14,1) ;
\node[above] at (13,1.4) {$\frac{m}{2} \cdot u$ bits} ;

\draw [thick, decoration={brace, amplitude=5pt, mirror, raise=4pt}, decorate] (0,0) -- (16,0) ;
\node[below] at (8,-0.5) {$M\cdot \frac{m}{2} \cdot u$ bits} ;

\draw (12,-0.5) -- (8,-2.9) ;
\draw (14,-0.5) -- (16,-2.9) ;

\draw[thick] (11.25,-3) -- (8,-3) -- (8,-4) -- (11.25,-4) ;
\draw[thick] (12.75,-3) -- (16,-3) -- (16,-4) -- (12.75,-4) node[right] at (16,-3.5) {$\frac{m}{2}$ blocks};
\draw[thick, dashed] (11.25,-3) -- (12.75,-3) ;
\draw[thick, dashed] (11.25,-4) -- (12.75,-4) ;
\draw[thin] (9.5,-3) -- (9.5,-4) ;
\draw[thin] (11,-3) -- (11,-4) ;
\draw[thin] (13,-3) -- (13,-4) ;
\draw[thin] (14.5,-3) -- (14.5,-4) ;

\draw [thick, decoration={brace, amplitude=5pt, mirror, raise=4pt}, decorate] (8,-4) -- (16,-4) ;
\node[below] at (12,-4.5) {$\frac{m}{2} \cdot u$ bits} ;

\draw [thick, decoration={brace, amplitude=5pt, raise=4pt}, decorate] (13,-3) -- (14.5,-3) ;
\node[above] at (13.75,-2.6) {$u$ bits} ;

\end{tikzpicture}
\caption{Encoding an input integer as a polynomial over $\R$}
\label{dkss:fig:encode}
\end{figure}

Finally, set the upper $M$ outer coefficients to zero
to stretch $a(x)$ to degree-bound $2M$.
Figure~\ref{dkss:fig:encode} depicts this process.

\subsection{Performing the FFT}
\label{dkss:sec:forwfft}

Section \ref{fft} described a radix-2 Cooley-Tukey FFT.
The DKSS algorithm uses an FFT with a higher radix, but still the same basic concept.
A Cooley-Tukey FFT works for any length that is a power of~2, here the length is $2M$ and it
can be split as $2M = 2m \cdot \mu$, with $\mu = 2M / 2m$.

The DKSS algorithms uses a radix-$\mu$ decimation in time Cooley-Tukey FFT (cf.\ \cite[sec.~4.1]{Duhamel1990}),
that is, it first does $\mu$ FFTs of length $2m$, then multiplies the
results by \index{Twiddle factors}``twiddle factors''
and finally performs $2m$ FFTs of length $\mu$.
We can exploit the fact that the length-$2m$ FFT uses $\alpha$ as
root of unity, since multiplications with powers of $\alpha$ can be performed
\index{Cyclic shift}%
as cyclic shifts and are thus cheap.

We now describe the process formally.
By construction, $a(x) \in \R[x]$ is a polynomial with degree-bound $2M$ and
$\rho \in \R$ is a principal $2M$-th root of unity.
Bear in mind that $\rho^{2M/2m} = \rho^\mu = \alpha$.
Since $\alpha^m \equiv -1$, $\alpha$ is a primitive $2m$-th root of unity in $\R$.
We can compute the length-$2M$ DFT of $a(x)$ with $\rho$ as root of unity in three steps:

\begin{enumerate}[label=\roman*.,ref=\roman*]

\item \label{innerdft}  
Perform \emph{inner DFTs}.%
\footnote{Please note that the inner and outer \emph{DFTs} have no relation to the inner or outer \emph{coefficients}.}

Figure~\ref{dkss:mot:input} shows the input vector $a$, which contains the coefficients
of the polynomial $a(x)$. The arrow indicates the ordering of the elements for the DFT.
\bigskip

\begin{minipage}{\linewidth}
\centering
\begin{tikzpicture}
{\small
\matrix [matrix of math nodes,left delimiter=[,right delimiter={]}] (m) {
   a_0   & a_1    & \ldots & a_{\mu-1}  & a_{\mu}   & a_{\mu+1} & \ldots & a_{(2m-1)\mu-1}  & a_{(2m-1)\mu}  & a_{(2m-1)\mu+1} & \ldots & a_{2m\mu-1} \\
};}
\node[outer sep=2pt] (braceleft) at (m.north west) {};
\node[outer sep=2pt] (braceright) at (m.north east) {};
\draw [thick, decoration={brace, amplitude=5pt, raise=8pt}, decorate] (braceleft) -- (braceright)
node [pos=0.5,anchor=south,yshift=12pt] {$2M = 2m \cdot \mu$ elements};
\begin{scope}[on background layer]
   \draw[->,line width=2.5mm,gray!15] (m-1-1.west) -- (m-1-12.east);
\end{scope}
\end{tikzpicture}

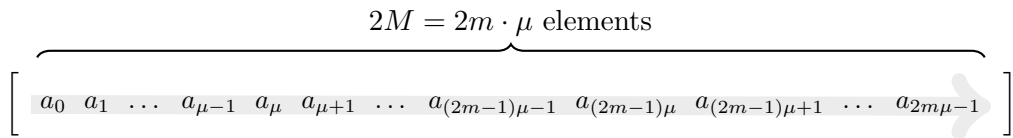
\captionof{figure}{Input vector $a$}
\label{dkss:mot:input}
\end{minipage}

\bigskip
Rewrite the input vector $a$ as $2m$ rows of $\mu$ columns and perform FFTs
on the columns, see Figure~\ref{dkss:mot:rewrite}. The boxes
hold the values of vectors called $e_\ell$, while the arrows indicate the ordering
of their elements.
\bigskip

\begin{minipage}{\linewidth}
\centering
\begin{tikzpicture}
\matrix [matrix of math nodes,left delimiter=[,right delimiter={]}] (m) {
   a_0            & a_1                & \ldots & a_{\mu-1}    \\
   a_{\mu}        & a_{\mu+1}          & \ldots & a_{2\mu-1}   \\
   \vdots         & \vdots             & {}     & \vdots       \\
   a_{(2m-1)\mu}  & a_{(2m-1)\mu+1}    & \ldots & a_{2m\mu-1}  \\
};
\node[outer sep=2pt] (upbraceleft) at (m.north west) {};
\node[outer sep=2pt] (upbraceright) at (m.north east) {};
\node[outer sep=-2pt] (lbraceup) at (m.north east) {};
\node[outer sep=-2pt] (lbracedown) at (m.south east) {};
\draw [thick, decoration={brace, amplitude=5pt, raise=8pt}, decorate] (upbraceleft) -- (upbraceright)
   node [pos=0.5,anchor=south,yshift=11pt] {$\mu$ columns};
\draw [thick, decoration={brace, amplitude=5pt, raise=13pt}, decorate] (lbraceup) -- (lbracedown)
   node [pos=0.5,anchor=west,xshift=17pt] {$2m$ rows};
\begin{scope}[on background layer]
   \draw[->,line width=2.5mm,gray!15] (m-1-1.north) -- (m-4-1.south);
   \draw[->,line width=2.5mm,gray!15] (m-1-2.north) -- (m-4-2.south);
   \draw[->,line width=2.5mm,gray!15] (m-1-4.north) -- (m-4-4.south);
\end{scope}
\node[matrixbox, fit=(m-1-1.north west) (m-4-1.south west) (m-4-1.south east),draw] {};
\node[below, anchor=north, yshift=-5pt] at (m-4-1.south) {$=e_0$};
\node[matrixbox, fit=(m-1-2.north west) (m-4-2.south west) (m-4-2.south east),draw] {};
\node[below, anchor=north, yshift=-5pt] at (m-4-2.south) {$=e_1$};
\node[matrixbox, fit=(m-1-4.north west) (m-4-4.south west) (m-4-4.south east),draw] {};
\node[below, anchor=north, yshift=-5pt] at (m-4-4.south) {$=e_{\mu-1}$};
\end{tikzpicture}

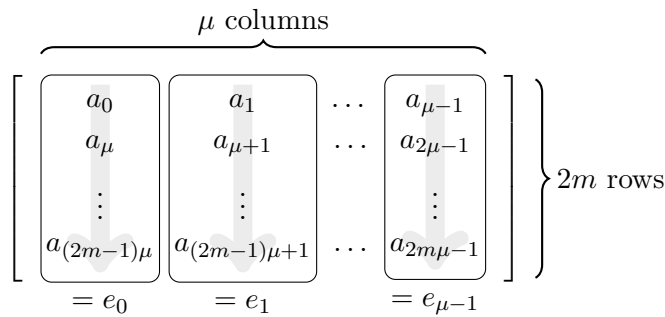
\captionof{figure}{Input vector $a$ written as $\mu$ column vectors of $2m$ elements}
\label{dkss:mot:rewrite}
\end{minipage}
\smallskip

We now define polynomials $\bar{a}_v(x)$, which are residues of
modular division%
. We will show that they
can be calculated by performing DFTs on the $e_\ell$.

Let $v \in [0 : 2m-1]$ and define polynomials $\bar{a}_v(x) \in \R[x]$ with degree-bound $\mu$ as
\[  \bar{a}_v(x) \coloneqq a(x) \bmod (x^{\mu} - \alpha^v).  \numberthis\label{dkss:abar}  \]

Denote $a_j \in \R$ the $j$-th coefficient of $a(x)$, let $\ell \in [0 : \mu-1]$ and define $e_\ell(y) \in \R[y]$ as
\[  e_\ell(y) \coloneqq \sum_{j=0}^{2m-1} a_{j\mu+\ell} \cdot y^j.  \label{defeell}\numberthis  \]
That is, the $j$-th coefficient of $e_\ell(y)$ is the $(j\mu+\ell)$-th coefficient of $a(x)$,
and $e_\ell(y)$ is a polynomial over $\R$ with degree-bound $2m$.

To calculate $\bar{a}_v(x)$, write it out:
\begin{align*}
\bar{a}_v(x)
    = {}&   a(x) \bmod (x^\mu - \alpha^v)   \\
    = {}&   (a_0 + a_1 x + \ldots + a_{\mu-1} x^{\mu-1} +   \\
      {}&   a_\mu x^{\mu} + a_{\mu+1} x^{\mu+1} + \ldots + a_{2\mu-1} x^{2\mu-1} +   \\
      {}&   a_{2\mu} x^{2\mu} + a_{2\mu+1} x^{2\mu+1} + \ldots + a_{3\mu-1} x^{3\mu-1} +   \\
      {}&   \ldots + a_{2M-1} x^{2M-1}) \bmod (x^\mu - \alpha^v).  \\
\intertext{Since $x^{\mu} \equiv \alpha^v \imod{x^{\mu} - \alpha^v}$, replace $x^{\mu}$ with $\alpha^v$ and get}
\bar{a}_v(x)
    = {}&   a_0 + a_1 x + \ldots + a_{\mu-1} x^{\mu-1} +   \\
      {}&   a_\mu \alpha^v + a_{\mu+1} \alpha^v x + \ldots + a_{2\mu-1} \alpha^v x^{\mu-1} +   \\
      {}&   a_{2\mu} \alpha^{2v} + a_{2\mu+1} \alpha^{2v} x + \ldots + a_{3\mu-1} \alpha^{2v} x^{\mu-1} +   \\
      {}&   \ldots + a_{2M-1} \alpha^{(2m-1)v} x^{\mu-1}.
\end{align*}

Denote $\bar{a}_{v,\ell}$ the $\ell$-th coefficient of $\bar{a}_v(x)$.
Adding up coefficients of matching powers of $x$ yields
\begin{align*}
\bar{a}_{v,\ell}  &= \sum_{j=0}^{2m-1} a_{j\mu + \ell} \cdot \alpha^{jv}.   \\
\intertext{Compare this to \eqref{defeell} to see that}
\bar{a}_{v,\ell}  &= e_\ell(\alpha^v).
\end{align*}
So to find the $\ell$-th coefficient of each $\bar{a}_v(x)$ we can perform a length-$2m$ DFT of $e_\ell(y)$,
using $\alpha$ as root of unity. Call these the \emph{inner DFTs}. If we let
$\ell$ run through its $\mu$ possible values, we get the coefficients of all $\bar{a}_v(x)$.
Figure~\ref{dkss:mot:after} shows the result of the inner DFTs.
\bigskip

\begin{minipage}{\linewidth}
\centering
\begin{tikzpicture}
\matrix [matrix of math nodes,left delimiter=[,right delimiter={]}] (m) {
   \bar{a}_{0,0}     & \bar{a}_{0,1}      & \ldots    & \bar{a}_{0,\mu-1}     \\
   \bar{a}_{1,0}     & \bar{a}_{1,1}      & \ldots    & \bar{a}_{1,\mu-1}     \\
   \vdots            & \vdots             & {}        & \vdots                \\
   \bar{a}_{2m-1,0}  & \bar{a}_{2m-1,1}   & \ldots    & \bar{a}_{2m-1,\mu-1}  \\
};
\coordinate (mleft1)  at ($(m.north west)!(m-1-1.west)!(m.south west)$);
\coordinate (mleft2)  at ($(m.north west)!(m-2-1.west)!(m.south west)$);
\coordinate (mleft4)  at ($(m.north west)!(m-4-1.west)!(m.south west)$);
\coordinate (mright1) at ($(m.north east)!(m-1-1.west)!(m.south east)$);
\coordinate (mright2) at ($(m.north east)!(m-2-1.west)!(m.south east)$);
\coordinate (mright4) at ($(m.north east)!(m-4-1.west)!(m.south east)$);
\coordinate (left1)  at ($(mleft1)!5pt!(mright1)$);
\coordinate (left2)  at ($(mleft2)!5pt!(mright2)$);
\coordinate (left4)  at ($(mleft4)!5pt!(mright4)$);
\coordinate (right1) at ($(mright1)!5pt!(mleft1)$);
\coordinate (right2) at ($(mright2)!5pt!(mleft2)$);
\coordinate (right4) at ($(mright4)!5pt!(mleft4)$);
\begin{scope}[on background layer]
   \draw[->,line width=2.5mm,gray!15] (left1) -- (right1);
   \draw[->,line width=2.5mm,gray!15] (left2) -- (right2);
   \draw[->,line width=2.5mm,gray!15] (left4) -- (right4);
\end{scope}
\node[matrixboxtwo, fit=(left1) (right1),draw] {};
\node[below, anchor=east,xshift=-12pt] at (left1) {$\bar{a}_0=$};
\node[matrixboxtwo, fit=(left2) (right2),draw] {};
\node[below, anchor=east,xshift=-12pt] at (left2) {$\bar{a}_1=$};
\node[matrixboxtwo, fit=(left4) (right4),draw] {};
\node[below, anchor=east,xshift=-12pt] at (left4) {$\bar{a}_{2m-1}=$};
\end{tikzpicture}
\captionof{figure}{Result of inner DFTs as $2m$ row vectors of $\mu$ elements}
\label{dkss:mot:after}
\end{minipage}
\smallskip

Multiplications by powers of $\alpha$ can be performed as cyclic shifts. Since $\alpha^m \equiv -1$,
coefficients of powers $\ge m$ wrap around with changed sign.
This works much in the same way
as the integer $2$ in Schönhage and Strassen's multiplication algorithm in Section \ref{smul}.

\item  
\label{badmuls}
Perform \emph{bad multiplications}.

What \DKSS{} call \emph{bad multiplications} is known as multiplications by twiddle factors
in the Cooley-Tukey FFT.

Our goal is to compute the DFT of $a(x)$ with $\rho$ as $2M$-th root of unity, that is,
to compute $a(\rho^i)$, for $i \in [0:2M-1]$.
Express $i$ as $i = 2m \cdot f + v$ with $f \in [0 : \mu-1]$ and $v \in [0 : 2m-1]$.
Then
\[  a(\rho^i) = a(\rho^{2m \cdot f + v}) = \bar{a}_v(\rho^{2m \cdot f + v}),  \numberthis\label{dkss:eq:bardash}  \]
because according to \eqref{dkss:abar}
\[  \bar{a}_v(\rho^{2m \cdot f + v}) = a(\rho^{2m \cdot f + v}) \bmod (\underbrace{(\rho^{2m \cdot f + v})^{\mu} - \alpha^v}_{\eqqcolon \xi})  \]
\begin{align*}
\text{with } \quad
\xi  &=  (\rho^{2m \cdot f + v})^{\mu} - \alpha^v  \\
     &=  (\underbrace{\rho^{2M}}_{=1})^f \cdot (\underbrace{\rho^{\mu}}_{=\alpha})^v - \alpha^v  \\
     &=  \alpha^v - \alpha^v  \\
     &=  0. \, \footnotemark
\end{align*}
We already computed the polynomials $\bar{a}_v(x)$ in Step \ref{innerdft} above.
In order to efficiently compute $\bar{a}_v(\rho^{2m \cdot f+v})$, we define
\[  \widetilde{a}_v(x) \coloneqq \bar{a}_v(x \cdot \rho^v),  \numberthis\label{dkss:eq:tildedef}\]
so that if
$\widetilde{a}_v(x)$ is evaluated at $x = \rho^{2m \cdot f}$ we get
$\widetilde{a}_v(\rho^{2m \cdot f}) = \bar{a}_v(\rho^{2m \cdot f + v})$.
\footnotetext{Since $a = b \imod c$ means $a-b = kc$, for some $k \in \Z$, $a = b \imod 0$ means equality.}

Computing $\widetilde{a}_v(x)$ can be done by computing its coefficients
$\widetilde{a}_{v,\ell} = \bar{a}_{v,\ell} \cdot \rho^{v \ell}$, with $\ell \in [0:\mu-1]$.
Since coefficients are themselves polynomials, use Kronecker-Schönhage substitution
as described in Section \ref{dkss:sec:compmul} to efficiently multiply them.

\item  
Perform \emph{outer DFTs}.

Now all that is left is to evaluate the $\widetilde{a}_v(x)$, $v \in [0 : 2m-1]$, at $x = \rho^{2m \cdot f}$,
for $f \in [0 : \mu-1]$.
In Step \ref{badmuls} we arranged $\widetilde{a}_v(x)$ in such a way that
this evaluation is nothing but a
length-$\mu$ DFT of $\widetilde{a}_v(x)$ with $\rho^{2m}$ as root of unity.
Call these the \emph{outer DFTs}. They are depicted in Figure~\ref{dkss:mot:outer}.
\bigskip

\begin{minipage}{\linewidth}
\centering
\begin{tikzpicture}
\matrix [matrix of math nodes,left delimiter=[,right delimiter={]}] (m) {
   \widetilde{a}_{0,0}     & \widetilde{a}_{0,1}      & \ldots    & \widetilde{a}_{0,\mu-1}     \\
   \widetilde{a}_{1,0}     & \widetilde{a}_{1,1}      & \ldots    & \widetilde{a}_{1,\mu-1}     \\
   \vdots            & \vdots             & {}        & \vdots                \\
   \widetilde{a}_{2m-1,0}  & \widetilde{a}_{2m-1,1}   & \ldots    & \widetilde{a}_{2m-1,\mu-1}  \\
};
\coordinate (mleft1)  at ($(m.north west)!(m-1-1.west)!(m.south west)$);
\coordinate (mleft2)  at ($(m.north west)!(m-2-1.west)!(m.south west)$);
\coordinate (mleft4)  at ($(m.north west)!(m-4-1.west)!(m.south west)$);
\coordinate (mright1) at ($(m.north east)!(m-1-1.west)!(m.south east)$);
\coordinate (mright2) at ($(m.north east)!(m-2-1.west)!(m.south east)$);
\coordinate (mright4) at ($(m.north east)!(m-4-1.west)!(m.south east)$);
\coordinate (left1)  at ($(mleft1)!5pt!(mright1)$);
\coordinate (left2)  at ($(mleft2)!5pt!(mright2)$);
\coordinate (left4)  at ($(mleft4)!5pt!(mright4)$);
\coordinate (right1) at ($(mright1)!5pt!(mleft1)$);
\coordinate (right2) at ($(mright2)!5pt!(mleft2)$);
\coordinate (right4) at ($(mright4)!5pt!(mleft4)$);
\begin{scope}[on background layer]
   \draw[->,line width=2.5mm,gray!15] (left1) -- (right1);
   \draw[->,line width=2.5mm,gray!15] (left2) -- (right2);
   \draw[->,line width=2.5mm,gray!15] (left4) -- (right4);
\end{scope}
\node[matrixboxtwo, fit=(left1) (right1),draw] {};
\node[below, anchor=east,xshift=-12pt] at (left1) {$\widetilde{a}_0=$};
\node[matrixboxtwo, fit=(left2) (right2),draw] {};
\node[below, anchor=east,xshift=-12pt] at (left2) {$\widetilde{a}_1=$};
\node[matrixboxtwo, fit=(left4) (right4),draw] {};
\node[below, anchor=east,xshift=-12pt] at (left4) {$\widetilde{a}_{2m-1}=$};
\end{tikzpicture}
\captionof{figure}{Outer DFTs on $2m$ row vectors of $\mu$ elements}
\label{dkss:mot:outer}
\end{minipage}
\smallskip

If $M \ge m$ this is done by a recursive call to the FFT routine and according to \eqref{dkss:eq:bardash} and \eqref{dkss:eq:tildedef}
computes $\widetilde{a}_v(\rho^{2m \cdot f}) = \bar{a}_v(\rho^{2m \cdot f + v}) = a(\rho^{2m \cdot f + v}) = a(\rho^i)$.

If $M < m$, just computing an inner DFT with $\alpha^{2m/2M}$ as $(2m/2M)$-th root of unity is sufficient.

\end{enumerate}

\subsection{Componentwise Multiplication}
\label{dkss:sec:compmul}  
Multiply coefficients $\widetilde{a}_i$ by $\widetilde{b}_i$ to compute $2M$ product coefficients $\widehat{c}_i \coloneqq \widehat{a}_i\widehat{b}_i$.
Since coefficients are from $\R$ and are thus themselves polynomials,
we use Kronecker-Schönhage substitution (cf.\ \cite[sec.~2]{Schoenhage1982}, \cite[sec.~1.3 \& 1.9]{Brent2011})
to multiply them and reduce polynomial multiplication to integer multiplication.
Then we can use the DKSS algorithm recursively.

\begin{mydef}[Kronecker-Schönhage substitution]
\hspace*{0pt}\refstepcounter{dummy}\label{kssubst}\index{Kronecker-Schönhage\enskip substitution}
Kronecker-Schönhage substitution reduces \linebreak polynomial multiplication to
integer multiplication.
Since $\R = \P[\alpha]/(\alpha^m+1)$ consists of polynomials with degree-bound $m$, whose
coefficients are in $\P = \Zpc$,
each coefficient can be stored in $d \coloneqq \lceil \log p^z \rceil$ bits.
Coefficients are to be multiplied, so $2d$~bits per coefficient product must be allocated
to prevent overflow.
Furthermore, multiplication of two polynomials with degree-bound $m$ leads
to $m$ summands for the middle coefficients,
thus another $\log m$ bits per coefficient are required.
\end{mydef}

This substitution converts elements
of $\R$ into integers that are $m(2d+\log m)$ bits long. Then these integers are multiplied
and from the result the product polynomial is recovered.

\subsection{Backwards FFT}
\label{dkss:sec:backwfft}  
The backwards FFT works exactly like the forward FFT described in Step~\ref{dkss:step:forwfft}.
We use in fact an inverse FFT and
reordering and scaling of the resulting coefficients is handled in the next step.

\subsection{Carry Propagation}
\label{dkss:sec:carryadd}  
In Step \ref{dkss:step:encode}, we encoded an input number $a$ into the polynomial $a(x)$ by
putting $um/2$~bits into each outer coefficient and from there distributing $u$ bits into each
of the $m/2$~lower inner coefficients.
When decoding the product polynomial $c(x)$ into the number $c$, we must use the same
weight as for encoding, so we evaluate the inner coefficients at $\alpha = 2^u$
and the outer coefficients at $x = 2^{um/2}$. Of course, on a binary computer this
evaluation can be done by bit-shifting and addition.

We must take the ordering of the resulting coefficients into account.  In Section \ref{fftpolymul}
we defined a backwards transform to get results that are properly ordered.
However, for simplicity of implementation, we use again a forward transform
and access its resulting coefficients in different order.

Furthermore, all result coefficients are scaled by a factor of $2M$, so we have to divide them
by $2M$ prior to addition.

\section{Run-time Analysis}
\label{dkss:runtime}

\subsection{Analysis of each Step}

Our goal is to find an upper bound to the bit complexity $T(N)$ needed to multiply two
nonnegative $N$-bit integers using the implementation of DKSS multiplication to get their $2N$-bit product.
We estimate the run-time of each step individually.

\begin{enumerate}
\item  
Choosing $M$ and $m$ does only depend on the length of the input and can be done in constant time.

\item  
Computing $u$ takes constant time as does finding $p$, since we precomputed all values for $p$
for the supported hardware. Thus, this step has cost $O(1)$ as well.

\item  
In this step we compute a $2M$-th root of unity $\rho \in \R$ from a known
generator $\zeta$ of $\mathbb{F}^*_p$.
$T_\P$ denotes the time to multiply two arbitrary numbers in $\P$.
First, we use Hensel lifting to calculate $\zeta_z$ in $z-1$ lifting steps.
In each step we have to calculate
\[  \zeta_{s+1} \coloneqq \zeta_s - (\zeta_s^{p-1} - 1) \cdot ((p-1)\zeta_s^{p-2})^{-1}.  \]
This can be done with 1 exponentiation, 3 multiplications, 4 subtractions and 1~modular inverse.

To exponentiate, we use \index{Binary exponentiation}\emph{binary exponentiation} \cite[ch.~4.6.3]{Knuth1997},
which requires $O(\log p)$ multiplications in $\P$, and to find the modular inverse
we use the \index{Extended Euclidean algorithm}\emph{extended Euclidean algorithm} \cite[ch.~4.5.2]{Knuth1997}
with $O(\log p^z)$ steps, where each step costs $O(\log p^z)$.
After lifting, we calculate $\omega = \zeta_z^h$, where $h < p / 2M$.

Together, the cost $T_\omega$ to calculate $\omega$ is
\begin{align*}
T_{\omega} = {} & (z-1) \big( O(\log p)T_\P + 3 T_\P + 4 O(\log p^z) + O(\log p^z)O(\log p^z) \big) \: +  \\
             {} & O(\log p)T_\P  \\
           = {} & O(\log p \cdot T_\P + \log^2 p^z).
\end{align*}

After that, we perform Lagrange interpolation: according to \eqref{dkss:lagrange}
it consists of $m$~additions of polynomials in $\R$,
each of which is computed by $m-1$ multiplications of degree-1 polynomials with polynomials in $\R$
plus $m-1$ modular inverses in $\P$.

Thus, the run-time for Lagrange interpolation is
\begin{align*}
T_{Lagrange}         &= m (mO(\log p^z) + (m-1) (2mT_\P + O(\log p^z \cdot \log p^z)))  \\
                     &= O(m^2 (\log p^z + mT_\P + \log^2 p^z))  \\
                     &= O(m^2 (mT_\P + \log^2 p^z)).
\end{align*}

Ordinary multiplication can multiply $n$-bit integers in
run-time $O(n^2)$, hence $T_\P$ can be bounded by $O(\log^2 p^z)$.
Using \eqref{dkss:peqo} we estimate $p^z = O(N^{Lz} / \log^{2Lz} N)$ and
recall that $m=O(\log N)$.
We get as total time to compute $\rho$:
\begin{align*}
T_{\rho}             &= T_\omega + T_{Lagrange}  \\
                     &= O \big( \log p \cdot T_\P + \log^2 p^z) \big) + O \big(m^2 (mT_\P + \log^2 p^z) \big)  \\
                     &= O \big( \log p \cdot O(\log^2 p^z) + \log^2 p^z + m^2 (mO(\log^2 p^z) + \log^2 p^z) \big)  \\
                     &= O \big( \log p \cdot \log^2 p^z + m^2 (m \log^2 p^z + \log^2 p^z) \big)  \\
                     &= O \big( (\log p + m^3) \log^2 p^z \big)  \\
                     &= O \big( (\log (N^{L} / \log^{2L} N) + \log^3 N) \log^2 (N^{Lz} / \log^{2Lz} N) \big)  \\
                     &= O ( \log^3 N \cdot \log^2 N )  \\
                     &= O ( \log^5 N ).
\end{align*}

\item  
Encoding input numbers as polynomials can be done in time proportional to the length of
the numbers, that is, in time $O(N)$.

\item  
As we will see, the FFT is one of the two most time-consuming steps; the other one being
the multiplication of sample values.
Let us first evaluate the run-time of a length-$2M$ FFT over $\R$, denoted by $T_D(2M)$.
We analyze the run-time of each step of the FFT individually. $T_\R$ denotes the time
needed to multiply two arbitrary elements of $\R$ and will be specified later.
\begin{enumerate}[label=\roman*.,ref=\theenumi.\roman*]

\item  
The first step performs $\mu = 2M/2m$ inner FFTs over $\R$ of length $2m$.
To calculate one DFT we need to perform $2m \log(2m)$ additions and $m \log(2m)$ multiplications by powers of $\alpha$,
cf.\ \eqref{fftcost}. A single addition costs $O(m \log p^z)$, since an element of $\R$ is
a polynomial over $\P=\Zpc$ with degree-bound $m$.
Since multiplication by a power of $\alpha$ can be done with a cyclic shift, its run-time
is of the same order as that of an addition.  So the run-time to compute one inner DFT is
\[ 3m \log(2m) \cdot O(m \log p^z) = O(m^2 \log m \cdot \log p^z), \]
and the run-time to compute all $2M/2m$ inner DFTs in this step is
\[ 2M/2m \cdot O(m^2 \log m \cdot \log p^z) = O(M m \log m \cdot \log p^z). \]

\item  
\label{dkss:runtime:badmuls}
Here, we prepare the $2m$ polynomials $\bar{a}_v(x)$ for the outer DFTs.
For each $v \in [0 : 2m-1]$, the polynomial $\bar{a}_v(x)$ has $\mu = 2M/2m$ coefficients,
which makes a total of $2M$ multiplications in $\R$ by powers of $\rho$ to compute all $\widetilde{a}_v(x)$.
The same number of multiplications is needed to compute the powers of $\rho$.
So this step has a total run-time of $4M \cdot T_\R$.

\item  
This last step computes $2m$ outer DFTs.
The FFT routine is invoked recursively to do this.
The total run-time for this step is the time for $2m$ DFTs of length $2M/2m$ and hence
\[ 2m \cdot T_D(2M/2m). \]
The recursion stops when the FFT length is $\le 2m$, that is, after $\log_{2m}(2M)$ levels.

\end{enumerate}

The total run-time $T_D(2M)$ of the FFT is
the sum of the run-times of all three steps, that is,
\begin{align*}
T_D(2M) &= O(M m \log m \cdot \log p^z) + 4M \cdot T_\R + 2m \cdot T_D(2M/2m)  \\
        &= \log_{2m}(2M) \cdot \big(O(M m \log m \cdot \log p^z) + 4M \cdot T_\R \big).  \numberthis \label{dkss:dftcost}
\end{align*}

\item  
Each of the $2M$ coefficient pairs $\widehat{a_i}$, $\widehat{b_i}$ can be multiplied in time $T_\R$.
Thus, the run-time for this step is $2M \cdot T_\R$.

\item  
The backwards FFT has the same cost as the forward FFT, see \eqref{dkss:dftcost}.

\item  
Decoding the polynomials back into integers and performing carry propagation can be done
with $2Mm$ additions of length $\log p^z$, hence with cost
\begin{align*}
T_{decode}        &= O(2Mm\log p^z)  \\
                  &= O \Big( \frac{N}{\log^2 N} \log N \cdot \log \frac{N^{Lz}}{\log^{2Lz} N} \Big)  \\
                  &= O \Big( \frac{N}{\log N} \log N^{Lz} \Big)  \\
                  &= O(N).
\end{align*}

\end{enumerate}

\subsection{Putting Everything Together}
\label{dkss:sec:together}

To conclude our evaluation of the run-time, we need to upper bound the value of $T_\R$,
the time needed to multiply two arbitrary elements of $\R$.
For that purpose, we use Kronecker-Schönhage substitution as described in Section~\ref{dkss:sec:compmul}.

\index{Kronecker-Schönhage\enskip substitution}
\begin{mythm}[Kronecker-Schönhage substitution]
Multiplication in $\R$ can be reduced to integer multiplication of length $O(\log^2 N)$ bits.
\end{mythm}

This substitution converts elements
of $\R$ into integers of $m(2d+\log m)$ bits, with $d=\lceil \log p^z \rceil$, multiplies the integers
and from the integer result recovers the product polynomial.
To see how large these integers get in terms of $N$, we use \eqref{dkss:peqo} and obtain
\begin{align*}
m(2d+\log m)   &= O(\log N \cdot (2 \lceil \log p^z \rceil + \log \log N))  \\
               &= O(\log N \cdot (\log p^z + \log \log N))  \\
               &= O(\log N \cdot (\log (N^{Lz} / \log^{2Lz} N) + \log \log N))  \\
               &= O(\log N \cdot Lz \log N)  \\
               &= O(\log^2 N).  \numberthis\label{dkss:kslen}
\end{align*}
$T(N)$ denotes the time to multiply two $N$-bit integers, so $T_\R = T(O(\log^2 N))$.  \texttombstone

Adding up the run-time estimates of all steps we get the total run-time $T(N)$ for DKSS multiplication:
\begin{align*}
T(N)  ={}& O(1) + O(1) + O(\log^5 N) + O(N) + 2T_D(2M) + 2M T_\R + T_D(2M) + O(N) \\
      ={}& 3 T_D(2M) + 2M T_\R + O(N)  \\
      ={}& 3 \big(\log_{2m}(2M) (O(M m \log m \cdot \log p^z) + 4M T_\R)\big) + 2M T_\R + O(N)  \\
      ={}& O \big(\log_{2m}(2M) (M m \log m \cdot \log p^z + M T_\R) + M T_\R + N \big)  \\
      ={}& O \big(M \log_{2m}(2M) (m \log m \cdot \log p^z + T(O(\log^2 N))) + M T(O(\log^2 N)) + N \big).  \\
\intertext{In terms of $N$ that is}
T(N)  ={}& O \Big(\frac{N}{\log^2 N} \frac{\log(2N/\log^2 N)}{\log(2 \log N)} \big(\log N \cdot \log \log N \cdot \log \Big( \frac{N^{Lz}}{\log^{2Lz} N} \Big) + {}  \\
         & T(O(\log^2 N)) \big) + \frac{N}{\log^2 N} T(O(\log^2 N)) + N \Big)  \\
      ={}& O \Big(\frac{N}{\log^2 N} \frac{\log N}{\log \log N} \big(\log N \cdot \log \log N \cdot \log N + T(O(\log^2 N)) \big) \: +  \\
         & \frac{N}{\log^2 N} T(O(\log^2 N)) + N \Big)  \\
      ={}& O \Big(N \log N + \frac{N}{\log N \cdot \log \log N} T(O(\log^2 N)) + \frac{N}{\log^2 N} T(O(\log^2 N)) + N \Big)  \\
      ={}& O \Big(N \log N + \frac{N}{\log N \cdot \log \log N} \cdot T(O(\log^2 N)) \Big).  \numberthis\label{dkss:costrec}
\end{align*}

\subsection{Resolving the Recursion}

To solve the recursion, we will need the following estimation.
Observe that for any real $x \ge 4$ it holds that
\[
\frac{\log (\lambda \log^2 x)}{\log \log x} =
\frac{\log (\lambda (\log x)^2)}{\log \log x} =
\frac{\log \lambda + 2 \log \log x}{\log \log x} \le
\log \lambda + 2.  \numberthis\label{dkss:est1}
\]
The following notation is introduced to abbreviate the upcoming nested logarithms:
define $f_{0}(N) \coloneqq N$ and $f_{i}(N) \coloneqq \lambda \log^2 f_{i-1}(N)$, for $i \in \N$ and some $\lambda$.
Furthermore, let $\tau \ge 4$ be the smallest length where the algorithm is used, otherwise a simpler algorithm is used.
Now we express the run-time from \eqref{dkss:costrec} with explicit constants,
assuming that $\lambda \log^2 N = f_{1}(N) \ge \tau$ and unroll the recursion once:
\begin{align*}
T(N) \le {}&   \mu \big(N \log N + \frac{N}{\log N \cdot \log \log N} T(\lambda \log^2 N) \big)  \\
     =   {}&   \mu N \log N \Big(1 + \frac{T(\lambda \log^2 N)}{\log^2 N \cdot \log \log N} \Big)  \\
     \le {}&   \mu N \log N \Big(1 + \frac{\mu (\lambda \log^2 N) \log (\lambda \log^2 N)}{\log^2 N \cdot \log \log N}(1 + \frac{T(\lambda \log^2 (\lambda \log^2 N))}{\log^2 (\lambda \log^2 N) \cdot \log \log (\lambda \log^2 N)}) \Big)  \\
     =   {}&   \mu N \log N \Big(1 + \frac{\mu \lambda \log (\lambda \log^2 N)}{\log \log N}(1 + \frac{T(\lambda \log^2 f_{1}(N))}{\log^2 f_{1}(N) \cdot \log \log f_{1}(N)}) \Big).  \\
\intertext{Using \eqref{dkss:est1} leads to}
T(N) \le {}&   \mu N \log N \Big(1 + \underbrace{\mu \lambda (\log \lambda + 2)}_{\eqqcolon \eta}(1 + \frac{T(\lambda \log^2 f_{1}(N))}{\log^2 f_{1}(N) \cdot \log \log f_{1}(N)}) \Big)  \\
     =   {}&   \mu N \log N \Big(1 + \eta + \eta \cdot \frac{T(\lambda \log^2 f_{1}(N))}{\log^2 f_{1}(N) \cdot \log \log f_{1}(N)} \Big).  \\
\intertext{Assuming $\lambda \log^2 f_{1}(N) = f_{2}(N) \ge \tau$ we unroll once more and get}
T(N) \le {}&   
               \mu N \log N \Big(1 + \eta + {} \\
           &   \eta \cdot \frac{\mu \lambda \log^2 f_{1}(N) \cdot \log (\lambda \log^2 f_{1}(N))}{\log^2 f_{1}(N) \cdot \log \log f_{1}(N)} (1 + \frac{T(\lambda \log^2 f_{2}(N))}{\log^2 f_{2}(N) \cdot \log \log f_{2}(N)}) \Big).  \\
\intertext{Again canceling out and using \eqref{dkss:est1} gives}
T(N) \le {}&   \mu N \log N \Big(1 + \eta + \eta \underbrace{\mu \lambda (\log \lambda + 2)}_{= \eta} (1 + \frac{T(\lambda \log^2 f_{2}(N))}{\log^2 f_{2}(N) \log \log f_{2}(N)}) \Big)  \\
     =   {}&   \mu N \log N \Big(1 + \eta + \eta^2 + \eta^2 \cdot \frac{T(\lambda \log^2 f_{2}(N))}{\log^2 f_{2}(N) \log \log f_{2}(N)} \Big)  \\
     =   {}&   \mu N \log N \Big(\sum_{i=0}^2 \eta^i + \eta^2 \cdot \frac{T(\lambda \log^2 f_{2}(N))}{\log^2 f_{2}(N) \log \log f_{2}(N)} \Big).  \\
\intertext{Obviously, after unrolling $j \in \N_0$ levels of recursion and assuming $f_{j}(N) \ge \tau$ we get}
T(N) \le {}&   \mu N \log N \Big(\sum_{i=0}^j \eta^i + \eta^j \cdot \frac{T(\lambda \log^2 f_{j}(N))}{\log^2 f_{j}(N) \log \log f_{j}(N)} \Big).  \numberthis\label{dkss:est2}
\end{align*}

\bigskip
The remaining question is now: how many levels of recursion are there for a given $N$?
To find out, we look for a lower bound for $N$ after $j$ levels of recursion.

Equation \eqref{dkss:est2} applies if $f_j(N) \ge \tau$. If $j \ge 1$ we can reduce $f_j(N)$ once and get
\begin{align*}
f_j(N)                     \ge  {}&   \tau  \\
\lambda \log^2 f_{j-1}(N)  \ge  {}&   \tau  \\
\log^2 f_{j-1}(N)          \ge  {}&   \tau / \lambda  \\
\log f_{j-1}(N)            \ge  {}&   \sqrt{\tau / \lambda}  \\
f_{j-1}(N)                 \ge  {}&   2^{\sqrt{\tau / \lambda}}.  \numberthis\label{dkss:est3a}  \\
\intertext{A second reduction works quite like the first, assuming $j \ge 2$:}
\lambda \log^2 f_{j-2}(N)  \ge  {}&   2^{\sqrt{\tau / \lambda}}  \\
\log f_{j-2}(N)            \ge  {}&   \sqrt{2^{\sqrt{\tau / \lambda}} / \lambda}  \\
f_{j-2}(N)                 \ge  {}&   2^{\sqrt{2^{\sqrt{\tau / \lambda}} / \lambda}}.
\end{align*}
Transforming the exponent we get
\[
\sqrt{2^{\sqrt{\tau/\lambda}} / \lambda} =
\sqrt{(2^{\sqrt{1 / \lambda}})^{\sqrt{\tau}} / \lambda} =
\sqrt{(2^{\sqrt{1 / \lambda}})^{\sqrt{\tau}}} / \sqrt{\lambda} =
{\underbrace{(2^{\sqrt{1 / \lambda}})^{\frac{1}{2}}}_{\eqqcolon \beta}}{}^{\cdot \sqrt{\tau}} / \sqrt{\lambda} =
\beta^{\sqrt{\tau}} / \sqrt{\lambda}.
\]
Now use that and reduce again, assuming $j \ge 3$:
\begin{align*}
f_{j-2}(N)                 \ge  {}&   2^{\beta^{\sqrt{\tau}} / \sqrt{\lambda}}  \\
\lambda \log^2 f_{j-3}(N)  \ge  {}&   2^{\beta^{\sqrt{\tau}} / \sqrt{\lambda}}  \\
f_{j-3}(N)                 \ge  {}&   2^{\sqrt{2^{\beta^{\sqrt{\tau}} / \sqrt{\lambda}} / \lambda}}.
\end{align*}
Transforming the exponent again gives
\[
\sqrt{2^{\beta^{\sqrt{\tau}} / \sqrt{\lambda}} / \lambda} =
\sqrt{(2^{\sqrt{1 / \lambda}})^{\beta^{\sqrt{\tau}}} / \lambda} =
\sqrt{(2^{\sqrt{1 / \lambda}})^{\beta^{\sqrt{\tau}}}} / \sqrt{\lambda} =
{\underbrace{(2^{\sqrt{1 / \lambda}})^{\frac{1}{2}}}_{=\beta}}{}^{\cdot \beta^{\sqrt{\tau}}} / \sqrt{\lambda} =
\beta^{\beta^{\sqrt{\tau}}} / \sqrt{\lambda},
\]
which yields
\begin{align*}
f_{j-3}(N)                 \ge  {}&   2^{\beta^{\beta^{\sqrt{\tau}}} / \sqrt{\lambda}}.  \numberthis\label{dkss:est3}
\end{align*}
So we see that with each unroll step of $f_j(N)$ we get another exponentiation by $\beta$
in the exponent.
\begin{mydef}[Iterated Exponentials]
\hspace*{0pt}\refstepcounter{dummy}\label{iterexp}\index{Iterated exponentials}%
Let $a$, $x \in \mathbb{R}$, $n \in \N_0$ and denote $\exp_a(x) = a^x$, then
\[
\exp_a^n(x) \coloneqq
\begin{dcases}
x                       & \text{if } n = 0 \\
\exp_a(\exp_a^{n-1}(x)) & \text{if } n > 0
\end{dcases}
\]
is called \emph{iterated exponentials} or \emph{power tower}. For example, $\exp_a^3(x) = a^{a^{a^x}}$.
This notation is inspired by Euler's $\exp(x)$ function and \emph{functional iteration} in \cite[p.~58]{Cormen2009}.
\end{mydef}
With the help of iterated exponentials we can write \eqref{dkss:est3} as
\begin{align*}
f_{j-3}(N)                 \ge  {}&   2^{\exp_{\beta}^2(\sqrt{\tau}) / \sqrt{\lambda}},  \\
\intertext{and if we reduce $f_j(N)$ fully we get}
N = f_{0}(N)               \ge  {}&   2^{\exp_{\beta}^{j-1}(\sqrt{\tau}) / \sqrt{\lambda}}.  \numberthis\label{dkss:est4}
\end{align*}
We are close to the goal, which we can attain with help of the following
\begin{mydef}[Iterated Logarithm {\cite[p.~58]{Cormen2009}}]
\hspace*{0pt}\refstepcounter{dummy}\label{logstar}\index{Iterated logarithm}%
Let $a$, $x \in \mathbb{R}_{>0}$, then the \emph{iterated logarithm} is defined as
\[
\log^*_a(x) \coloneqq
\begin{dcases}
0                       & \text{if } x \le 1 \\
\log^*_a(\log_a x) + 1  & \text{if } x > 1
\end{dcases}  \numberthis\label{logstareq}
\]
and is the inverse of $\exp_a^n(1)$, that is, $\log^*_a(\exp_a^n(1)) = n$.
The iterated logarithm is the number of $\log_a$-operations needed to
bring its argument to a value $\le 1$. As usual, $\log^* x \coloneqq \log^*_2 x$.
\end{mydef}
Now we use the iterated logarithm on \eqref{dkss:est4} and get
\begin{align*}
N                                                                       \ge {}&  2^{\exp_{\beta}^{j-1}(\sqrt{\tau}) / \sqrt{\lambda}}  \\
\log N                                                                  \ge {}&  \exp_{\beta}^{j-1}(\sqrt{\tau}) / \sqrt{\lambda}      \\
\sqrt{\lambda} \log N                                                   \ge {}&  \exp_{\beta}^{j-1}(\sqrt{\tau})                       \\
\log^*_{\beta}(\sqrt{\lambda} \log N)                                   \ge {}&  j-1 + \log^*_{\beta}\sqrt{\tau}                       \\
\log^*_{\beta}(\sqrt{\lambda} \log N) + 1 - \log^*_{\beta}\sqrt{\tau}   \ge {}&  j.  \numberthis\label{dkss:est5}
\end{align*}
We can replace $\log^*_{\beta} x$ by $O(\log^*_2 x)$. To see why, observe that if
$\beta$ could be expressed as some power tower of 2, say, $\beta = 2^{2^{2}}$, that is, $\log^* \beta = 3$, then a power tower of $\beta$ is less
than one of 2 with thrice the length, because $\beta^\beta = (2^{2^2})^{\beta} < 2^{2^{(2^{\beta})}}$.
Hence, $\log^*_{\beta} x \le \log^* x \cdot \log^* \beta = O(\log^* x)$,
since $\beta$ is constant.

Since only $N$ is a variable, this finally leads to the estimate
\begin{align*}
j  &\le  \log^*_{\beta}(\sqrt{\lambda} \log N) + 1 - \log^*_{\beta}\sqrt{\tau}  \\
   &= O(  \log^*_{\beta}(\sqrt{\lambda} \log N))  \\
   &=O(  \log^*_{\beta} N)  \\
   &=  O(\log^* N).  \numberthis\label{dkss:est6}
\end{align*}
Now, we can pick up \eqref{dkss:est2} again. We assume $\eta \not= 1$,
$f_j(N) \ge \tau$, but $f_{j+1}(N) < \tau$ and hence in analogy to \eqref{dkss:est3a},
$f_j(N) < 2^{\sqrt{\tau / \lambda}}$. Then we get
\begin{align*}
T(N) \le {}&   \mu N \log N \Big(\sum_{i=0}^j \eta^i + \eta^j \cdot \frac{T(\lambda \log^2 f_{j}(N))}{\log^2 f_{j}(N) \log \log f_{j}(N)} \Big)  \\
     \le {}&   \mu N \log N \Big(\frac{\eta^{j+1} - 1}{\eta - 1} + \eta^j \cdot \frac{T(f_{j+1}(N))}{\log^2 f_{j}(N) \log \log f_{j}(N)} \Big).  \\
\intertext{Capturing the constants into Big-O's yields}
T(N) =   {}&   \mu N \log N (O(\eta^{j+1}) + O(\eta^j) )  \\
T(N) =   {}&   O(N \log N \cdot \eta^{j+1})  \\
     =   {}&   N \log N \cdot \eta^{O(\log^* N)}.  \\
\intertext{Expressing $\eta$ and the constant from $O(\ldots)$ as $2^\kappa$, for some constant $\kappa$, we write}
T(N) =   {}&   N \log N \cdot (2^\kappa)^{O(\log^* N)}  \\
     =   {}&   N \cdot \log N \cdot 2^{O(\log^* N)}.  \numberthis\label{dkss:est9}
\end{align*}

\section{Differences to DKSS Paper}
\label{dkss:diffs}

The intention of this thesis is to assess the speed of an implementation of DKSS multiplication
on a modern computer.  Its architecture imposes certain limits on its software. For example,
the amount of memory that can be addressed is limited by the size of the processor's index registers.
A more compelling limit is that the universe contains only a finite amount of matter and energy,
as far as we know.  A computer will need at least one electron per bit and thus,
even if we could harness all (dark) matter and energy for memory bits, any storable number
could surely not exceed $2^{300}$ bits in length.

Another limit creeps in with the speed of the machine: there is no practical use to provide a solution
that will run several thousand years or more to complete.  An estimation of the run-time to multiply
numbers with $2^{65}$ bits leads to a minimum of 7000 years on the test machine.

This led me to assume a maximum length of input numbers. Since the implementation runs
on a 64-bit CPU, the number's length is de facto limited to $8 \cdot 2^{64} / 4 = 2^{65}$ bits.
And since the length is limited, we can precompute some constants needed in the algorithm,
namely the prime $p$ and a generator $\zeta$ of $\mathbb{F}^*_p$.
I did this for values of $p$ with 2 to 1704~bits in length.

\DKSS{} went to great lengths to show that suitable primes $p$ can be found at run-time
and to make their construction work, they use $p^z$ as modulus, $z > 1$, as we have seen in
Sections \ref{dkss:sec:findp} and \ref{dkss:sec:findrho}.

Furthermore, they encode input numbers as $k$-variate polynomials, where the degree
in each variable is $< 2M$.  That is, outer polynomials are in $\R[X_1, \ldots, X_k]$.
When it comes to the FFT, they fix one variable, say $X_k$, and treat the outer
polynomials as univariate polynomials over $\S \coloneqq \R[X_1, \ldots, X_{k-1}]$.
Note that $\rho$ is a principal $2M$-th root of unity in $\S$ as well.
Then they perform FFT multiplication of a univariate polynomial over $\S$.
The componentwise multiplication uses FFT multiplication recursively, because now
two $(k-1)$-variate polynomials have to be multiplied.

Since the only need for $k$-variate polynomials was to show that $p$ can be found at
run-time, I was able to use $k=1$ and use univariate polynomials in the implementation. Furthermore,
it was easy to precompute $p$ to greater sizes, so there was no need for $z > 1$
and thus I dropped Hensel lifting to find $\zeta_z$ as well.

\bigskip
I changed some variable names from \cite{De2013} to avoid confusion with other
variables of the same name or to improve clarity.  If the reader is familiar with the original paper,
here is a small overview of changed names:

\begin{center}
\begin{tabular}{ l || c | c }
Description                                     & DKSS paper   & This thesis \\
\hline
Exponent of prime $p$ in modulus                & $c$          & $z$  \\
Number of variables for outer polynomials       & $k$          & (dropped, $k=1$)  \\
Factor in progression for finding prime $p$     & $i$          & $h$  \\
Residue polynomials in DFT                      & $a_j$        & $\bar{a}_v$  \\
Index variable in DFT                           & $k$          & $f$  \\
Radix of FFT                                    & $2M/2m$      & $\mu$  \\
\end{tabular}
\end{center}

\chapter{Implementation of DKSS Multiplication}
\label{chapter4}
\fancyhead[RE,LO]{Chapter 4. \emph{Implementation of DKSS Multiplication}}
\fancyhead[LE,RO]{\thepage}
\index{DKSS multiplication!implementation (\DMUL{})}

In this chapter my implementation of DKSS multiplication is presented.
Parameter selection is discussed and exemplary source code is shown,
together with a description of tests performed to assert the software's correctness.
Then, measured execution time, memory requirements and source code size is examined.
I discuss the results of profiling and lastly,
extrapolate run-time for increasing input lengths.

\section{Parameter Selection}
\label{dkss:impl:param}

The description of parameter selection in Section \ref{dkss:desc}
leaves some freedom on how exactly to calculate
$M$, $m$, $u$ and $p$. Recall that we are performing FFTs of polynomials
with degree-bound $2M$ in $\R[x]$, where $\R = \P[\alpha]/(\alpha^m+1)$ and $\P=\Zpc$.
We call coefficients in $\R[x]$ \emph{outer coefficients} and coefficients in $\P[\alpha]$
\emph{inner coefficients}. Both input numbers have $N$ bits, parameter $u$ is the number of bits
of the input number that go into each inner coefficient and $z$ is constant.

I aimed at a monotonically increasing graph of execution time, that is, growing input lengths
lead to growing execution times.  Parameter selection that leads to a
rough graph suggests that better parameters could be selected.

This led me to choose the prime $p$ first.
Section \ref{dkss:sec:findp} mentions lower bounds for $p^z$. Recall that $M \approx N / \log^2 N$
and $m \approx \log N$. I use
\[  p^z \ge \half Mm 2^{2u} \approx \half N^5 / \log N.  \numberthis\label{tightpzbound}  \]
Furthermore, I decided to round up the number of bits of $p$ to the
next multiple of the word size.  Since both allocated memory and
cost of division (for modular reductions) depend on the number of
words, it seemed prudent to make the most out of it.  Benchmarks show that
this was a good choice, see Figure~\ref{dkssmulgraph} for a graph of timings.

DKSS multiplication uses $\P = \Zpc$ with $z > 1$ to lower run-time in the asymptotic
case by lowering the upper bound for finding the prime $p$. But that doesn't apply
here, since the machine
this implementation runs on enforces upper limits of the length of numbers.
So despite the description of the process of Hensel lifting in Section \ref{dkss:sec:findrho},
I did not implement it, because precomputation of larger prime
numbers was the easier choice (see Linnik's Theorem on page~\pageref{zsix}).
Furthermore, the special build of Proth prime numbers could be exploited in the future
to speed up modular reductions.

Having chosen $p$, I then select the largest $u$ that is able to hold the whole $2N$ bits of the
result.
It follows from \eqref{tightpzbound} that $\log (p^z) \ge \log (Mm) + 2u - 1$.  Since $\log(p^z)$
is chosen first, I try to maximize $u$.  The larger $u$ is, the less
coefficients are needed. After finding an $u$ that fits, I try to minimize
the product $Mm$, because the smaller $Mm$ is, the smaller the FFT length and the
memory requirements are.

Lastly, I set $M$ and $m$ and try to maintain the quotient $M/m \approx N / \log^3 N$
that follows from the description in Section \ref{dkss:sec:choosem}.
On the other hand, factors can be moved around between $M$ and $m$,
since in selection of $u$ and $p$ only the product $Mm$
is needed.  I did some short tests on selecting
$M/m \approx k \cdot N / \log^3 N$ for some $k$, but it seemed that $k = 1$
was overall a good choice.

\section{A Look at the Code}
\label{dkss:lookcode}

If the parameters are given (namely $M$, $m$, $u$, $p^z$ and $\rho$),
the main idea of DKSS multiplication lies in the structure of the ring $\R$
and the way the DFT is computed: inner DFTs, bad multiplications and
outer DFTs.

To give an impression of the implementation, following is the FFT main routine.
Language
constructs (like templates and \code{typedef}s), debug code and assertions were stripped
to improve readability. As mentioned in Section \ref{memmgt}, \code{tape_alloc}
is a stack-like memory allocator. It takes the number of \code{word}s requested as argument.

\begin{lstlisting}[language=C++]
void dkss_fft(
   word* a,             // input vector
   unsigned M,
   unsigned m,
   unsigned oclen,      // outer coeff length = m * iclen
   unsigned iclen,      // inner coeff length >= bit_length(p^z) / bits(word)
   word* pz,            // modulus p^z
   word* rho_pow,       // powers [0 : m-1] of \rho
   unsigned base_pow)   // quotient of order of top level \rho and now
{
   if (M <= m) {        // use inner DFT right away
      tape_alloc tmp(oclen);              // allocate some memory from the "tape"
      word* t = tmp.p;                    // length oclen
      unsigned log_M = int_log2(M);
      fft_shuffle(a, log_M + 1, oclen);   // pre-shuffle the values
      dkss_inner_fft_eval(a, M, m, oclen, iclen, pz, t);
      return;
   }

   unsigned mu = M / m;                      // \mu = 2M/2m
   unsigned log2m = int_log2(m) + 1;
   tape_alloc tmp(2*M * oclen + 2*m * oclen + oclen);
   word* abar = tmp.p;                       // length 2*M*oclen, \bar{a}
   word* el = abar + 2*M * oclen;            // length 2*m*oclen, e_\ell
   word* t = el + 2*m*oclen;                 // length oclen, temp storage

   // perform inner DFTs
   // i guess it's better to copy elements instead of using pointers and work
   // in-place, because this way cache thrashing can only occur once when
   // copying and not on every access.
   for (unsigned l=0; l<mu; ++l) {           // cycle through all values for l
      // assemble e_l(y):
      // the j-th coeff of e_l(y) is the (j*mu+l)-th coeff of a(x)
      // for the FFT evaluation, we assemble e_l(y) already in shuffled order
      word* a_jxl = a + l*oclen;             // points to a_l
      for (unsigned j=0; j<2*m; ++j) {
         word* el_j = el + bit_rev(j, log2m) * oclen;
         copy(el_j, a_jxl, oclen);
         a_jxl += mu * oclen;                // point to next a_{j*\mu+l}
      }

      // perform inner DFT on e_l(y) with alpha as 2m-th root of unity
      dkss_inner_fft_eval(el, m, m, oclen, iclen, pz, t);

      // l-th coeffs of all a_v(x) is e_l(alpha^v), i.e. v-th coeff of DFT(e_l)
      // this copies transformed elements back into place
      word* el_v = el;
      word* abar_vl = abar + l*oclen;
      for (unsigned v=0; v<2*m; ++v) {
         copy(abar_vl, el_v, oclen);
         el_v += oclen;
         abar_vl += mu * oclen;
      }
   }

   // perform bad muls and outer DFTs
   word* abar_v = abar;
   word* rho_vl = t;                         // just for the name
   const index top_mu = mu * base_pow;       // top level mu
   unsigned psh = int_log2(top_mu);          // shift count
   // cycle through all a_v to perform bad muls and outer DFTs
   for (unsigned v=0; v<2*m; ++v) {
      // skip first loop iteration: v == 0, i.e. abar_{v,l} *= rho^0 = 1
      word* abar_vl = abar_v;
      unsigned vlbase = 0;
      for (unsigned l=1; l<mu; ++l) {        // cycle thru all values for l
         vlbase += v * base_pow;
         abar_vl += oclen;
         unsigned pi = vlbase & ((1 << psh) - 1);     // vlbase % top_mu
         unsigned pe = vlbase >> psh;                 // vlbase / top_mu
         // select right rho_pow and do cyclic shift
         modpoly_mul_xpow_mod_mp1(rho_vl, rho_pow + pi*oclen, pe, m, iclen, pz);
         // abar_{v,l} *= rho^{vl}
         modpoly_mul_mod_mp1(abar_vl, abar_vl, rho_vl, m, iclen, pz);
      }

      // now abar_v contains \tilde{a}_v.  ready to do outer DFT: recursive call
      dkss_fft(abar_v, mu/2, m, oclen, iclen, pz, rho_pow, base_pow * 2*m);

      // copy back to 'a' array
      word* a_fxv = a + v * oclen;
      word* abar_vf = abar_v;
      for (unsigned f=0; f<mu; ++f) {
         copy(a_fxv, abar_vf, oclen);
         abar_vf += oclen;
         a_fxv += 2*m * oclen;
      }

      abar_v += mu * oclen;
   }
}
\end{lstlisting}

The listing shows one of the few optimizations I was able to implement:
in the run-time analysis in Section \ref{dkss:runtime}, Step \ref{dkss:runtime:badmuls}
we counted $2M$ multiplications by powers of $\rho$ and another $2M$ multiplications
to compute those powers.
I was able to reduce the number of multiplications for the latter from $2M$ to $\mu = 2M/2m$.

I used the fact that $\rho^{2M/2m} = \rho^{\mu} = \alpha$:
if $i \in [0 : 2M-1]$, set $r \coloneqq \lfloor i / \mu \rfloor$ and $s \coloneqq i \bmod \mu$
and thus $i = r \mu + s$.

Therefore it holds that $\rho^i = \rho^{r \mu + s} = \rho^{\mu \cdot r} \rho^s = \alpha^r \rho^s$.
We can obtain $\rho^i$ with an additional cyclic shift by precomputing all $\rho^s$, $s \in [0 : \mu-1]$.
In benchmarks, this almost halved the run-time.

In the above listing function \code{dkss_inner_fft_eval()} is called.
This function doesn't differ much from the \QMUL{} FFT evaluate function \code{qmul_evaluate()}
on page~\pageref{qmullisting}, except that this time functions instead of operators are used
to add and subtract elements, and multiplications by powers of the root of unity
are done by cyclic shifts.
Following is the listing of \code{dkss_inner_fft_eval()}:

\lstset{aboveskip=0.5cm}
\begin{lstlisting}[language=C++]
void dkss_inner_fft_eval(
   word* e,                // input vector
   unsigned n_half,        // half of FFT length
   unsigned m,
   unsigned oclen,         // outer coeff length = m * iclen
   unsigned iclen,         // inner coeff length >= bit_length(pz) / bits(word)
   word* pz,               // p^z
   word* t)                // temp storage
{
   if (n_half == 1) {
      // lowest layer: butterfly of two outer coeffs,
      // i.e. add and sub of two inner polynomials
      word* e2 = e + oclen;                  // second inner polynomial
      copy(t, e2, oclen);
      modpoly_sub(e2, e, t, m, iclen, pz);   // e2 = e - t
      modpoly_add(e, e, t, m, iclen, pz);    // e = e + t
      return;
   }

   dkss_inner_fft_eval(e, n_half/2, m, oclen, iclen, pz, t);
   dkss_inner_fft_eval(e + n_half*oclen, n_half/2, m, oclen, iclen, pz, t);

   unsigned inc = m / n_half;                // increment for each loop
   word* e1 = e;                             // first inner polynomial
   word* e2 = e + n_half*oclen;              // second inner polynomial
   unsigned pow = 0;
   for (unsigned i=0; i<n_half; ++i) {
      // w = omega_n^i, t = w*e2
      modpoly_mul_xpow_mod_mp1(t, e2, pow, m, iclen, pz);  // cyclic shift by pow
      modpoly_sub(e2, e1, t, m, iclen, pz);  // e2 = e1 - t
      modpoly_add(e1, e1, t, m, iclen, pz);  // e1 = e1 + t
      e1 += oclen;
      e2 += oclen;
      pow += inc;
   }
}
\end{lstlisting}

\section{Asserting the Code's Correctness}

Development included writing a lot of test code.  Every
major function has some \emph{unit tests} following it.  The unit tests usually contain
fixed data to be processed by the function to be tested
and compare its output to results that are known to be
correct, since they were computed by other means: Python
programs were used to compute the correct results for FFTs in polynomial quotient
rings, a different method for multiplication was used to test DKSS multiplication,
and sometimes the correct results were more or less obvious and could be hard-coded by hand.

\bigskip
Additionally, functions contain \emph{assertions} (like {\cpp}'s \code{assert()}),
which are assumptions that are written together with the (proper) code
and are checked at run-time. Often, these are pre- and post-conditions of functions.
Some \code{assert}s call functions that were solely written for use in assertions,
like a test for primitivity of a root.

To have the best of both worlds, code can be compiled in \emph{Debug} or
\emph{Release mode} with Visual Studio.  Release builds have all
\code{assert}s disabled and are compiled with optimizations for maximum speed, while Debug builds
feature assertion checking, but code optimization is disabled to aid debugging.
Test code is usually run in Debug mode, while benchmarks are run in Release mode.

\bigskip
Furthermore, after development of DKSS multiplication was completed, it was
integrated into my framework of long integer routines that is maintained
as a private project.  This framework is used
for primality testing of Mersenne numbers \index{Mersenne numbers}%
(numbers of the form $2^p-1$).
Of course, it can not compare
to the \emph{Great Internet Mersenne Prime Search} \cite{GIMPS},
the distributed effort to find new Mersenne prime numbers that is going
on since 1996 and has found the last eleven record prime numbers.

Nevertheless, I have been checking Mersenne numbers
for primality for over two years now and a database exists of the low
64~bits of the result (called the \emph{residue}) for each Mersenne number.
The primality test used for Mersenne numbers is the \emph{Lucas-Lehmer test}\index{Lucas-Lehmer test}
\cite[ch.~4.2.1]{Crandall2005}. It consists of
a loop of a long integer square, a subtraction by~2 and a modular reduction.
The nature of this test causes even single-bit errors to proliferate, so any error
would most likely alter the residue as well.
Since it is hard to test all code paths with unit tests
this makes it a good way to test a multiplication routine.

As a \emph{system test} DKSS multiplication was used in Mersenne number primality tests and
its results were compared against existing results.
The first 35 Mersenne primes (the largest being $2^{\num{1398269}}-1$) were
correctly identified as such.  Furthermore, all Mersenne numbers $2^p-1$ with $p < \num{120607}$
and various other sizes were tested and the residues matched.

\section{Execution Time}
\label{dkss:impl:exectime}

Our main interest is to find out how fast DKSS multiplication is in comparison
to other, well established algorithms.  Except for small and medium lengths,
Schönhage-Strassen multiplication was the fastest algorithm that used all-integer
methods in practice so far.  I compare both implementations \DMUL{}
and \SMUL{} to one another.

Figure~\ref{dkssmulgraph} shows graphs of \DMUL{} and \SMUL{} execution time
(and Figure~\ref{exectimes} shows some of the raw data).
The cyan-colored and the magenta-colored graph show
execution time if $p$ was not rounded up to the next
multiple of the word size, and if $p$ was in fact rounded up, respectively (cf.~Section \ref{dkss:impl:param}).
It is always faster to use a rounded up $p$ than to use the ``original'' value.

\begin{figure}
\bigskip
\centering

\begin{tikzpicture}
\begin{loglogaxis}[
   width=13cm,
   axis y line*=left,
   xlabel={Input words},
   ylabel={Execution cycles},
   legend pos=north west]
   xmin=100, xmax=100000000,
   ymin=10000, ymax=55000000000000,

\pgfplotsset{cycle list shift=3}

\addplot[color=green] table[x index=0, y index=1] {smul-speed-short.txt};
\addlegendentry{\SMUL{}}
\addplot[mark=|,color=cyan] table[x index=0, y index=1] {dkssmul-speed-allp.txt};
\addlegendentry{\DMUL{} all $p$}
\addplot[mark=diamond,color=magenta] table[x index=0, y index=1] {dkssmul-speed.txt};
\addlegendentry{\DMUL{}}

\end{loglogaxis}
\begin{loglogaxis}[
   width=13cm,
   axis y line*=right,
   hide x axis,
   xmin=100, xmax=100000000,
   ymin=0.0000029, ymax=16176,
   ylabel={Execution time [s]},
]
\end{loglogaxis}

\end{tikzpicture}

\caption{Execution time of \DMUL{}}
\label{dkssmulgraph}
\end{figure}

As can be seen clearly, \DMUL{} is much slower (about 30~times) than \SMUL{} (printed in green)
over the whole range of tested input lengths.
From this graph it is hard to see if \DMUL{} is gaining on \SMUL{}.
Section~\ref{dkss:impl:augur} discusses the quotient of run-times and the location
of a crossover point in detail.

The stair-like graph stems from the fact that execution time almost totally
depends on the FFT length $2M$ and the size of elements of $\R = \P / (\alpha^m+1)$
with $\P = \Zpc$.
Since both $M$ and $m$ are powers of 2, many different
input lengths lead to the same set of parameters.

The graph shows that execution time is almost the same for the beginning and the
end of each step of the stair. The only part that depends directly on $N$ is the
encoding of the input numbers and decoding into the resulting product.
But the time needed to do the FFT clearly dominates overall execution time.

In contrast to \DMUL{}, the \SMUL{} execution time graph is much smoother. In fact,
it is reproduced without marks that would otherwise only obscure the graph,
because there are a total of \num{12969}~data points available, of which
465~representative points are shown.

Obviously, \DMUL{} parameter selection could be improved, since sometimes larger input
numbers lead to faster execution times.  Either, more research on parameter
selection or a calibration process should smooth this out.

\section{Memory Requirements}
\label{dkss:impl:memory}

\DMUL{} memory requirements are dominated by three times the size of the polynomials:
input $a(x)$ and $b(x) \in \R[x]$ and the $\bar{a}_v(x)$.
The result $c(x)$ requires no further memory, since storage
of one of the input polynomials can be reused. An improved implementation could
save the $\bar{a}_v(x)$ directly back into the polynomial without need for temporary storage,
thus saving one third of memory requirements. To accomplish that a fast matrix transposition
\index{Matrix transposition}%
is needed, which in itself is not trivial (cf.~\cite[exercise 1.3.3-12]{Knuth1997a}).

The polynomials each have $2M$ coefficients in $\R = \P[\alpha] / (\alpha^m+1)$,
where $\P = \Zpc$.  Hence, each polynomial needs $2Mm \lceil \log p^z \rceil$ bits.
With $M \approx N / \log^2 N$, $m \approx \log N$ and $p^z \approx \half N^5 / \log N$
(see \eqref{tightpzbound}) that results in
\begin{align*}
2Mm \lceil \log p^z \rceil   & \approx   2 N / \log^2 N \cdot \log N \cdot \log(\half N^5 / \log N)   \\
               & =         2 N / \log N \cdot (-1 + 5 \log N - \log \log N)   \\
               & \approx   10 N.
\end{align*}
The listing of function \code{dkss_fft()} in Section \ref{dkss:lookcode} shows
that more memory, namely another $(2m + 1) \cdot \lceil \log p^z \rceil \approx 10 \log^2 N$ bits,
is allocated, but compared to $10 N$ bits for each polynomial that is of no big
consequence.  The same applies to the $2M/2m$ precomputed powers of $\rho$,
each with a length
of $m \lceil \log p^z \rceil$ bits. Together, they only need \mbox{$2M/2m \cdot m \lceil \log p^z \rceil = M \lceil \log p^z \rceil$}~bits,
that is, a $2m$-th part of the memory of one polynomial.
Hence, if both input numbers have $N$ bits, total memory needed by \DMUL{} is
\[  M_{\DMUL{}}(N) \approx 30 N \text{ bits}.  \]

Let us now compare the memory requirements of \DMUL{} to \SMUL{}. According to \eqref{smul:mem},
$M_{\SMUL{}}(N') = 4N'$ bits. I wrote ``$N'$'', since in Chapter \ref{smul} ``$N$'' describes
the length of the \emph{product}, hence
$N' = 2N$ to adjust notation to this chapter. Ergo, the approximate amount of
temporary memory for \SMUL{} is $M_{\SMUL{}}(N) = 4N' = 8N$ bits.

\begin{figure}
\centering
\small
\begin{tabular}{ r | r | r | r | r | r }
       Input length &       \DMUL{} memory &\DMUL{} &       \SMUL{} memory &\SMUL{} &      Q   \\
            (words) &              (bytes) &blow-up &              (bytes) &blow-up &          \\
\hline
     \num{    3648} &   \num{     803584 } &  27.54 &   \num{     251848 } &   8.63 &   3.19   \\
     \num{    7168} &   \num{    1623040 } &  28.30 &   \num{     501704 } &   8.75 &   3.24   \\
     \num{   14336} &   \num{    3228672 } &  28.15 &   \num{     962728 } &   8.39 &   3.35   \\
     \num{   28160} &   \num{    6439936 } &  28.59 &   \num{    1855528 } &   8.24 &   3.47   \\
     \num{   56320} &   \num{   12862464 } &  28.55 &   \num{    3693672 } &   8.20 &   3.48   \\
     \num{  110592} &   \num{   25707520 } &  29.06 &   \num{    7240136 } &   8.18 &   3.55   \\
     \num{  221184} &   \num{   51422464 } &  29.06 &   \num{   14331384 } &   8.10 &   3.59   \\
     \num{  434176} &   \num{  102819072 } &  29.60 &   \num{   28372232 } &   8.17 &   3.62   \\
     \num{  868352} &   \num{  205612288 } &  29.60 &   \num{   56716552 } &   8.16 &   3.63   \\
     \num{ 1703936} &   \num{  406915072 } &  29.85 &   \num{  111269224 } &   8.16 &   3.66   \\
     \num{ 2752512} &   \num{  616798080 } &  28.01 &   \num{  178538880 } &   8.11 &   3.45   \\
     \num{ 5505024} &   \num{ 1233557376 } &  28.01 &   \num{  361056896 } &   8.20 &   3.42   \\
     \num{10878976} &   \num{ 2467113216 } &  28.35 &   \num{  705184592 } &   8.10 &   3.50   \\
     \num{21757952} &   \num{ 4934174976 } &  28.35 &   \num{ 1477143728 } &   8.49 &   3.34   \\
     \num{42991616} &   \num{ 9765277440 } &  28.39 &   \num{ 2819507024 } &   8.20 &   3.46   \\
     \num{85983232} &   \num{19530403584 } &  28.39 &   \num{ 5638486864 } &   8.20 &   3.46   \\
\end{tabular}
\caption{Memory requirements of \DMUL{} and \SMUL{}}
\label{compmem}
\end{figure}

Figure \ref{compmem} shows an overview of actual memory consumption for selected input sizes
for both \DMUL{} and \SMUL{}.
The lengths chosen are the most favorable lengths for \DMUL{}.
At those lengths, the coefficients of the polynomials in \DMUL{} are fully filled
with bits from input numbers $a$ and~$b$ (as much as possible, as the
upper half of each polynomial still has to be zero to leave room for the product).
Increasing the lengths by one would lead to the least favorable lengths that
need about double the memory for almost the same input length.

The column ``\DMUL{} blow-up'' shows the quotient of \DMUL{}
memory requirements and the size of \emph{one} input factor in bytes.
The column ``\SMUL{} blow-up'' shows the same quotient for \SMUL{}.
The column ``Q'' shows the quotient of \DMUL{} and \SMUL{} memory requirements.
Column ``\DMUL{} blow-up'' nicely fits the approximated memory of $30 N$ as well
as column ``\SMUL{} blow-up'' supports the approximated memory requirements of $8 N$.

\section{Source Code Size}
\label{dkss:impl:codesize}

Given the description of the DKSS algorithm in Chapter \ref{chapter3}, the implementation
is relatively straight-forward. About one third of the newly written code is needed for
performing polynomial arithmetic: addition, subtraction, comparison, cyclic shifting and output
and furthermore, using Kronecker-Schönhage substitution, multiplication, squaring and exponentiation.
The other two thirds are taken up by the core DKSS routines, code to compute the primes $p$
and other supporting code.

Underlying the DKSS code are routines that had to be written,
but are not otherwise mentioned here, since they are not an immediate part of DKSS multiplication, like:
factoring of long integers into primes and Lucas primality test \cite[sec.~4.1]{Crandall2005}
(for the computation of primes $p$ for rings $\P$),
extended Euclidean algorithm (to compute modular inverses in Hensel lifting and Lagrange interpolation),
a {\cpp} class for long numbers (to handle non-time-critical calculations easily),
a faster division with remainder (see \cite[ch.~4.3.1, p.~272]{Knuth1997} and \cite[ch.~1.4.1]{Brent2011}).
Other code that was used had already been written before: basic arithmetic, benchmarking code for speed tests,
the Lucas-Lehmer test for primality for Mersenne numbers and a database of Mersenne number primality test results.

To give an idea about the size of the source code of DKSS multiplication,
the following table shows the counts of lines of code.
The second column (``Total source lines'') contains the count including
test and debug code, assertions, comments and empty lines, while the third column
excludes those and only counts lines of code that actually do work in a production version (``Pure code lines'').
The big difference in numbers is mostly because of test code.
The above mentioned underlying routines are not included in the counts.

\begin{center}
\begin{tabular}{ p{7cm} || r | r }
Description                                     & Total source lines    & Pure code lines \\
\hline
Polynomial arithmetic                           & 958                   & 295       \\
Core DKSS multiplication                        & 1374                  & 336       \\
Precomputation of primes $p$                    & 139                   & 86        \\
Other supporting code                           & 279                   & 157       \\
\hline
Total program code                              & 2750                  & 874       \\
Table of precomputed primes $p$                 & 1707                  & 1705      \\
\hline
Total                                           & 4457                  & 2579      \\
\end{tabular}
\end{center}

``Table of precomputed primes $p$'' contains an array of 1703 prime numbers of the form $h \cdot 2^n + 1$
for each bit length from 2 to 1704, with the smallest odd  $h$.  Data from this array is needed
for \DMUL{}, but it doesn't really qualify as code, because it's only a list of constants.
Since only values for $p$ are used that are a multiple of 64~bits long and input numbers are limited
by the 64-bit address space of the CPU, a list with 6 values
for $p$ would have done as well.

\pagebreak 
Compare this to the line counts of the implementation of Schönhage-Strassen multiplication:

\begin{center}
\begin{tabular}{ p{7cm} || r | r }
Description                                     & Total source lines    & Pure code lines \\
\hline
Core \SMUL{} multiplication                     & 805                   & 323       \\
Fast cyclic shifts                              & 518                   & 253       \\
Other supporting code                           & 414                   & 237       \\
\hline
Total                                           & 1737                  & 813       \\
\end{tabular}
\end{center}

The row ``Fast cyclic shifts'' shows a special feature of the \SMUL{} implementation: I went to
great lengths to write fast cyclic shift code that takes advantage of different shift counts
(like word- or byte-aligned). The original function for cyclic shifts had only 4~lines!

\section{Profiling}
\label{dkss:impl:profile}

To get a feeling for which parts of \DMUL{} use up the most computing time,
I did some \emph{profiling} of the code.  Visual Studio's built-in profiling
did not perform very accurately and I had some technical difficulties.  So instead
I used a small self-made solution: I timed the execution of certain code
parts manually.

This is not a thorough investigation, but just serves to gain a better
understanding where \emph{hot spots} of execution lie.  Thus, I have
chosen just five different input lengths for measurement.

\bigskip
In a first run, I measured the execution times for FFT setup (precomputation
of $\rho$ and its powers), the time needed for all three FFTs, pointwise multiplications
and encode/decode/normalization of the result.

\begin{figure}[h]
\centering
\small
\begin{tabular}{ r | r | r | r | r }
Input length         & FFT setup & \code{dkss_fft()} & Pointwise       & En/decode \&   \\
(words)              &           &                   & multiplications & normalize      \\
\hline
\num{    3648}       & 18.00 \%  & 58.60 \%          & 16.55 \%        & 6.85 \%        \\
\num{   28160}       &  2.26 \%  & 79.46 \%          & 12.88 \%        & 5.40 \%        \\
\num{  221184}       &  0.66 \%  & 84.40 \%          & 10.53 \%        & 4.41 \%        \\
\num{10878976}       &  0.39 \%  & 88.56 \%          & 8.19 \%         & 2.86 \%        \\
\num{42991616}       &  0.27 \%  & 87.71 \%          & 9.32 \%         & 2.70 \%        \\
\end{tabular}
\caption{Profiling percentages for \DMUL{}}
\label{profall}
\end{figure}

Figure \ref{profall} shows the results.  I only present percentages of execution time.
From this table several conclusions can be drawn:
\begin{itemize}
\item Computation of $\rho$ and its powers, something which has to be done before
the FFT starts, takes a diminishing share of time as the input gets longer. When numbers are
in the millions of words long, it doesn't carry any significant weight in the overall run-time.
This was to be expected.

\item The same holds in principle for encoding, decoding and normalizing of the polynomials.
It's more expensive than computing $\rho$ and its powers,
but with a decreasing share of the total cost. This too, was
to be expected.

\item Even the pointwise multiplications seem to be getting less prominent in the
overall cost. Maybe this shows that parameters could be selected better? More research
is needed here.

\item The one part which is taking a growing share of the total cost is the DKSS FFT itself.
I cannot assess from this data whether the share will be ever growing or reaches a plateau.
Still, most of the execution time is spent here, so this is why we look more closely into
its run-time.
\end{itemize}

In Figure \ref{proffft} we see the percentages of execution time that are needed by
the constituent parts of the DKSS FFT. It is performed by computing inner DFTs,
bad multiplications and outer DFTs, which for their part are
calculated by recursively calling the FFT routine
and therefore again calculating inner DFTs and bad multiplications. The respective columns
contain the execution time summed up over all levels of recursion. This table is
normalized, so that total time of \code{dkss_fft()} is 100~\%.

\begin{figure}[h]
\centering
\small
\begin{tabular}{ r | r | r | r }
Input length (words) & Inner FFT   & Bad multiplications & Rest      \\
\hline
\num{    3648}       & 22.24 \%    & 76.09 \%            & 1.67 \%   \\
\num{   28160}       & 16.98 \%    & 80.90 \%            & 2.12 \%   \\
\num{  221184}       & 16.20 \%    & 81.23 \%            & 2.57 \%   \\
\num{10878976}       & 10.21 \%    & 87.66 \%            & 2.13 \%   \\
\num{42991616}       &  9.50 \%    & 89.36 \%            & 1.14 \%   \\
\end{tabular}
\caption{Profiling percentages for \code{dkss_fft()}}
\label{proffft}
\end{figure}

The column titled ``Rest'' contains some call overhead, the copying of
$a_{j\mu+\ell}$ into $e_\ell$ and the copy back of the $\bar{a}_v$ coefficients into
the $a$ array. I suspected that cache thrashing would slow this process down a lot,
but these results show that this is not the case.

From this more specific analysis we learn that most of the time in \code{dkss_fft()}
is used up by bad multiplications and their share is growing. That sure is a hot spot.
So we will have a look into bad multiplications, which are multiplications
of two arbitrary elements of $\R$.

\bigskip
Figure \ref{profbadmuls} shows a breakdown of execution time for multiplications
of elements of $\R$. Multiplications are done by Kronecker-Schönhage substitution:
encode polynomials as integers, multiply the integers, decode them back to polynomials,
perform the ``wrap around'', that is, the modulo $(\alpha^m+1)$ operation, and
perform the modulo $p^z$ operation on the inner coefficients.
Again, total time of bad multiplications was normalized to 100~\%.

\begin{figure}[h]
\centering
\small
\begin{tabular}{ r | r | r || r | r | r }
Input length    & $m$ & Words per        & Integer        & Modular    & Rest       \\
(words)         &     & inner coefficient& multiplication & reduction  &            \\
\hline
\num{    3648}  &  16 & 2                & 47.33 \%       & 38.92 \%   & 13.75 \%   \\
\num{   28160}  &  16 & 2                & 46.72 \%       & 39.69 \%   & 13.59 \%   \\
\num{  221184}  &  16 & 2                & 46.57 \%       & 39.80 \%   & 13.63 \%   \\
\num{10878976}  &  16 & 3                & 57.37 \%       & 33.14 \%   &  9.49 \%   \\
\num{42991616}  &  32 & 3                & 66.48 \%       & 26.67 \%   &  6.84 \%   \\
\end{tabular}
\caption{Profiling percentages for bad multiplications}
\label{profbadmuls}
\end{figure}

Since Kronecker-Schönhage substitution depends on $\R$,
it only depends on parameters $m$ and $p^z$, but not $M$ nor $u$.
The first three rows have the same values for $m$ and $p^z$, so it fits the theory well
that the percentages are more or less the same.

Time needed for modular reduction is not negligible and a better means
than modulo division might save some time here (\index{Proth prime}\emph{Fast mod operation for Proth moduli},
\cite[p.~457]{Crandall2005}). But the trend seems to be that for growing lengths
the share of execution time needed for modular reductions is shrinking.

In column ``Rest'' the times for encoding and decoding between polynomials
and integers are lumped together.
This seems to be quite slow and a more careful implementation could speed it up,
but again, that percentage will only drop as
input numbers get longer.

From this profiling analysis we have learned that bad multiplications are really bad!
Up to 90~\% of execution time is spent there and its share is growing.
In order to reduce overall execution time,
we should reduce the number of bad multiplications and/or make them cheaper.
Maybe better parameter selection could reduce execution time here, which is left
open for future research.

\section{Gazing into the Crystal Ball}
\label{dkss:impl:augur}

One goal of this thesis is to compare the speed of a \DMUL{} implementation with
an \SMUL{} implementation.
As was described in Section~\ref{dkss:impl:exectime}, \SMUL{} is still much faster for
the lengths tested.

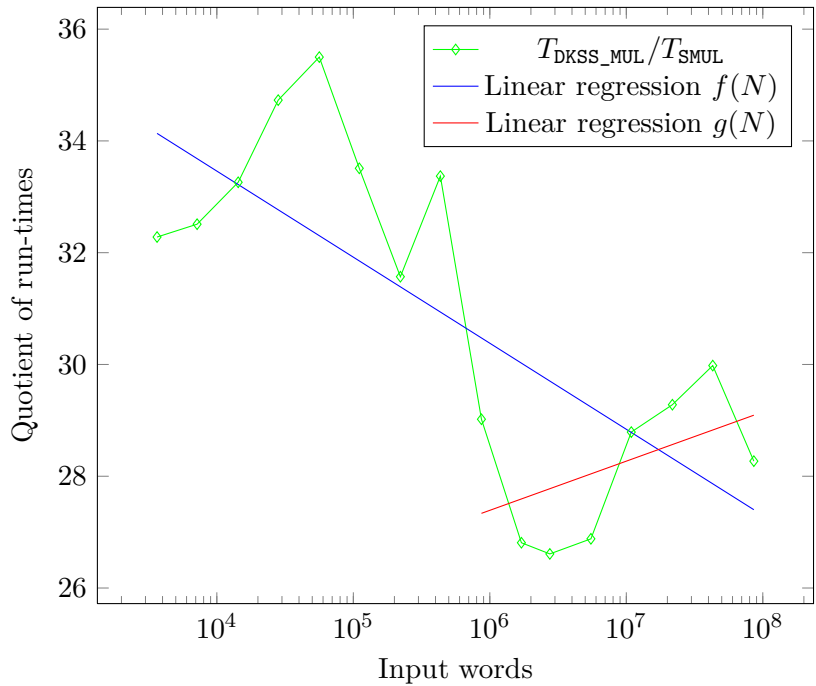
\begin{figure}
\bigskip
\centering

\begin{tikzpicture}
\begin{semilogxaxis}[
   width=11cm,
   xlabel={Input words},
   ylabel={Quotient of run-times},
   legend pos=north east]

\pgfplotsset{cycle list shift=3}

\addplot[mark=diamond,color=green] table[x index=0, y index=3] {dkssmul-quotient.txt};
\addlegendentry{$T_{\DMUL{}} / T_{\SMUL{}}$}
\addplot[domain=3648:85983232, blue] { -1.54*ln(x)/ln(10) + 39.62 };  
\addlegendentry{Linear regression $f(N)$}
\addplot[domain=868352:85983232, red] { 0.88*ln(x)/ln(10) + 22.11 };  
\addlegendentry{Linear regression $g(N)$}

\end{semilogxaxis}
\end{tikzpicture}

\caption{Quotient of \DMUL{} and \SMUL{} run-times vs.\ input length}
\label{quotientgraph}
\end{figure}

In addition, it would be interesting to estimate the input length where \DMUL{} starts to be
faster than \SMUL{}.
To do that, we look again at the most favorable lengths for \DMUL{},
that is, the lower right points of the steps in Figure~\ref{dkssmulgraph},
where the execution time graph for \DMUL{}  is nearest to the \SMUL{} graph.
Figure~\ref{exectimes} lists execution times at those points and the quotient of these times.
Figure~\ref{quotientgraph} shows a graph of the quotient of
execution times vs.\ input length.

\begin{figure}
\centering
\small
\begin{tabular}{ r | r | r | r | r | r }
Length               &          \DMUL{} time &   \DMUL{} &          \SMUL{} time &  \SMUL{} & Quotient \\
(words)              &              (cycles) & (min:sec) &              (cycles) &(min:sec) &          \\
\hline
     \num{    3648}  &  \num{    133948102}  &  0:00.039 &    \num{     4149866} & 0:00.001 &   32.28  \\
     \num{    7168}  &  \num{    288821718}  &  0:00.085 &    \num{     8884604} & 0:00.003 &   32.51  \\
     \num{   14336}  &  \num{    636214972}  &  0:00.187 &    \num{    19131288} & 0:00.006 &   33.26  \\
     \num{   28160}  &  \num{   1373645624}  &  0:00.404 &    \num{    39547108} & 0:00.012 &   34.73  \\
     \num{   56320}  &  \num{   2908271180}  &  0:00.855 &    \num{    81912772} & 0:00.024 &   35.50  \\
     \num{  110592}  &  \num{   6013189608}  &  0:01.769 &    \num{   179448020} & 0:00.053 &   33.51  \\
     \num{  221184}  &  \num{  13430829526}  &  0:03.950 &    \num{   425460492} & 0:00.125 &   31.57  \\
     \num{  434176}  &  \num{  29461464342}  &  0:08.665 &    \num{   882781300} & 0:00.260 &   33.37  \\
     \num{  868352}  &  \num{  62917787338}  &  0:18.505 &    \num{  2167722116} & 0:00.638 &   29.02  \\
     \num{ 1703936}  &  \num{ 122680187946}  &  0:36.082 &    \num{  4576352552} & 0:01.346 &   26.81  \\
     \num{ 2752512}  &  \num{ 199424397176}  &  0:58.654 &    \num{  7495493476} & 0:02.205 &   26.61  \\
     \num{ 5505024}  &  \num{ 410390455672}  &  2:00.703 &    \num{ 15269441152} & 0:04.491 &   26.88  \\
     \num{10878976}  &  \num{ 892949727060}  &  4:22.632 &    \num{ 31013681856} & 0:09.122 &   28.79  \\
     \num{21757952}  &  \num{1917703330120}  &  9:24.030 &    \num{ 65485660216} & 0:19.260 &   29.28  \\
     \num{42991616}  &  \num{3965210546518}  & 19:26.238 &    \num{132248494436} & 0:38.897 &   29.98  \\
     \num{85983232}  &  \num{8145120758260}  & 39:55.624 &    \num{288089862672} & 1:24.732 &   28.27  \\
\end{tabular}
\caption{Execution times of \DMUL{} and \SMUL{}}
\label{exectimes}
\end{figure}

At first sight, there is an apparent trend in the quotient of execution times.
Looking at Figure~\ref{quotientgraph} we might, as a first approximation, assume a
linear relationship between $\log_{10} N$ and the quotient of execution times.
Linear regression with a least squares estimation leads to the line
$f(N) = -1.54 \cdot \log_{10} N + 39.62$, which has a cor\-re\-la\-tion coefficient of $-0.723$.
Solving $f(N)=1$ leads to $N \approx 10^{25} \approx 2^{83}$ bits.

On the other hand, analysis of \SMUL{} execution times in Section \ref{smul:speed}
showed that \SMUL{} reaches its ``final'' speed only above input lengths of about 512~Kwords
(cf.~Figure~\ref{smulcgraph}). So
it seems that Figure~\ref{quotientgraph} not so much shows the speed-up through improved
speed of \DMUL{}, but the slow-down of \SMUL{} because of diminishing positive effects
of caching. If we do linear regression with data points starting at input length
512~Kwords only, we get $g(N) = 0.88 \cdot \log_{10} N + 22.11$, that is, the quotient
would be growing!  Obviously, this type of analysis is not very reliable.

\bigskip
As we did for \SMUL{} in Section~\ref{smul:speed}, we can use the measured data points to try to model
the run-time of \DMUL{}. Writing \eqref{dkss:est9} with an explicit constant we get
\[  T_\delta(N) \le N \cdot \log N \cdot 2^{\delta \cdot \log^* N}.  \numberthis\label{dkss:modeltime}\]
Calculating the constant $\delta$ from each data point and plotting all of them
gives the graph in Figure~\ref{dkss:dmulcgraph}, with average $\delta \approx \avgdmulc$.
In contrast to Figure~\ref{smulcgraph}, no effect of caching is apparent.
We only have few data points, so this model of run-time is not very resilient.
Yet, the modeled run-time matches the measured values $\pm 10~\%$ and even within $\pm 5~\%$ for
input lengths $\ge \num{28160}$ words.
With the few data points we have, it seems to be the best we can do.

\pagebreak 
Taking \eqref{smul:modeltime} and \eqref{dkss:modeltime} we can solve
\begin{align*}
T_\delta(N)                         & \le    T_\sigma(N)   \\
N \log N \cdot 2^{\delta \log^* N}  & \le    \sigma N \log N \cdot \log \log N   \\
2^{\delta \log^* N}                 & \le    \sigma \log \log N   \\
\delta \log^* N                     & \le    \log \sigma + \log \log \log N.   \\
\intertext{For large $N$ we substitute $\nu \coloneqq \log \log N$ and with
\eqref{logstareq} get}
\delta (\log^*(\log \log N) + 2)    & \le    \log \sigma + \log \log \log N   \\
\delta (\log^* \nu + 2)             & \le    \log \sigma + \log \nu.   \numberthis\label{dsmulcross}
\end{align*}
Solving \eqref{dsmulcross} numerically yields the enormous solution of $\nu \ge \eventwoexpexp$ and hence
$N \ge 10^{10^{\scriptstyle\evenexpexp}}$ bits!
An optimistic estimation of the number of bits for computer memory available in this universe
is $10^{100}$. So this crossover point is \emph{orders of orders}
of magnitude higher than any machine could hold that anyone could ever build.

Even if \DMUL{} was only about 2~times slower than \SMUL{}, the crossover point would still be
\label{unreachable}at $N \approx 10^{300}$ bits and thus unreachable.

\begin{figure}
\bigskip
\centering

\begin{tikzpicture}
\begin{semilogxaxis}[
   width=11cm,
   xlabel={Input bits},
   ylabel={Constant $\delta$},
   legend pos=north east]

\addplot[mark=diamond,color=magenta] table[x index=0, y index=1] {dmulc-graph.txt};
\addlegendentry{\DMUL{} constant $\delta$}
\node[pin=87:Level~2 cache size,draw=black] at (axis cs:74348,1.277) {};
\node[pin={[pin distance=1.5cm]272:Level~3 cache size},draw=black] at (axis cs:2347583,1.259) {};
\addplot[domain=29696:5502926848, blue] { \avgdmulc };  
\addlegendentry{average $\delta$}

\end{semilogxaxis}
\end{tikzpicture}

\caption{\DMUL{} constant $\delta$}
\label{dkss:dmulcgraph}
\end{figure}

\FloatBarrier

\chapter{Conclusion}
\label{chapter5}
\fancyhead[RE,LO]{Chapter 5. \emph{Conclusion}}
\fancyhead[LE,RO]{\thepage}

\DKSS{} describe a new procedure to multiply very large integers efficiently
(cf.\ Chapter~\ref{chapter3}, implemented as \DMUL{}).
The currently widely used all-integer multiplication algorithm for large numbers is by Schön\-hage and Stras\-sen \cite{SS1971}
(my implementation is called \SMUL{}, cf.\ Section~\ref{smul}).
The run-time of \DMUL{} is in a better complexity class than that of \SMUL{},
meaning that if input numbers are long enough, \DMUL{} will be faster than \SMUL{}.
Both algorithms were implemented and their run-time (Section~\ref{dkss:impl:exectime})
and memory consumption (Section~\ref{dkss:impl:memory}) were compared
(on a PC with 32~GB memory and a 3.4~GHz processor).

The results indicate that Schönhage and Strassen's
multiplication algorithm is the better choice for a variety of reasons:
\begin{enumerate}
\item \SMUL{} is faster than \DMUL{}.

Benchmarks show that \SMUL{} is still
about 26~to~36 times faster than \DMUL{} (Section \ref{dkss:impl:exectime} and
especially Figures~\ref{dkssmulgraph} and~\ref{quotientgraph}).
The estimate of the input length at which \DMUL{} is faster than \SMUL{}
(Section~\ref{dkss:impl:augur}) is $N \ge 10^{10^{\scriptstyle\evenexpexp}}$ bits
(which is larger than googolplex), but even if
\SMUL{} was only 2~times faster than \DMUL{}, the crossover point would be so large
that it could never be reached.

\item \SMUL{} requires less memory than \DMUL{}.

If both input numbers are $N$ bits long, \DMUL{}
requires about $30N$ bits of temporary memory, where \SMUL{} requires only about $8N$ bits
(Sections~\ref{dkss:impl:memory} and~\ref{smul:runtime}).
The memory requirements of \SMUL{} can not be lowered significantly, but
there is an obvious possibility to lower \DMUL{} memory
consumption to its lower limit of about $20N$ bits that was not implemented.

\item \SMUL{} is easier to implement than \DMUL{}.

A simple implementation of \SMUL{} needs about 550~lines of
{\cpp} code, where \DMUL{} requires about 900~lines plus at least 6~lines of constants and
more supporting routines, see Section~\ref{dkss:impl:codesize}. An improved and faster version of \SMUL{}
requires about 800~lines of code.
\end{enumerate}

It should be mentioned here that the \SMUL{} implementation is better optimized
than \DMUL{}.  The reason for that is that Schönhage-Strassen multiplication is now studied and in wide use
for many years and its details are well understood. I have spent considerable time
to improve its implementation.  In contrast, DKSS multiplication is still quite young and to my knowledge
this is the first implementation of it.  Section \ref{conc:outlook} describes
several possible improvements to \DMUL{} that could be realized.  Still, in my appraisal
none of them has the potential to speed up \DMUL{} so much that it becomes faster
than \SMUL{} in the range of input lengths that was examined here or even in ranges
that might be accessible in the future.

\section{Outlook}
\label{conc:outlook}

In the course of writing, I encountered several possible areas for improvement.
I list them here and try
to assess their potential to improve run-time.
\begin{itemize}
\item Find optimum values of parameters $M$, $m$, $u$ and $p^z$ for any given $N$.

Figure~\ref{dkssmulgraph} still shows some
areas where longer input numbers lead to shorter execution times.
Furthermore, Section~\ref{dkss:impl:profile} shows some developments in
percentages of run-times that could suggest that a better choice of parameters is possible.
More research is needed to understand how to choose the fastest set of parameters.

\item Cache computation of $\rho$ and its powers.

This is an obvious possibility to save execution time, but it cannot save a great share
when numbers get longer. Figure~\ref{profall} shows
how the percentage of execution time of ``FFT setup'' diminishes as numbers get longer.
This has no potential to lower the crossover point.

\item Add support for ``sparse integers'' in the underlying multiplication.

\index{Kronecker-Schönhage\enskip substitution}%
\DMUL{} reduces multiplication of long integers to multiplications in $\R$, a
polynomial ring. When it comes to multiplication of two elements of $\R$, they
are again converted to integers (via Kronecker-Schönhage substitution,
see Section~\ref{dkss:sec:compmul}) and have
to be padded with zeros.  About half of the words of each
factor are zero and a future multiplication routine could exploit
that. Profiling in Section~\ref{dkss:impl:profile} showed that up to 85~\%
of execution time is spent with multiplication of elements of $\R$ and a rising
percentage of that is used by the underlying integer multiplication.
I optimistically estimate the potential of this idea to speed up \DMUL{}
to be almost a factor of 2.

\item Count the number of non-zero coefficients in Kronecker-Schönhage substitution.

We have to pad the polynomial coefficients for Kronecker-Schönhage substitution (cf.\ Section~\ref{dkss:sec:compmul})
with zeros, partly because multiple coefficient products are summed up and we
must prevent that sum from overflowing. By counting the number of non-zero coefficients
prior to multiplying them, we could upper bound the number of products.
I estimate one or two bits of padding per coefficient product could be saved, but since
coefficients are themselves at least 64~bits long, their product is at least 128~bits,
so the potential saving can be no more than about 1--2~\% and shrinks when
numbers and thus coefficients get longer.

\item Implement \code{dkss_fft()} with less extra memory but matrix transposition instead.

\index{Matrix transposition}%
This is definitely an improvement that should be implemented, because it brings
down the memory requirements from about $30 N$ bits to about $20 N$ bits (cf.\ Section~\ref{dkss:impl:memory}).
Yet, from the numbers obtained by profiling, I estimate the potential saving
in run-time to be only a few percent at best. Furthermore, it seems that
efficient matrix transposition by itself is non-trivial.

\item Exploit the build of Proth prime numbers $p$.

\index{Proth prime}%
The modulus of $\P$ is a prime number of the form $h \cdot 2M + 1$,
where $h$ is a small positive odd integer and $M$ is a power of~2.
Maybe modular reductions can be sped up by the technique listed in
\cite[p.~457]{Crandall2005}. This has the potential to save a great part of the
cost of modular reductions, which showed to cost about 22~\% of run-time in profiling.

\end{itemize}

If all potential savings listed above could be achieved, this would speed up \DMUL{}
by a factor of about~2.5. Not included in this factor is a better parameter selection.
But even if that and other, yet unthought-of, im\-prove\-ments lead to another speed-up by a factor of~2,
\DMUL{} would still be at least 4.8~times slower than \SMUL{} and need about
2.5~times more memory. As explained on page \pageref{unreachable}, even then the
crossover point could never be reached.

\appendix 

\chapter{Technicalities}
\label{tech}
\fancyhead[RE,LO]{Appendix A. \emph{Technicalities}}
\fancyhead[LE,RO]{\thepage}

Tests and benchmarks were run on a machine with an Intel Core i7-3770 processor
(Ivy Bridge microarchitecture) with
3.40 GHz clock rate.  Hyper-threading, enhanced SpeedStep and Turbo Boost were disabled
to enhance accuracy of timings.
The mainboard is an ASUS P8Z77-V with 32~GB PC-1600 dual channel DDR3 memory.

The CPU has four cores, of which only one core was used while benchmarking. That is,
the other cores were not switched off, but no other CPU-intensive
process was running, except for the operating system itself.
To improve cache performance, the process affinity was fixed to processor~2, which seems to get less
interrupt and DPC load than processor~0.

The CPU has level~1 caches per core of both 32~KB for data and 32~KB for instructions,
unified level~2 caches of 256~KB per core and a unified level~3 cache of 8~MB for all cores.
Caches lines are 64~bytes long and all caches are 8-way set associate, except the level~3 cache,
which is 16-way set associative.

The operating system used was Windows~7 Ultimate with Service Pack~1 in 64-bit mode.

For benchmarking, the priority class of the process was set to the highest non-realtime value, that is, \code{HIGH_PRIORITY_CLASS}.
The thread priority was also the highest non-realtime value, \code{THREAD_PRIORITY_HIGHEST}.
Together, that results in a base priority level of~13.

Timings were taken by use of Windows' \code{QueryThreadCycleTime()} function that counts only
CPU cycles spent by the thread in question.
It queries the CPU's \emph{Time Stamp Counter} (TSC) and its resolution is extremely good:
even though the CPU instruction \code{RDTSC} is not serializing
(so some machine language instructions might be executed out-of-order), the
accuracy should be of the order of a 100 cycles at worst, most likely better.

As development environment Microsoft's Visual Studio 2012, v11.0.61030.00 Update~4 was used
which includes the \cpp{} compiler v17.00.61030. Code was compiled with options \code{/Ox}
(full optimization), \code{/Ob2} (expand any suitable inline function), \code{/Oi}
(enable intrinsic functions), \code{/Ot} (favor fast code) and \code{/GL}
(whole program optimization).
\backmatter

\label{Bibliography}
\fancyhead[RE,LO]{\emph{Bibliography}}
\fancyhead[LE,RO]{\thepage}
\bibliographystyle{myalpha}  
\bibliography{chris}  
\cleardoublepage

\label{Index}
\fancyhead[RE,LO]{\emph{Index}}
\fancyhead[LE,RO]{\thepage}
\printindex

\end{document}